\documentclass[aps,prc,twocolumn,floatfix,showpacs,nofootinbib,longbibliography]{revtex4-1}

\usepackage{graphicx,amsfonts,amssymb,amsbsy}
\usepackage{amsmath,amsthm,latexsym}
\usepackage{pstricks}
\usepackage{float}
\usepackage{ulem}

\usepackage[colorlinks]{hyperref}
\hypersetup{linkcolor=red,citecolor=blue,urlcolor=blue}

\begin{document}

\title{Electroweak decay of quark matter within dense astrophysical combustion flames}

\author{J. A. Rosero-Gil$^1$}
\author{G. Lugones$^2$}

\affiliation{$^1$ Centro Brasileiro de Pesquisas F\'isicas, 
Rua Dr. Xavier Sigaud 150, 22290-180, Rio de Janeiro, RJ, Brazil}
\affiliation{$^2$ Centro de Ci\^{e}ncias Naturais e Humanas, Universidade Federal do ABC, \\ Av. dos Estados 5001, CEP 09210-580, Santo Andr\'{e}, SP, Brazil}

\begin{abstract}
We study the weak interaction processes taking place within a combustion flame that converts dense hadronic matter into quark matter in a compact star.  Using the Boltzmann equation we follow the evolution of a small element of just deconfined quark matter all along the flame interior until it reaches chemical equilibrium at the back boundary of the flame. We obtain the reaction  rates and neutrino emissivities  of  all  the  relevant  weak  interaction processes without making any assumption about the neutrino degeneracy. We analyse systematically the role the initial conditions of unburnt hadronic matter, such as density, temperature, neutrino trapping and composition, focusing on typical astrophysical scenarios such as cold neutron stars, protoneutron stars, and post merger compact objects. We find that the temperature within the flame rises significantly in a timescale of 1 nanosecond.  The increase in $T$ strongly depends on the initial strangeness of hadronic matter and tends to be more drastic at larger densities. Typical final values range between $20$ and $60 \, \mathrm{MeV}$. The nonleptonic process $u + d \rightarrow u + s$ is always dominant in cold stars, but in hot objects the process $u + e^{-} \leftrightarrow d + {\nu_e}$ becomes relevant, and in some cases dominant,  near chemical equilibrium.  The rates for the other processes are orders of magnitude smaller. We find that the neutrino emissivity per baryon is very large, leading to a total energy release per baryon of $2-60 \, \mathrm{MeV}$ in the form of neutrinos along the flame. We discuss some astrophysical consequences of the results.
\end{abstract}

\maketitle

\section{Introduction}
\label{intro}

The transition from hadronic matter to quark matter may have a key role in several highly energetic astrophysical phenomena such as core collapse supernova explosions \cite{Benvenuto:1989qr,Zha:2020gjw,Fischer:2017lag}, gamma ray bursts \cite{Lugones:2002vj,Ouyed:2001ts,Ouyed_2020}, binary neutron star (NS) mergers \cite{Bauswein:2018bma,Most:2018eaw} and phase-transition-induced collapse of NSs \cite{Lin:2005zda,Abdikamalov:2008df,Cheng:2009zr}. Although the potential relevance of this conversion has been recognised for decades, there are still several unresolved issues  such the exact mechanism that triggers the transition \cite{Iida:1998pi,Olesen:1993ek,Bombaci:2004mt,Lugones:2015bya,Bombaci:2009jt} and its  propagation mode to the rest of the star \cite{Lugones:1994xg,Herzog:2011sn,Pagliara:2013tza,Ouyed:2017nuy}. This is of key importance to assess potentially observable astrophysical signatures.

According to theoretical models, the transition to quark matter in a compact star would begin with the nucleation of a small deconfined seed inside the stellar core when the density of hadronic matter goes beyond a critical density \cite{Olesen:1993ek, Iida:1998pi, Lugones:1997gg, Benvenuto:1999uk, Bombaci:2004mt, Lugones:2005ni, Bombaci:2006cs, Bombaci:2009jt, Carmo:2013oba, Lugones:2015bya}.
The formation of such seed would occur in two steps. First, a small lump of hadrons deconfines in a strong interaction time scale, leaving a small quark drop that is not in equilibrium under weak interactions. Then, weak interactions drive the system to chemical equilibrium in a weak interaction time scale. 
The energy released in such conversion can ignite hadronic matter in the neighborhood of the initial seed, creating a self sustained combustion process that may convert to quark matter the core of the star and even the whole star if quark matter is absolutely stable (see \cite{Lugones:2015bya} and references therein). During the conversion process, a combustion front (\textit{flame}) separating the unburnt hadronic matter from the burnt quark matter travels outwards along the star \cite{Lugones:1994xg, Lugones:2002vj, Keranen:2004vj, Niebergal:2010ds, Herzog:2011sn, Fischer:2010wp, Pagliara:2013tza}.

\begin{figure}[tb]
\includegraphics[scale=0.33]{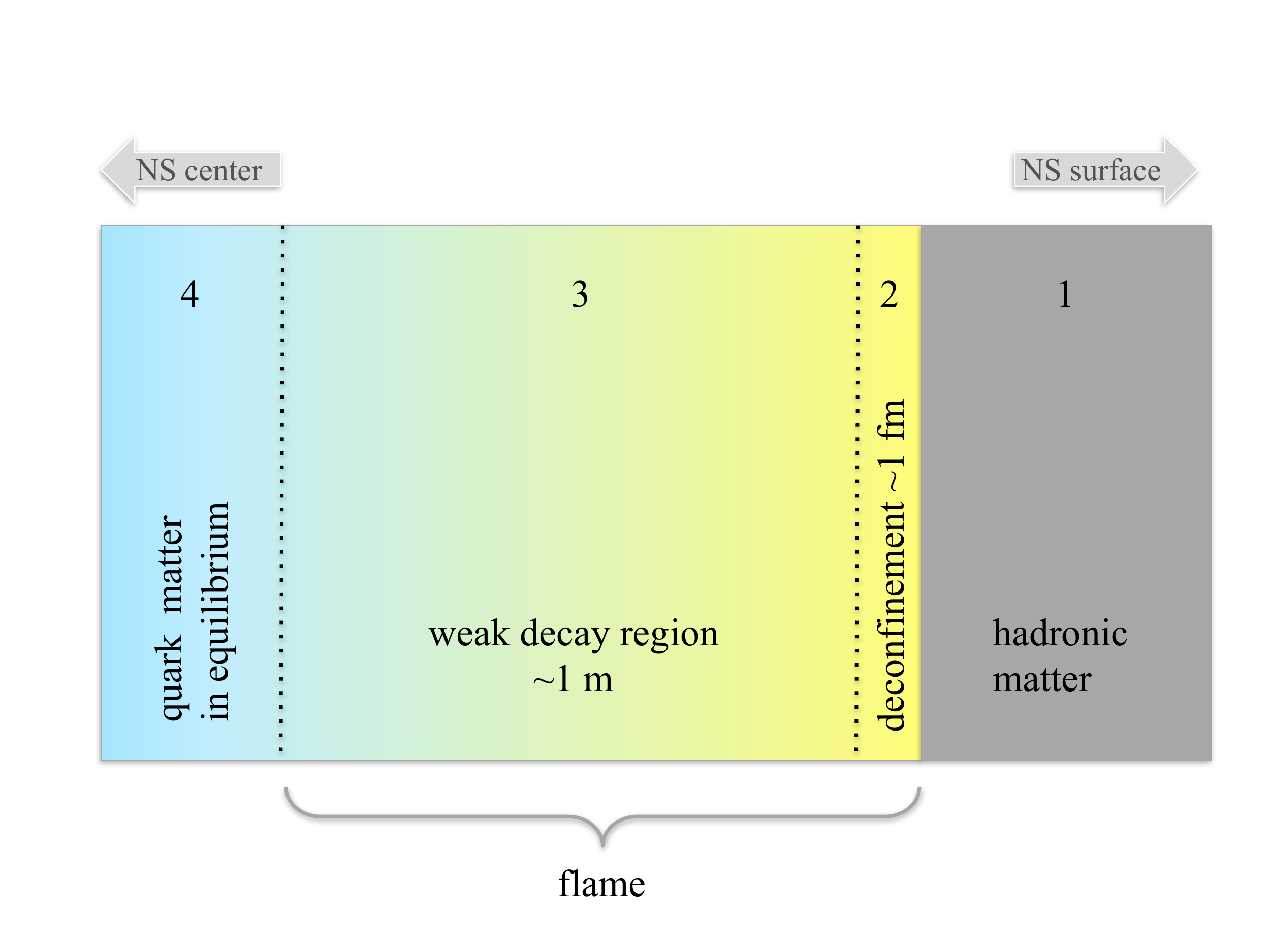}
\caption{Sketch of a flame converting hadronic matter into quark matter in a compact star.  The flame propagates to the right. Hadronic matter in region 1 deconfines in region 2.  Weak interactions drive deconfined quark matter to thermodynamic equilibrium in region 3. In region 4, quark matter is in full thermodynamic equilibrium. The thickness of the flame is estimated in the case of a detonation (fast combustion). For deflagrations (slow combustion) the flame thickness may be smaller. } 
\label{fig:flame}
\end{figure}

The structure of the flame is depicted in Fig. \ref{fig:flame}.   The combustion front propagates to the right with a  velocity that depends on the combustion mode, deflagration or detonation. In the fastest case, the velocity is of the order of $c/\sqrt{3}$, where $c$ is the speed of light  \cite{Lugones:1994xg}. In such a case, at the flame front there is a region of thickness $l_{\mathrm{strong}} \sim \tau_{\mathrm{strong}} \times c/\sqrt{3} \approx 10^{-23} \mathrm{s} \times  c/\sqrt{3} \sim 1 \, \mathrm{fm}$ where hadronic matter deconfines. Behind it, there is a \textit{weak decay region}  of thickness $l_{\mathrm{weak}} \sim \tau_{\mathrm{weak}} \times c/\sqrt{3} \approx 10^{-8} \mathrm{s} \times  c/\sqrt{3} \sim 1 \, \mathrm{m}$ where quark matter approaches chemical equilibrium through weak interactions {($\tau_{\mathrm{weak}} \sim 10^{-8} \mathrm{s}$ is a rough estimate taken from Refs. \cite{dai1995conversion,Dai:1995uj,Dai:1993nq,Anand:1997vk} but its value is subject to density and temperature variations as we shall see later)}. If the propagation velocity is smaller, these thicknesses are proportionally smaller. 

In this work, we focus on a fluid element that is initially located inside the deconfinement zone of the flame (see region 2 of Fig. \ref{fig:flame}). In this region, hadrons have just deconfined and matter is made up of quarks and leptons out of chemical equilibrium. The abundance of each quark and lepton species in region 2 is the same as in region 1, with the only difference that in region 1 the quarks are confined within hadrons and in region 2 they are deconfined. As time passes, the separation surface between regions 1 and 2 moves to the right in Fig. \ref{fig:flame}. In a reference frame in which the flame is at rest, we would see that our fluid element moves to the left in Fig. \ref{fig:flame} across the entire region 3. Along this path quarks and leptons interact with each other through weak reactions so that they come closer and closer to chemical equilibrium. Finally, the fluid element enters region 4, just at the moment when it reaches full chemical equilibrium. 

Our analysis will concentrate on the time evolution of the thermodynamic properties of the fluid element described above using the Boltzmann equation and calculating the rates of all the relevant weak reactions.  
We will pay special attention to the thermodynamic conditions at which the phase conversion occurs in typical astrophysical conditions. For example, in an old NS, accretion from a companion or rotational slowdown may trigger the conversion at the stellar core. In this case, the conversion begins in a low temperature environment without trapped neutrinos. On the other hand, the conversion may occur in a proto NS formed immediately after the gravitational collapse of the core of a massive star or in the massive compact object that may form after a binary NS merging. Such objects have very high temperatures in their interiors (typically few tens of MeV) and a large amount of trapped neutrinos, i.e. their mean free path is much shorter than the size of the star.  We will also analyse systematically the role of the initial hadronic composition, or equivalently, the quark and lepton concentrations in the deconfinement region of the flame.    

The paper is organized as follows. In Sec. \ref{sec:boltzmann} we summarize the equations that describe the time evolution of the particle abundances and the temperature inside the flame. In Sec. \ref{sec:eos} we describe the quark matter equation of state used inside the flame and the initial conditions assumed in the deconfinement region. In Sec. \ref{Sec-rate} we summarize the reaction rates and the neutrino emissivities in hot and dense quark matter without making any approximation about the degeneracy of neutrinos (the expressions are derived in Appendix \ref{reaction_rates}). In Sec. \ref{rens} we present our results for cold deleptonized NSs and in Sec. \ref{repns} for hot NSs with trapped neutrinos. In Sec. \ref{sec:conclusions} we present our conclusions.

\section{Boltzmann equation for quark matter decay} 
\label{sec:boltzmann}

\begin{table*}[tb]
\caption{Weak interaction processes considered in the present work and the corresponding expressions for $\left\langle |\mathcal{M}|^2 \right\rangle$ \cite{Iwamoto:1982zz}. $G_F$ is the Fermi weak coupling constant ($G_F/(c\hbar)^3=1.1664\times 10^{-5}$GeV$^{-2}$) and $\theta_C$ is the Cabibbo angle  ($\cos \theta_C = 0.973$). In all equations we will assume $\hbar = c = k_B = 1$.}
\label{table:reactions}
\centering
\begin{tabular*}{\linewidth}{l @{\extracolsep{\fill}} ll}
\hline\hline
\textrm{Label}	&	\textrm{Process} &   $\left\langle |\mathcal{M}|^2 \right\rangle$   \\
\hline
I	 &  $d\rightarrow u+e^- + \bar{\nu}_e$  &  $64 G_F^2 \cos^2 \theta_C (P_d \cdot P_{\bar{\nu}_e}) (P_u \cdot P_{e^-})$ \\		
II	 &  $s \rightarrow u+e^- + \bar{\nu}_e$ &  $64 G_F^2 \sin^2 \theta_C (P_s \cdot P_{\bar{\nu}_e}) (P_u \cdot P_{e^-})$  \\
III  &  $u+e^{-} \leftrightarrow  d+\nu_{e}$   &  $64 G_F^2 \cos^2 \theta_C (P_u \cdot P_{e^-}) (P_{\nu_e} \cdot P_d)$ \\
IV   &  $u+e^{-} \leftrightarrow  s+\nu_{e}$   &  $64 G_F^2 \sin^2 \theta_C (P_u \cdot P_{e^-}) (P_{\nu_e} \cdot P_s)$ \\
V  & $u_1 + d  \leftrightarrow  u_2 + s$ & $64G_F^2 \sin^2\theta_C  \cos^2\theta_C (P_{u_1} \cdot P_d) (P_{u_2} \cdot P_s)$ \\
\hline\hline
\end{tabular*}
\end{table*}

{The time evolution of quark matter composition inside the flame will be described by means of the Boltzmann equation. We will focus on a small fluid element within the flame whose size is much smaller than the flame thickness and we will neglect the effect of the gravitational external field. For simplicity, we will also neglect spatial gradients of the distribution function\footnote{{The order of magnitude of the term with $v \partial /\partial x$ in  the Boltzmann equation is $\sim v/L $, where $v$ is a typical fluid velocity and $L$ is the flame thickness. For fast combustions in neutron star matter we have $v \sim c_s \sim c$ (being $c_s$  the sound speed and $c$ the speed of light) and therefore  $v/L \sim 10^9 s^{-1}$. As we shall see later,  this is essentially of the same order of the term $\partial /\partial t \sim \tau^{-1}$, being $\tau \sim 10^{-9}-10^{-8} \, \mathrm{s}$ the timescale for attaining chemical equilibrium.  This means that the term with $v \partial /\partial x$ may be important in the case of fast combustions. However, for slow enough deflagrations $v$ is significantly smaller than $c$, and  the term with $v \partial /\partial x$ can be neglected safely.}}.} Matter in the flame is assumed to be composed by $u$, $d$, $s$ quarks,  electrons, electron neutrinos (and their antiparticles) interacting among themselves through weak interactions. Therefore, the Boltzmann transport equation for each particle species in the system reads:
\begin{eqnarray}
\frac{\partial f_i}{\partial t} = {\cal C} (f_i, f_j, ... ) ,   \qquad  i, j, ... = u, d, s, e^-, \nu_e,
\label{Boltzmann_equation}
\end{eqnarray} 
being
\begin{eqnarray}
{\cal C} (f_i, f_j, ... ) &=&  6 \int \frac{d^{3}p_{j}}{\left(2\pi\right)^{3}} ... \; W_{i, j, ...} \; {\cal S}(f_i, f_j,  ... ) ,
\end{eqnarray}
where $f_i$ is the distribution function of the particle species $i$, $W_{i, j, ...}$ is the transition probability for the processes creating and destroying such particles, and  ${\cal S}$ is a statistical blocking factor involving the Fermi-Dirac distribution functions of the ingoing and outgoing particles.
{In the case of a flame in a hot NS, the assumption of a Fermi-Dirac distribution for neutrinos is justified because they are trapped and their mean free path is very small.  For a flame in a cold NS the situation may be different. Close to the deconfinement zone shown in Fig. \ref{fig:flame}, matter is essentially transparent to neutrinos, but as the fluid element approaches the rear side of the flame, the temperature increases significantly and neutrinos may get trapped. The neutrino-transparent and neutrino-trapped regimes are the simplified extremes of a continuum,  which is realized at different regions of the flame. Between these extremes there is a semitransparent regime where the spectrum of neutrinos includes a low-energy population that escapes, a high-energy tail that is trapped, and an intermediate-energy range where the mean free path is of the order of the flame thickness. A full analysis of the neutrino spectrum in this case is complex and it is beyond the scope of the present work. For simplicity we will assume in this work that the neutrino distribution is always a  Fermi-Dirac one and let a more detailed analysis for future work. }

Integrating Eq. (\ref{Boltzmann_equation}) over the momentum of particle $i$, we have
\begin{eqnarray}
\frac{d Y_{i}}{d t}&=&\frac{1}{n_{B}}\sum_{P}\Gamma_{P} ,
\label{eq:evolution_abundances}
\end{eqnarray} 
where $n_B$ is the baryon number density, $n_i$ is the particle number density, $Y_i \equiv n_i/ n_B$ is the abundance of particle $i$, and $\Gamma_{P}$ is the rate of decay or production of the particle $i$  due to the process $P$ given by:
\begin{eqnarray}
\Gamma_{P} &=&  6\int\prod_{i=1}^{4}\left[\frac{d^{3}p_{i}}{\left(2\pi\right)^{3}}\right] W \times {\cal S} .
\end{eqnarray}
For the present problem, the transition probability $W$ has the form (see \cite{Colvero:2014zfa} and references therein):
\begin{equation}
  W =(2\pi)^4 \delta^4 \frac{\left\langle |\mathcal{M}|^2 
  \right\rangle}{2^4 E_1 E_2 E_3 E_4},
\end{equation}
where $\delta^4$ is a Dirac delta function for the conservation of four-momentum that will be specified later,  $P_i=(E_i,\mathbf{p}_i)$ is the four-momentum of any quark or lepton and $\left\langle |\mathcal{M}|^2 \right\rangle$ denotes the squared matrix element 
summed over final spins and averaged over the initial spins.  In Table \ref{table:reactions} we list 
all the processes under consideration and the corresponding expressions for $\left\langle |\mathcal{M}|^2 \right\rangle$. 

The time evolution of the temperature can be obtained by means of the first law of thermodynamics.  Since baryon number is a conserved quantity it is convenient to write the first law on a per baryon basis.  Let  $\epsilon$ be the total energy density  and $s$ the entropy per baryon. Then, we have:
\begin{eqnarray}
d\left(\frac{\epsilon}{n_B}\right)=-Pd\left(\frac{1}{n_B}\right) + Tds + \sum_{i}\mu_{i}dY_{i} .
\end{eqnarray}

{We will neglect the expansion of the fluid, i.e. the volume per baryon $v$ will be assumed to be essentially unchanged; $dv = d(1/n_B) =0$. We will also assume that the energy per baryon $\epsilon/n_B$ changes due to neutrinos that leave the system, i.e. $d(\frac{\epsilon}{n_B})/dt = \varepsilon_\nu$, where $\varepsilon_\nu$ is the neutrino emissivity per baryon due to the weak processes (see Sec. \ref{Sec-rate} for more details). Thus,  the first law of thermodynamics reads:
\begin{eqnarray}
T\frac{ds}{dt} + \sum_{i}\mu_{i}\frac{dY_{i}}{dt} = \varepsilon_\nu. 
\end{eqnarray}
The latter equation can ve rewritten as:
\begin{equation}
\begin{aligned}
T\left(\frac{\partial s}{\partial T}\right)_{\mu}\frac{dT}{dt}+T\sum_{i}\left(\frac{\partial s}{\partial\mu_{i}}\right)_{T}\frac{d\mu_{i}}{dt}+\sum_{i}\mu_{i}\frac{dY_{i}}{dt}=\varepsilon_\nu. 
\end{aligned}
\end{equation}
Using
\begin{eqnarray}
\frac{dn_{i}}{dt}&=&\left(\frac{\partial n_{i}}{\partial T}\right)_{\mu}\frac{dT}{dt}+\left(\frac{\partial n_{i}}{\partial\mu_{i}}\right)_{T}\frac{d\mu_{i}}{dt} ,
\end{eqnarray}
and defining 
\begin{eqnarray}
\beta  \equiv  T\left[\sum_{i}   \frac{ \big( \partial s / \partial \mu_ i \big)_T } {\big(\partial n_i / \partial \mu_i \big)_T }
\left(\frac{\partial n_{i}}{\partial T}\right)_{\mu}-\left(\frac{\partial s}{\partial T}\right)_{\mu}   \right]
\end{eqnarray}
we arrive to the equation that governs the time evolution of the temperature:
\begin{equation}
 \beta \frac{dT}{dt} =  \sum_{i} n_{B}T   \frac{ \big( \partial s / \partial \mu_ i \big)_T } { \big(\partial n_i / \partial \mu_i \big)_T }   \frac{dY_{i}}{dt}  -\sum_{i}\mu_{i}\frac{dY_{i}}{dt}+\varepsilon_{\nu},
 \label{evolution_of_T}
\end{equation}
}

Providing appropriate initial conditions (see Sec. \ref{sec:initial_conditions}), Eqs. \eqref{eq:evolution_abundances} and \eqref{evolution_of_T} allow us to describe the time evolution of a fluid element within the flame since deconfinement until the time at which full chemical equilibrium is attained.

\section{Equation of state and initial conditions}
\label{sec:eos}

\subsection{The quark matter equation of state}
\label{sec:MIT}

We describe quark matter by means of the MIT bag model for a system composed by $u$, $d$ and $s$ quarks, electrons and electron neutrinos with their antiparticles. In its simplest version the equation of state can be derived from a grand thermodynamic potential per unit volume of the form:
\begin{equation}
  \Omega = \sum_{i} \Omega_{i} + B,
  \label{eq:bag_omega} 
\end{equation}
where $\Omega_{i}$ is the thermodynamic potential for a gas of relativistic non-interacting fermions, the sum goes over $i=u,d,s,e^-,\nu_e$ and their antiparticles, and a  QCD vacuum pressure or bag constant $B$ is included in order to mimic long-range interactions among quarks. 

At finite temperature, the thermodynamic potential $\Omega_{i}$ is given by      
\begin{equation}
\Omega_{i}=\frac{g_{i}}{6\pi^{2}}\int_{0}^{\infty}k\frac{\partial E_{i}}{\partial k}\left(f_{i}+\bar{f}_{i}\right)k^{2}dk,
\label{eq:Inte13}
\end{equation}
where  $E_{i}(k)=\left(m_{i}^{2}+k^2\right)^{1/2}$ is the  single particle kinetic energy, $f_{i}$ and $\bar{f}_{i}$ are the Fermi-Dirac distribution functions for particles and antiparticles as function of temperature $T$ and chemical potential $\mu_i$. The degeneracy factor is $g_i = 2 (\mathrm{spin}) \times 3 (\mathrm{color}) =6$ for quarks, $g_i = 2 (\mathrm{spin})$ for electrons, and  $g_i = 1$ for (left-handed) neutrinos. In the above expression antiparticles were included through $\bar{f}_{i}= f_i(T,-\mu_{i})$ and all derived thermodynamic quantities  must be understood as net quantities, containing the contributions of both particles and 
antiparticles. 

In order to take into account the quark-quark interaction at short range, we will include an additional contribution to the thermodynamic potential of Eq. (\ref{eq:bag_omega}). QCD corrections to orders of $\alpha_c$ and $\alpha_c^{3/2}$ in perturbation theory have been derived in Ref. \cite{Kapusta:1979fh} for arbitrary temperatures, quark masses and chemical potentials. However, closed-form expressions are known only for approximate regimes. For degenerate massless quarks one finds the following correction to first order in $\alpha_c=g^2/4\pi$ \cite{Kalashnikov:1979dp}: 

\begin{eqnarray}
  \Omega_{(2),i} & = & -  \left[\frac{7}{60}\pi^2 T^4\left(\frac{50}{21}
  \frac{\alpha_c}{\pi}\right)\right.\nonumber\\
	&  & \left. + \left(\frac{1}{4\pi^2}\mu_i^4+\frac{1}{2}T^2\mu_i^2\right)\left(2
  \frac{\alpha_c}{\pi}\right)\right],
 \label{eq:Inte14}
\end{eqnarray}
for $i=u,d,s$. {Since the correction without approximations has a complex expression, we will use Eq. \eqref{eq:Inte14} for both degenerate and non-degenerate matter, and for massive quarks as well.}
Including this correction, the thermodynamic potential of Eq. (\ref{eq:bag_omega}) reads:
\begin{equation}
  \Omega = \sum_{i=u, d, s} \left( \Omega_{i} +\Omega_{(2),i} \right)   + \Omega_{e}  + \Omega_{\nu_e} + B,
 \label{eq:Inte15} 
\end{equation}
{where $\Omega_{i}$ is given by Eq. \eqref{eq:Inte13} and $\Omega_{(2),i}$ by Eq. \eqref{eq:Inte14}.} From $\Omega$ we can easily obtain the particle number density $n_i$, the mass-energy density $\epsilon_i$, the pressure $P_i$ and the entropy density $s_i$ using standard thermodynamic relationships.

In the specific case of  \textit{massless quarks}, a simple analytic expression is obtained {for the term in parenthesis in Eq. \eqref{eq:Inte15}: }
\begin{eqnarray}
\Omega_{i} +\Omega_{(2),i} 
&=&\frac{7}{60}\pi^{2}T^{4}\left(1-\frac{50\alpha_{c}}{21\pi}\right)\nonumber\\
&&+\left(\frac{1}{2}T^{2}\mu_{i}^{2}+\frac{1}{4\pi^{2}}\mu_{i}^{4}\right)\left(1-\frac{2\alpha_{c}}{\pi}\right),
\end{eqnarray}
{which will be used in this work for $u$ and $d$ quarks. For $s$ quarks we obtain $\Omega_s$ from  Eq. \eqref{eq:Inte13} and $\Omega_{(2),s}$ from Eq. \eqref{eq:Inte14}. }  

In the context of the MIT bag model, the strong coupling constant $\alpha_c$, the quark masses, and the bag constant $B$ are regarded as a free parameters.  Throughout this paper we use  $m_u = m_d = m_e = m_{\nu_e} = 0$, $m_s=150 \, \mathrm{MeV}$  and $\alpha_c=0,\, 0.47$. The value of $B$ is not needed in the calculations so we will let it undefined. As a consequence, our results are valid in principle for both absolutely stable (strange) quark matter and standard (non-absolutely stable) quark matter \cite{Farhi:1984qu}.

The equation of state depends on the temperature $T$ and on the chemical potentials $\mu_i$ of all the particle species ($i=u, d, s, e^-, \nu_e$). However, chemical potentials are not all independent. Local electric charge neutrality implies that
\begin{equation}
2 n_{u}- n_{d}- n_{s}- 3 n_{e} = 0.
\label{eq:charge_neutrality}
\end{equation}
Also, if we fix the baryon number density $n_B$ of the system, we have:
\begin{equation}
n_{B}=\tfrac{1}{3}\left(n_{u}+n_{d}+n_{s}\right).
\label{eq:baryon_number}
\end{equation}
These two equations allow to eliminate two chemical potentials when calculating the equation of state.

When the system is in equilibrium under weak interactions (such as  $d \rightarrow u + e^{-} + \overline{\nu}_{e}$,    $s \rightarrow u + e^{-} + \overline{\nu}_{e}$,   $d + u \leftrightarrow u + s$, etc.)  the chemical equilibrium conditions  $\mu_{d} = \mu_{u} + \mu_{e}$ and  $\mu_{s} = \mu_{d}$ hold, which allow to eliminate  two more chemical potentials. 
However, since we focus here on quark matter that is most of the time \textit{out of equilibrium} under weak interactions, such equations are not fulfilled.

\subsection{Initial conditions}
\label{sec:initial_conditions}

\begin{table*}[tbh]
  \centering
\caption{Initial particle number densities in quark matter just after deconfinement.}   
    \begin{tabular*}{\linewidth}{c @{\extracolsep{\fill}} ccc|cccccccccc}
    \hline
$n_B$ & $\xi$ & $\eta$  & $\kappa$ & $n_u$  & $n_d$  & $n_s$  & $n_e$  & $n_{\nu_{e}}$  \\
\, [fm$^{-3}$]   &  &   &   &  [fm$^{-3}$]  &  [fm$^{-3}$]   &  [fm$^{-3}$]  &  [fm$^{-3}$]   &  [fm$^{-3}$]  \\
    \hline \hline  
0.32 & 1.4 & 0    & 0     & 0.400 & 0.560 & 0.000 & 0.080 & 0.000 \\
0.32 & 1.4 & 0    & 0.01  & 0.400 & 0.560 & 0.000 & 0.080 & 0.003 \\
0.32 & 1.4 & 0.4  & 0     & 0.343 & 0.480 & 0.137 & 0.023 & 0.000 \\
0.32 & 1.4 & 0.4  & 0.01  & 0.343 & 0.480 & 0.137 & 0.023 & 0.003 \\
0.96 & 1.4 & 0    & 0     & 1.200 & 1.680 & 0.000 & 0.240 & 0.000 \\
0.96 & 1.4 & 0    & 0.01  & 1.200 & 1.680 & 0.000 & 0.240 & 0.012 \\
0.96 & 1.4 & 0.4 & 0      & 1.029 & 1.440 & 0.411 & 0.069
& 0.000 \\
0.96 & 1.4 & 0.4 & 0.01   & 1.029 & 1.440 & 0.411 & 0.069 & 0.010 \\
    \hline
    \end{tabular*}   
  \label{tab:dec_pns}
\end{table*}

As mentioned in the Introduction, the initial conditions in the deconfinement zone of the flame (region 2 of Fig. \ref{fig:flame}) are given by flavor conservation between the hadronic side and the quark side of the interface between regions 1 and 2. This condition can be written as \cite{Olesen:1993ek,Iida:1998pi,Lugones:1997gg,Benvenuto:1999uk,Bombaci:2004mt,Lugones:2005ni,Bombaci:2006cs}:
\begin{equation}
Y^H_i = Y^Q_i   \qquad  i=u,d,s,e^-, \nu_e \label{flavor}
\end{equation}
being $Y^H_i \equiv n^H_i / n^H_B$ and  $Y^Q_i \equiv n^Q_i / n^Q_B$ the abundances of each particle in the hadron and quark phase respectively. In other words, the just deconfined quark phase must have initially the same ``flavor'' composition than the $\beta$-stable hadronic phase from which it has been originated.
Notice that, since the hadronic phase is assumed to be electrically neutral, flavor conservation ensures automatically the charge neutrality of the just deconfined quark phase.

The conditions given in Eq. (\ref{flavor}) can be combined to obtain
\begin{equation}
n^Q_d = \xi ~ n^Q_u,  \quad
n^Q_s = \eta ~ n^Q_u,  \quad
n^Q_{\nu_e} = \kappa ~ n^Q_u.
\label{h3}
\end{equation}
\noindent The quantities $\xi \equiv Y^H_d /
Y^H_u$, $\eta \equiv Y^H_s / Y^H_u$  and $\kappa \equiv Y^H_{\nu_e} / Y^H_u$ are functions of the pressure and temperature, and they characterize the composition of the hadronic phase. These expressions are valid for any hadronic EOS. For hadronic matter containing $n$, $p$, $\Lambda$, $\Sigma^{+}$, $\Sigma^{0}$, $\Sigma^{-}$, $\Xi^{-}$ and $\Xi^{0}$, we have (cf. \cite{Olesen:1993ek, Iida:1998pi}):
\begin{eqnarray}
\xi &=& \frac{n_p  +  2  n_n  + n_{\Lambda} + n_{\Sigma^{0}} +  2
n_{\Sigma^{-}}  + n_{\Xi^{-}}}{2  n_p  +  n_n  +  n_{\Lambda} + 2
n_{\Sigma^{+}} + n_{\Sigma^{0}}  +  n_{\Xi^{0}}}, \label{xi} \\
\eta &=& \frac{n_{\Lambda}  + n_{\Sigma^{+}} + n_{\Sigma^{0}}  +
n_{\Sigma^{-}} + 2 n_{\Xi^{0}} + 2 n_{\Xi^{-}}}{2  n_p  +  n_n  +
n_{\Lambda} + 2 n_{\Sigma^{+}} + n_{\Sigma^{0}}  +  n_{\Xi^{0}}}, \\
\kappa &=& \frac{n^H_{\nu_e}}{2  n_p  +  n_n  +
n_{\Lambda} + 2 n_{\Sigma^{+}} + n_{\Sigma^{0}}  +  n_{\Xi^{0}}}.
\label{eta}
\end{eqnarray}
Combining Eqs. (\ref{h3}) together with baryon number conservation and charge neutrality, we obtain the following expressions for the initial particle number densities in terms of the baryon number density:
\begin{eqnarray}
n_u &=& \frac{3}{1 + \xi + \eta} n_B,     \\
n_d &=& \frac{3 \xi}{1 + \xi + \eta} n_B, \\ 
n_s &=& \frac{3 \eta}{1 + \xi + \eta} n_B,    \\
n_e &=& \frac{2-\xi-\eta}{1+\xi+\eta}n_{B},  \\
n_{\nu_e} &=& \frac{3 \kappa}{1 + \xi + \eta} n_B.
\end{eqnarray}
To keep the analysis as general as possible,  we do not consider in this work any specific hadronic EOS. Instead, we adopt different values of the parameters $\xi$, $\eta$ and $\kappa$ in order to explore the effect of the initial strangeness and neutrino trapping in the further evolution of quark matter. Specifically, we adopt the following parameters for the initial composition:
\begin{itemize}
    \item $n_B=0.32 \, \mathrm{fm}^{-3}$  and   $n_B =0.96 \, \mathrm{fm}^{-3}$,
    \item $\xi = 1.4$, 
    \item $\eta = 0$ (zero strangeness)  and  $\eta = 0.4$ (finite strangeness),
    \item $\kappa = 0$ (no neutrinos)  and  $\kappa = 0.01$ (neutrino trapping),
\end{itemize}
which are typical values \footnote{{As a guideline we considered the composition of  matter within the GM1 parametrization of a nuclear relativistic mean field model \cite{Glendenning:1991es}. For  $npe$ matter at $n_B=0.32 \, \mathrm{fm}^{-3}$, and assuming $0 <T \mathrm{[MeV]}< 40$  and $0 < \mu_{\nu_e} \mathrm{[MeV]}< 100$, we find $1.35 < \xi < 1.60$,  being $\kappa =0$  for $\mu_{\nu_e} = 0$ and $\kappa < 0.015$ for $\mu_{\nu_e} = 100~\mathrm{MeV}$. For $npe$ matter at $n_B=0.96 \, \mathrm{fm}^{-3}$ and varying the temperature and the neutrino chemical potential in the same ranges we find $1.25 < \xi < 1.35$,  with $\kappa =0$  for $\mu_{\nu_e} = 0$ and $\kappa < 0.005$ for $\mu_{\nu_e}  = 100~\mathrm{MeV}$. When hyperons are included in the GM1 parametrization (specifically when we consider the baryon octet) the value of $\xi$ stays within a $15 \%$ around $1.4$ and there are no significant changes in $\kappa$. The value of $\eta$ continues to be very low for $n_B=0.32 \, \mathrm{fm}^{-3}$, but it is $\sim 0.6$  for $n_B=0.96 \, \mathrm{fm}^{-3}$. 
In order to reduce as much as possible the number of cases, we adopt for both densities a representative value $\xi= 1.4$, with $\kappa =0$ for neutrino free matter and with $\kappa =0.01$ as a reasonable value for matter with trapped neutrinos. For strangeness, we adopt $\eta=0$ and a conservative but still high value $\eta=0.4$.  Notice that the case with $\eta=0.4$ is somewhat artificial for $n_B=0.32 \, \mathrm{fm}^{-3}$, but we will consider it for completeness.}} according to Refs. \cite{Olesen:1993ek,Iida:1998pi,Lugones:1997gg,Benvenuto:1999uk,Bombaci:2004mt,Lugones:2005ni,Bombaci:2006cs}. 
In Table \ref{tab:dec_pns}  we show the initial particle abundances for the models studied in the present paper.

\section{Reaction rates and neutrino emissivities in dense quark matter} \label{Sec-rate}

Once dense hadronic matter deconfines into quark matter, the initial state doesn't have in general the quark abundances that guarantee chemical equilibrium. Thus, weak interaction processes will take place and will drive the composition to an equilibrium configuration. 
Just after deconfinement of hadronic matter, the following processes may occur:
\begin{eqnarray}
\text{I} :\quad && d  \rightarrow  u+e^{-}+\bar{\nu}_{e}\label{d--uenu} , \\
\text{II} :\quad && s  \rightarrow  u+e^{-}+\bar{\nu}_{e}\label{s--uenu} , \\
\text{III} :\quad && u+e^{-}  \leftrightarrow  d+\nu_{e}\label{ue--dnu} , \\
\text{IV} :\quad && u+e^{-}  \leftrightarrow  s+\nu_{e}\label{ue--snu} , \\
\text{V} :\quad && u+d  \leftrightarrow  u+s\label{ud--us}   .
\end{eqnarray}
Depending on the astrophysical environment at which deconfinement is initiated, some of these processes may have a different relevance. For example, in a cold and deleptonized NS, neutrinos are free to escape from the system and neutrino captures in processes  III and IV do not happen{, but as temperature increases, these processes are more relevant and should be considered.}  
On the other hand, in just born PNSs or in hot compact stars potentially formed in a binary merger, the temperatures are very high (typically some tens of MeV)   and there is a large amount of trapped neutrinos, in the sense that they have a mean free path much shorter than the size of the star. 

Reaction rates and neutrino emissivities have been calculated in previous works for two different approximate cases \cite{Anand:1997vk}: (1) cold deleptonized matter, where quarks and electrons are degenerate, and (2) hot neutrino rich matter, where quarks, electrons and neutrinos were treated as degenerate. In this section, we generalize these results and obtain the reaction rates and neutrino emissivities assuming degenerate quarks and electrons, but without making any assumption about the degeneracy state of neutrinos. 
This approximation is implemented only in the squared  matrix  element $\left\langle |\mathcal{M}|^2 \right\rangle$  by replacing the particle momenta and energies by the corresponding Fermi momenta and chemical potentials. The approximation is not used  in the Fermi blocking factors, nor in the delta function.
We show below only the relevant results and present a detailed derivation in Appendix \ref{reaction_rates}.
{The error due to this assumption is small, as estimated in Appendix \ref{Comparison_rates} where we compare the approximate and the exact rates \cite{Madsen:1993xx} for the nonleptonic reaction,  which is the dominant one.}

The reaction rate for the decay process $d  \rightarrow  u+e^{-}+\bar{\nu}_{e}$ is 
\begin{eqnarray}
\Gamma_{\text{I}}= c_1 \int_{0}^{\infty}\frac{(\mu_{u}+\mu_{e}-\mu_{d}+ E_{\bar{\nu}_{e}})^{2}+\pi^{2}T^{2}}{2[e^{(\mu_{u}+\mu_{e}-\mu_{d}+E_{\bar{\nu}_{e}})/T}+1]}  \nonumber \\
\times \frac{I(\mu_{u},\mu_{e},\mu_{d},E_{\bar{\nu}_{e}})}{e^{(\mu_{\bar{\nu}_{e}}-E_{\bar{\nu}_{e}})/T}+1} dE_{\bar{\nu}_{e}},
\label{Gamma1}
\end{eqnarray}
where $c_1 = 3G_F^2  \cos^{2}\theta_{C} / (2\pi^5)$ and the angular integral $I$ given by Eq. \eqref{Iint} (see also Ref. \cite{Wadhwa:1995dv}). 

The rate $\Gamma_{\text{II}}$ for the process  $s  \rightarrow u+e^{-}+\bar{\nu}_{e}$ can be obtained  replacing $\mu_{d}$ by $\mu_{s}$ and $\cos^{2}\theta_{C}$ by $\sin^{2}\theta_{C}$ in the latter expression.

For the  process  $u+e^{-}  \leftrightarrow d+\nu_{e}$ we find
\begin{eqnarray}
\Gamma_{\text{III}}^{dir} = c_1 \int_{0}^{\infty}\frac{(\mu_{u}+\mu_{e}-\mu_{d}-E_{\nu_{e}})^{2}+\pi^{2}T^{2}}{2[e^{(-\mu_{u}-\mu_{e}+\mu_{d}+E_{\nu_{e}})/T}+1]} \nonumber\\
\times \frac{J(\mu_{u},\mu_{e},\mu_{d},E_{\nu_{e}})}{e^{(\mu_{\nu_{e}}-E_{\nu_{e}})/T}+1} dE_{\nu_{e}},
\label{Gamma3}
\end{eqnarray}
for the direct process (electron capture by $u$ quarks) and  $\Gamma_{\text{III}}^{rev} = e^{-\xi_{d}} \Gamma_{\text{III}}^{dir}$ for the reverse process (neutrino absorption by $d$ quarks) where $\xi_{d}= (\mu_u +\mu_e - \mu_d-\mu_{\nu_e})/T$. The angular integral $J$ is given in Eq. \eqref{JintA}. 

The rate $\Gamma_{\text{IV}}^{dir}$ for  $u+e^{-}  \rightarrow s+\nu_{e}$ can be obtained  replacing $\mu_{d}$ by $\mu_{s}$ and $\cos^{2}\theta_{C}$ by $\sin^{2}\theta_{C}$ in the  expression for $\Gamma_{\text{III}}^{dir}$. For the reverse process $s+\nu_{e} \rightarrow u+e^{-}$ we have  $\Gamma_{\text{IV}}^{rev} = e^{-\xi_s} \Gamma_{\text{IV}}^{dir}$ where $\xi_s = (\mu_u + \mu_e - \mu_s - \mu_{\nu_e})/T$.

Finally, for $u_1 + d \rightarrow u_2 + s$ we have
\begin{eqnarray}
\Gamma_{\text{V}}^{dir} = c_2 \int_{m_{s}}^{\infty}\frac{(\mu_{d}-E_{s})^{2}+\pi^{2}T^{2}}{2[e^{(\mu_{d}-E_{s})/T}+1]} \nonumber\\
\times \frac{J(\mu_{u},\mu_{d},\mu_{u},E_{s})}{e^{(\mu_{s}-E_{s})/T}+1}dE_{s},
\label{Gamma5}
\end{eqnarray}
where  $c_2 = 9 G_F^2 \sin^2 \theta_C \cos^{2}\theta_C / (2\pi^5)$. The rate for the reverse process $u+s \rightarrow u+d$ is given by  $\Gamma_{\text{V}}^{rev} = e^{-(\mu_d - \mu_s)/T} \Gamma_{\text{V}}^{dir}$.


The neutrino emissivity rate per baryon is given below for all the relevant processes. For $d  \rightarrow u+e^{-}+\bar{\nu}_{e}$ we have
\begin{eqnarray}
\varepsilon_{\text{I}} = c_1 \int_{-\infty}^{\mu_{\bar{\nu}_{e}}}\frac{(\mu_{u}+\mu_{e}-\mu_{d}+E_{\bar{\nu}_{e}})^{2}+\pi^{2}T^{2}}{2[e^{(\mu_{u}+\mu_{e}-\mu_{d}+E_{\bar{\nu}_{e}})/T}+1]}\nonumber\\
\varepsilon_{\text{I}} = c_1 \int_{-\infty}^{\mu_{\bar{\nu}_{e}}}\frac{(\mu_{u}+\mu_{e}-\mu_{d}+E_{\bar{\nu}_{e}})^{2}+\pi^{2}T^{2}}{2[e^{(\mu_{u}+\mu_{e}-\mu_{d}+E_{\bar{\nu}_{e}})/T}+1]}\nonumber\\
 \times \frac{I(\mu_{u},\mu_{e},\mu_{d},E_{\bar{\nu}_{e}})}{e^{(\mu_{\bar{\nu}_{e}}-E_{\bar{\nu}_{e}})/T}+1} E_{\bar{\nu}_{e}}dE_{\bar{\nu}_{e}} .
\end{eqnarray}
The emissivity $\varepsilon_{\text{II}}$ for  $s  \rightarrow u+e^{-}+\bar{\nu}_{e}$ can be obtained  replacing $\mu_{d}$ by $\mu_{s}$ and $\cos \theta_{C}$ by $\sin \theta_{C}$ in the previous expression. For $u+e^{-}  \rightarrow d+\nu_{e}$ we find
\begin{eqnarray}
\varepsilon_{\text{III}} = c_1 \int_{-\infty}^{\mu_{\nu_{e}}}\frac{(\mu_{u}+\mu_{e}-\mu_{d}-E_{\nu_{e}})^{2}+\pi^{2}T^{2}}{2[e^{(-\mu_{u}-\mu_{e}+\mu_{d}+E_{\nu_{e}})/T}+1]} \nonumber\\
\times  \frac{J(\mu_{u},\mu_{e},\mu_{d},E_{\nu_{e}})}{e^{(\mu_{\nu_{e}}-E_{\nu_{e}})/T}+1}E_{\nu_{e}}dE_{\nu_{e}} .
\end{eqnarray}
Similarly, the emissivity $\varepsilon_{\text{IV}}$ for  $u+e^{-}  \rightarrow s+\nu_{e}$ is obtained  replacing $\mu_{d}$ by $\mu_{s}$ and $\cos \theta_{C}$ by $\sin \theta_{C}$ in the previous formula.

The total neutrino and antineutrino emissivities in  cold deleptonized matter are $\varepsilon_\nu =\varepsilon_{\text{III}}+\varepsilon_{\text{IV}}$ and $\bar{\varepsilon}_{\bar{\nu}} = \varepsilon_{\text{I}} +\varepsilon_{\text{II}}$. For hot neutrino-rich matter, the total neutrino emissivity is $\varepsilon_\nu = \varepsilon_{\text{III}} (1-e^{-\xi_{d}})+\varepsilon_{\text{IV}} (1-e^{-\xi_{s}})$.

\section{Results for cold deleptonized Neutron Stars}
\label{rens}

\begin{figure}[tbh]
\begin{center}
\includegraphics[scale=0.22]{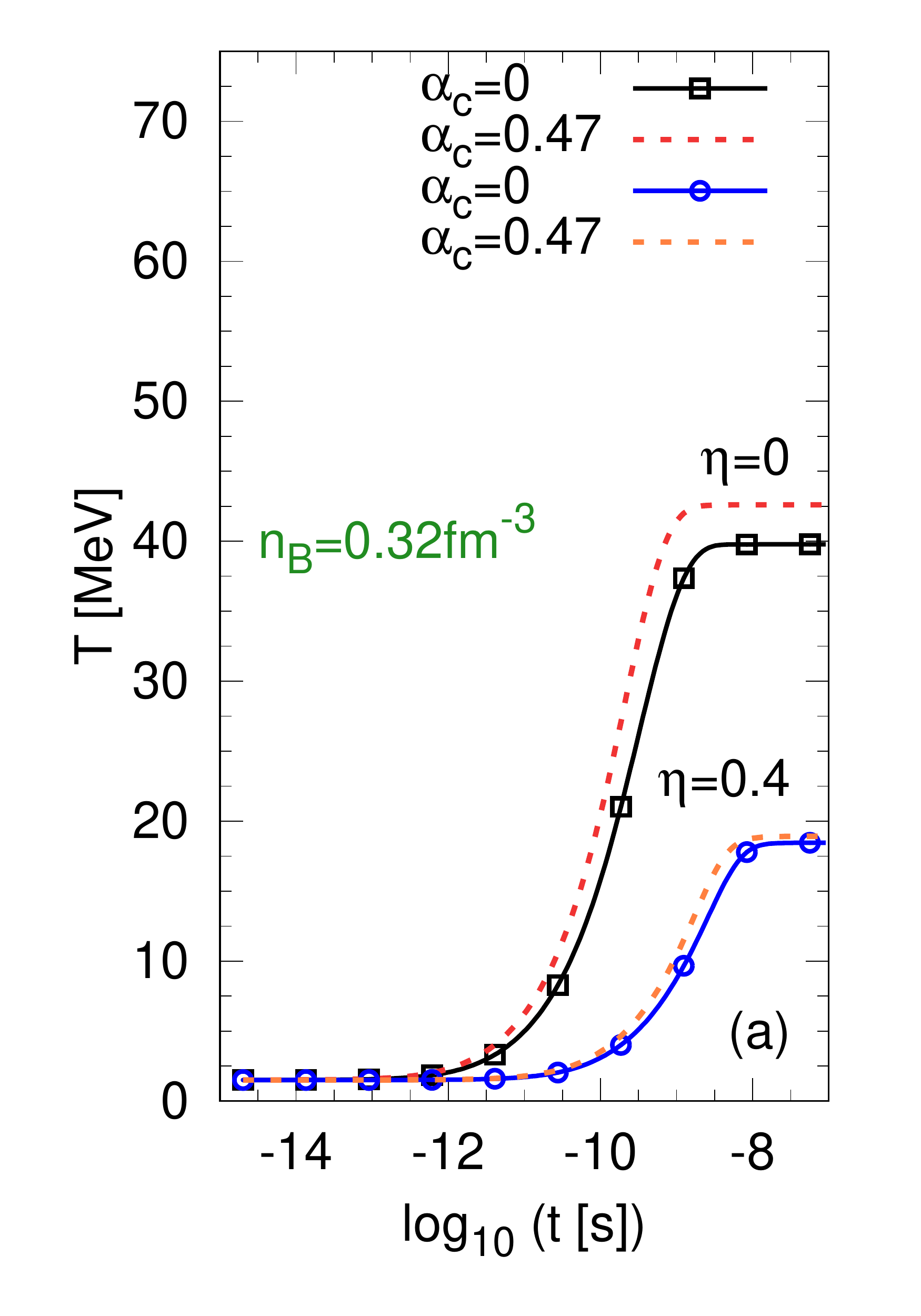}
\includegraphics[scale=0.22]{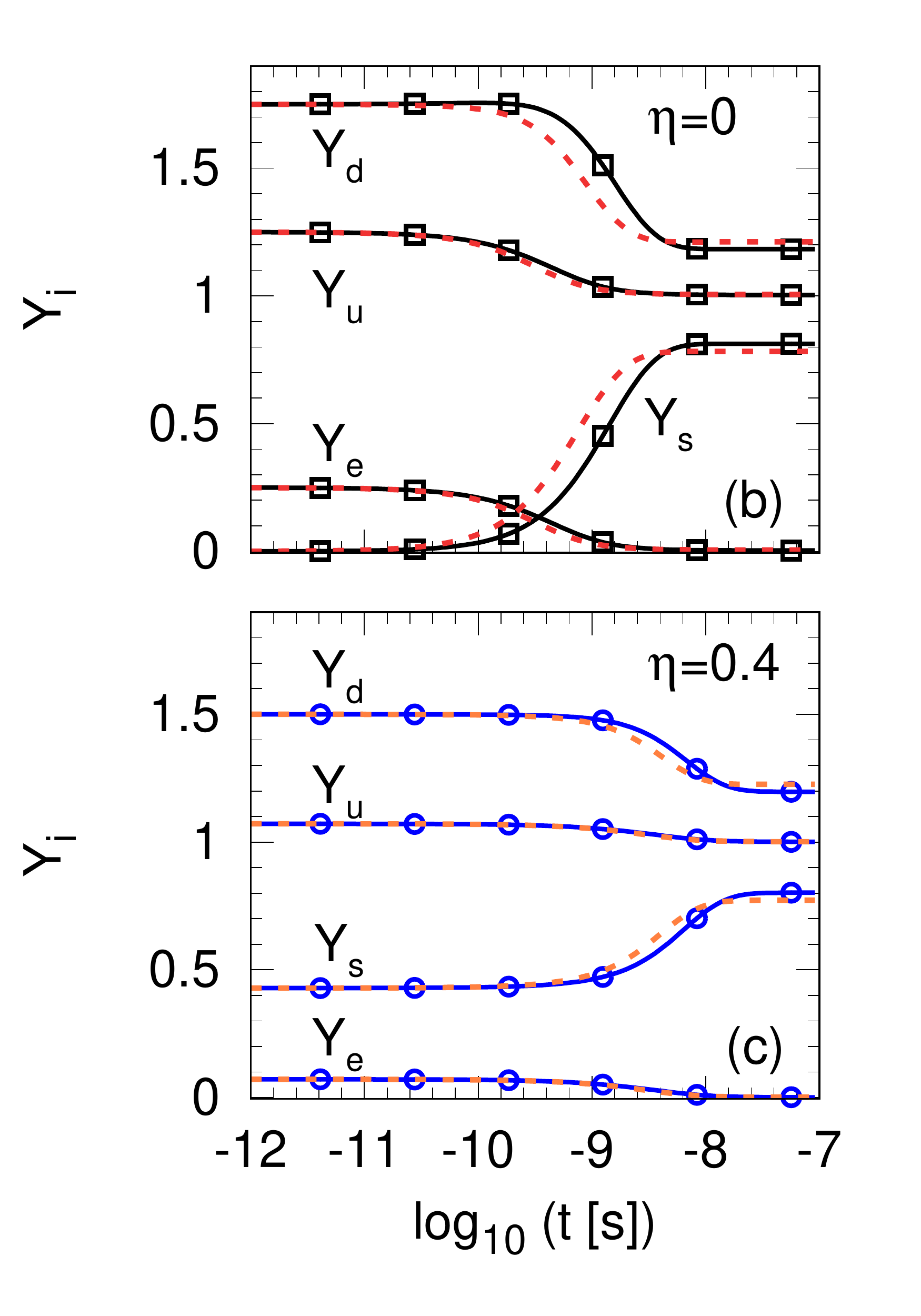}
\end{center}
\vspace{-0.6cm}
\caption{Time evolution of the temperature $T$ and the abundances $Y_i$ in a cold NS. We use $n_B=0.32 \, \mathrm{fm}^{-3}$,  $\alpha_c$ = 0 (solid lines),  and $\alpha_c$ = 0.47 (dashed lines). The initial temperature is $T_i = 1$ MeV and we use $\eta$ = 0 and 0.4 (see Table \ref{tab:dec_pns}). }
\label{TY-NS-eta0}
\end{figure}
%
\begin{figure}[tbh]
\begin{center}
\includegraphics[scale=0.22]{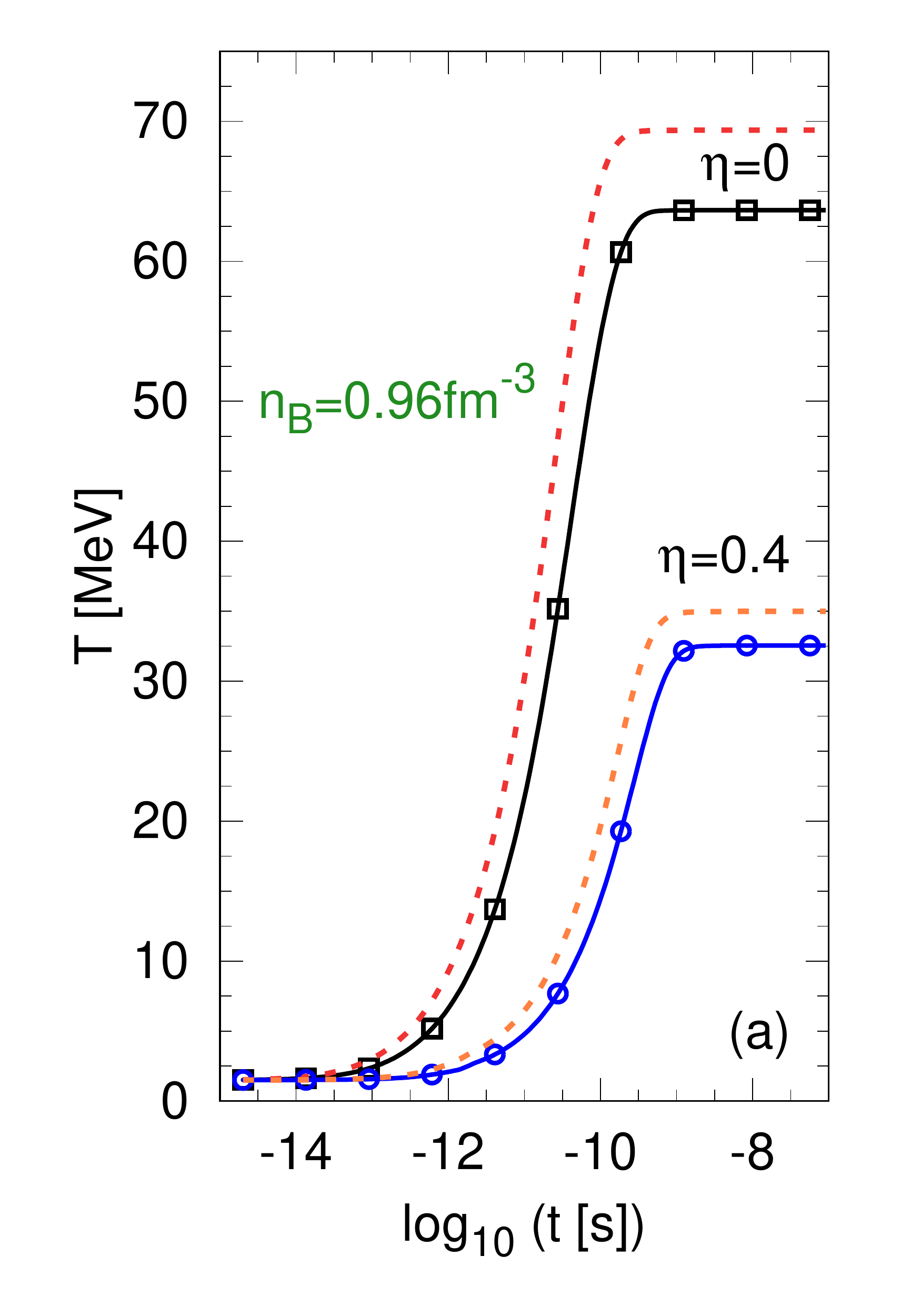}
\includegraphics[scale=0.22]{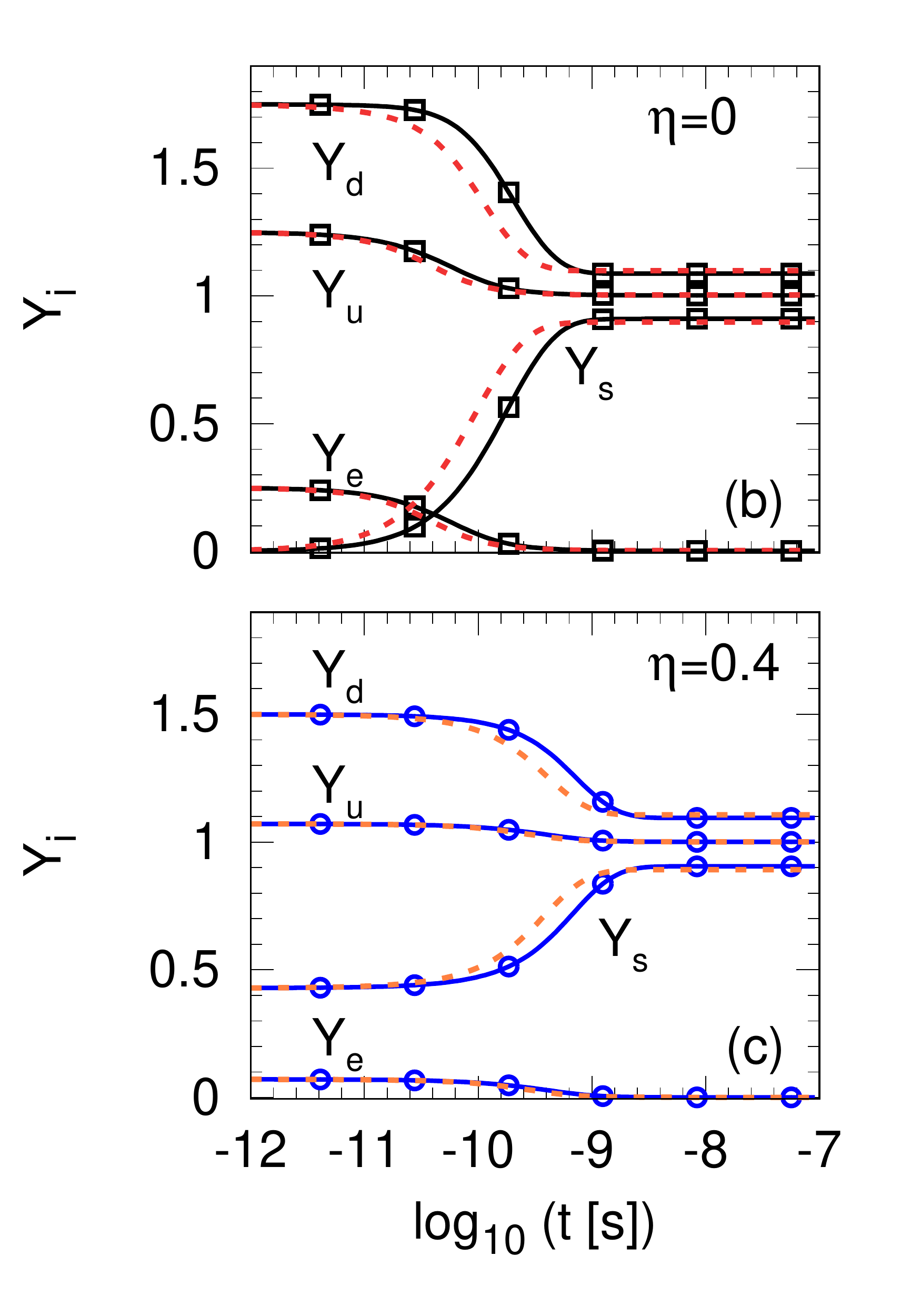}
\end{center}
\vspace{-0.6cm}
\caption{Same as in Fig. \ref{TY-NS-eta0} but for $n_B =0.96 \, \mathrm{fm}^{-3}$.}
\label{TY-NS-eta04}
\end{figure}

Let us assume that the hadron-quark combustion process occurs in a cold NS. Due to the energy released by weak reactions, quark matter within the flame becomes hot in a very short timescale. On the other hand, \textit{during flame propagation} we do not expect a significant heating of the hadronic matter ahead the combustion front because the flame propagates at a very high speed \cite{Lugones:2002vj,Niebergal:2010ds} and there isn't enough time for heat conduction to occur. Moreover, if the flame is supersonic there is no heat conduction at all to the hadronic layers during flame propagation. 
Thus, the evolution of quark matter abundances towards chemical equilibrium is governed by 
\begin{eqnarray}
\frac{dY_{u}}{dt}&=&\frac{1}{n_{B}}\left[\Gamma_{\text{I}}+\Gamma_{\text{II}}+\Gamma_{\text{III}}^{rev}-\Gamma_{\text{III}}^{dir}+\Gamma_{\text{IV}}^{rev}-\Gamma_{\text{IV}}^{dir}\right],  \\
\frac{dY_{d}}{dt}&=&\frac{1}{n_{B}}\left[-\Gamma_{\text{I}}-\Gamma_{\text{III}}^{rev}+\Gamma_{\text{III}}^{dir}-\Gamma_{\text{V}}^{dir}+\Gamma_{\text{V}}^{rev}\right] ,
\label{boltz1}
\end{eqnarray}
together with  Eq. (\ref{evolution_of_T}) for the time evolution of the temperature.
Electric charge neutrality and baryon number conservation (Eqs. \eqref{eq:charge_neutrality} and \eqref{eq:baryon_number}) can be used to relate $s$ quark and electron abundances with $u$ and $d$ quark abundances:
\begin{eqnarray}
&& Y_{s}=3-Y_{u}-Y_{d},\\
&& Y_{e}=Y_{u}-1.
\end{eqnarray}
Integrating the above equations numerically, we obtain the time evolution of the particle abundances and the temperature, as well as the neutrino emissivity as the system approaches to equilibrium.

\begin{figure}[tbh]
\begin{center}
\includegraphics[scale=0.31,angle=270]{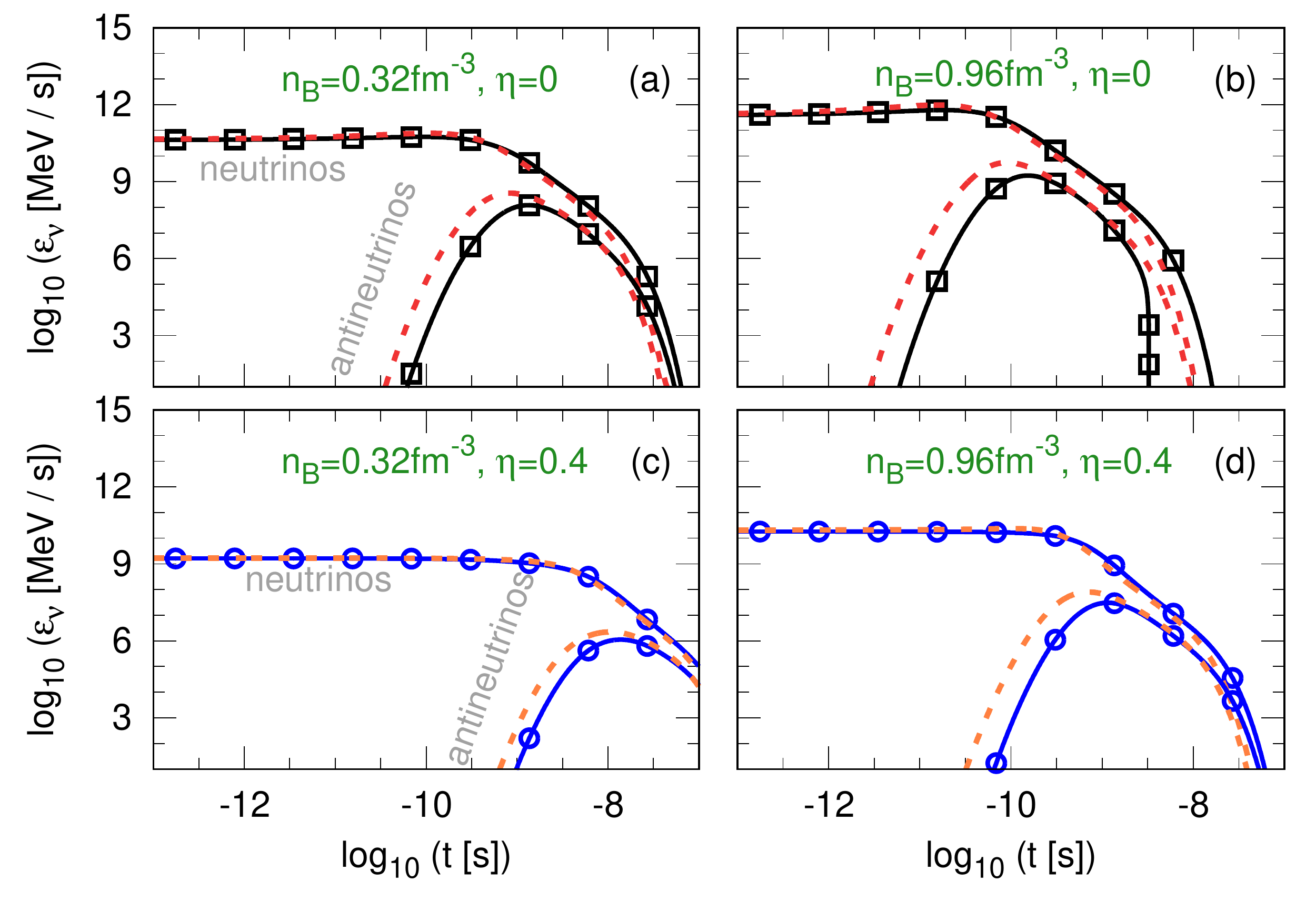}
\end{center}
\vspace{-0.6cm}
\caption{Neutrino and antineutrino energy loss rate per baryon as a function of time in a cold deleptonized NS. As in previous figures we assume  $\alpha_c$ = 0 (solid lines) and $\alpha_c$ = 0.47 (dashed lines).}
\label{Energ-NS-eta0}
\end{figure}

\begin{table*}[bth]
  \centering
\caption{Total energy per baryon released by quark matter in a cold NS in the form of neutrinos (${\cal E}_{\nu_{e}}$) and in the form of antineutrinos (${\cal E}_{\bar{\nu}_e}$). We also show ${\cal E}_{\nu_{e}} \times 10^{58}$   and   ${\cal E}_{\bar{\nu}_e} \times 10^{58}$ in order to have a rough estimate of the energy release in a typical NS with $10^{58}$ baryons. The initial and final values of the temperature are also presented.  We assumed $\alpha_c$= 0.}  
\begin{tabular*}{\linewidth}{c @{\extracolsep{\fill}} ccc|cccccc}
\hline
$n_B$ & $\xi$ & $\eta$  & $\kappa$  & $T_i$  & $T_f$  & ${\cal E}_{\nu_{e}}$ & ${\cal E}_{\nu_{e}} \times 10^{58}$ & ${\cal E}_{\bar{\nu}_e}$ & ${\cal E}_{\bar{\nu}_e} \times 10^{58}$ \\
\,[fm$^{-3}$]& & & &[MeV]&[MeV]&[MeV]&[ergs]  &[MeV]      &[ergs] \\
\hline \hline  
0.32 & 1.4 & 0  & 0   & 1 & 39.77  & 38.88 & $6.23 \times 10^{53}$  & 0.31 & $0.050 \times 10^{53}$ \\
0.32 & 1.4 & 0.4& 0   & 1 & 18.46  & 6.08 & $0.96 \times 10^{53}$  & 0.03 & $0.005 \times 10^{53}$ \\
0.96 & 1.4 & 0  & 0   & 1 & 63.65 & 60.05 & $9.62 \times 10^{53}$ & 0.50 & $0.081 \times 10^{53}$ \\
0.96 & 1.4 & 0.4& 0   & 1 & 32.55  & 9.94 & $1.59 \times 10^{53}$  & 0.07 & $0.011 \times 10^{53}$ \\
\hline
    \end{tabular*}   
  \label{tab:addlabelTeNS1}
\end{table*}

\begin{figure*}[tbh]
\begin{center}
\includegraphics[scale=0.22]{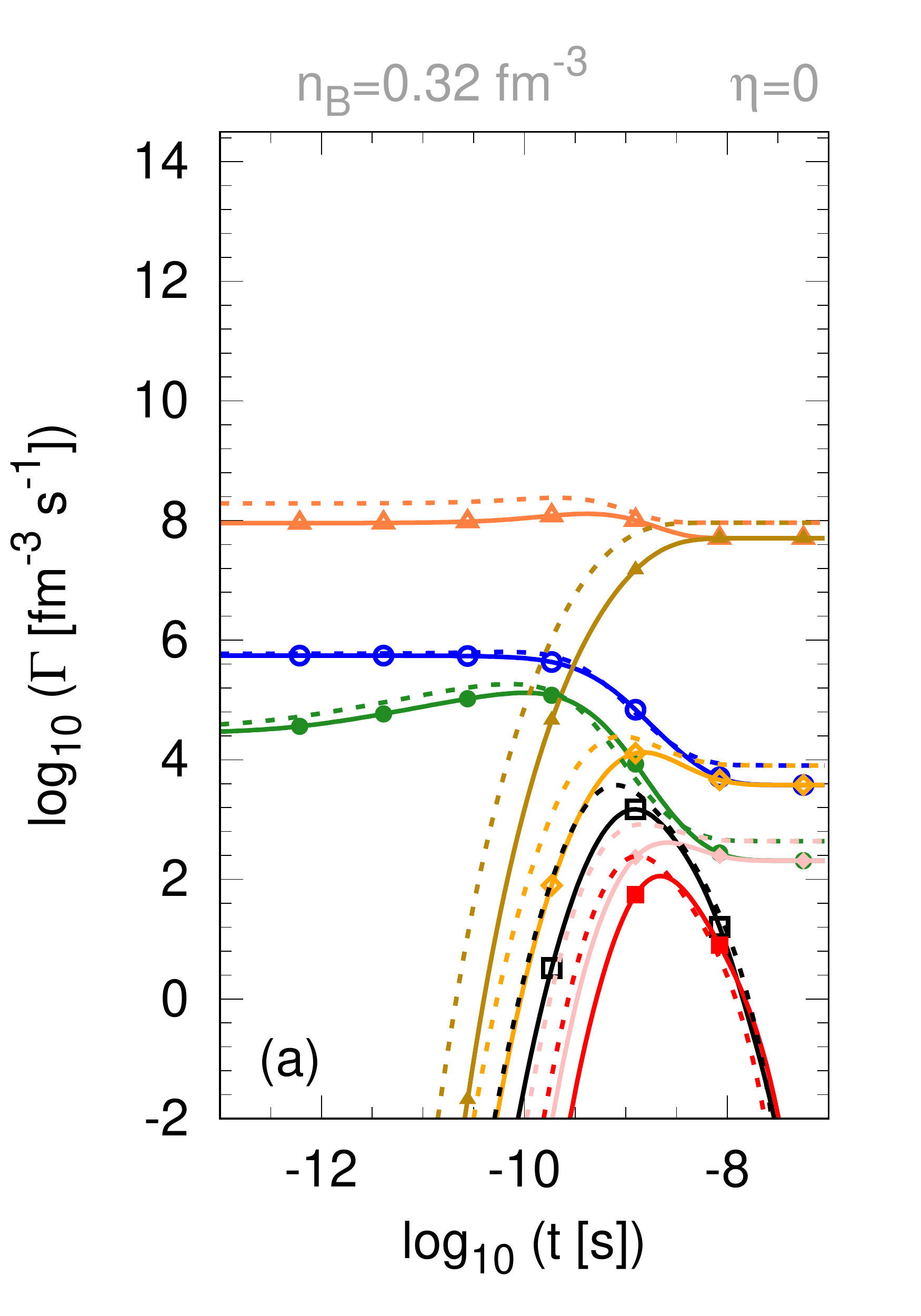}
\includegraphics[scale=0.22]{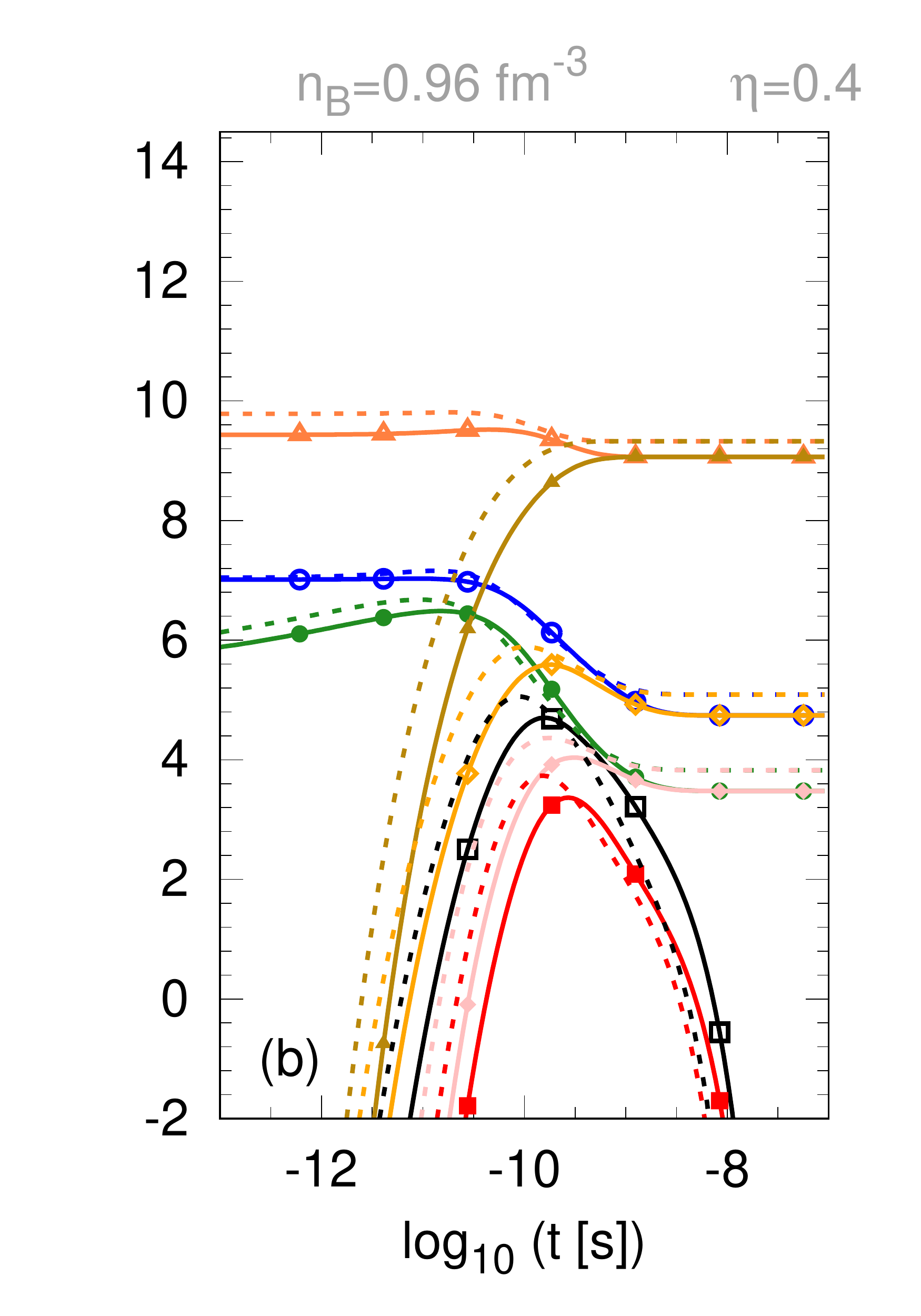}
\includegraphics[scale=0.22]{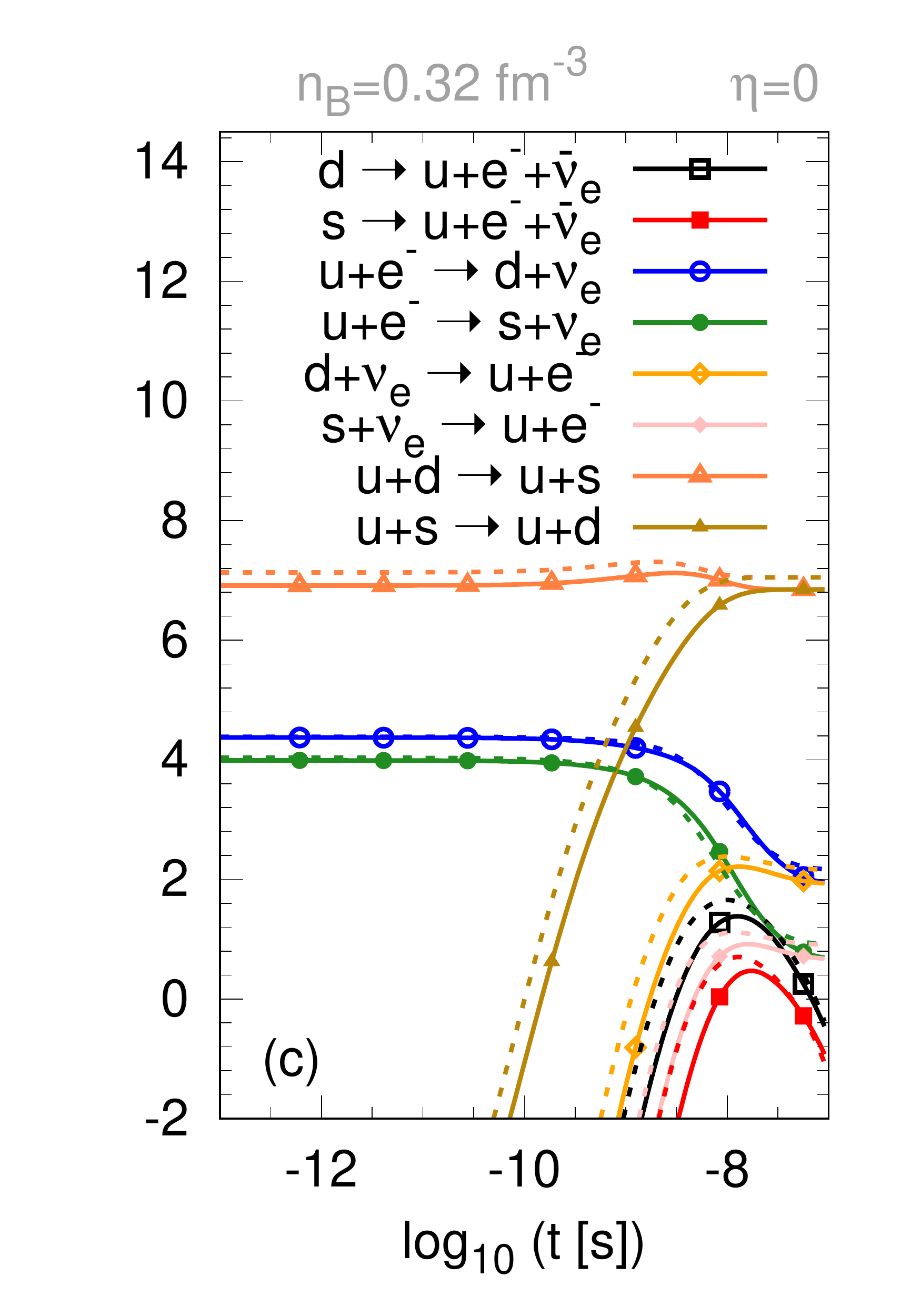}
\includegraphics[scale=0.22]{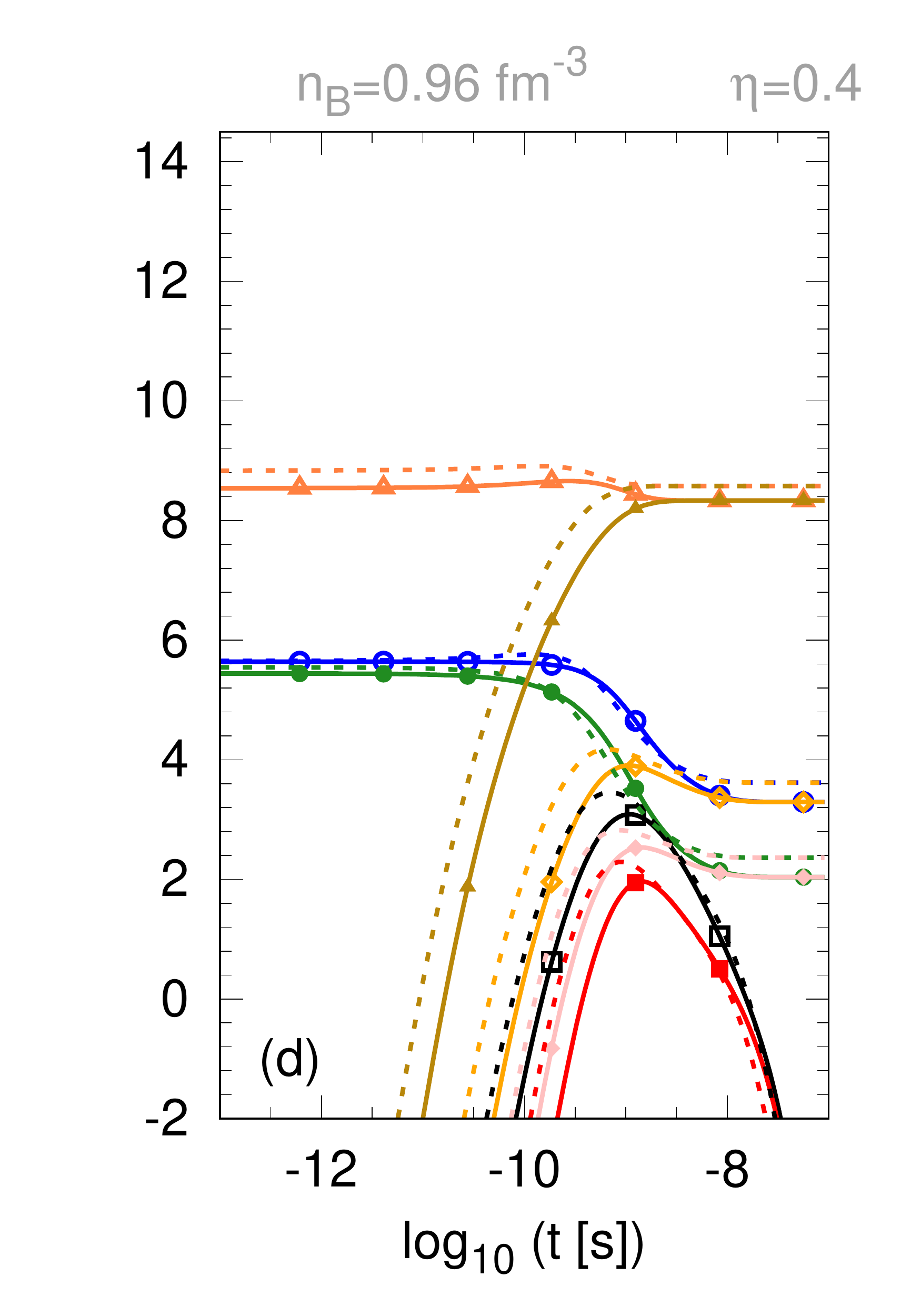}
\end{center}
\vspace{-0.6cm}
\caption{We show $\log_{10} ( \Gamma_i \mathrm{[fm^{-3} s^{-1}]} )$ as a function of time for all the relevant processes in a cold deleptonized  NS. The values of $n_B$ and $\eta$ are specified at the top of each panel. We assumed $\alpha_c = 0$ (solid lines) and $\alpha_c$ = 0.47 (dashed lines).}  
\label{ratesNS}
\end{figure*}

Figs. \ref{TY-NS-eta0} and \ref{TY-NS-eta04} show that after quark deconfinement there is  a significant increase in the temperature $T$ and in the strange quark abundance $Y_s$ in a timescale of $\sim 10^{-9} \, \mathrm{s}$  due to weak interaction decays that drive quark  matter to chemical equilibrium.  In this process, the abundances of  quarks $u$, $d$ and electrons decrease.  
The increase in $T$ strongly depends on the initial strangeness of hadronic matter: for vanishing initial strangeness the temperature increases considerably more than for large initial strangeness (see Table \ref{tab:addlabelTeNS1}). In fact, for transitions at $n_B=0.32 \, \mathrm{fm}^{-3}$ (see Fig. \ref{TY-NS-eta0}), $T$ goes from 1 MeV to about 40 MeV for $\eta$=0 (vanishing initial strangeness), and to about 20 MeV for $\eta$=0.4 (large initial strangeness). This occurs because for $\eta=0.4$ just deconfined matter is closer to chemical equilibrium than matter without strangeness. As a result, there is more energy release when $\eta=0$ and the final temperatures attained for $\eta=0$ are larger than for $\eta=0.4$.
Transitions at  higher densities present a more drastic temperature rise, as can be seen in Fig. \ref{TY-NS-eta04} for  $n_B =0.96 \, \mathrm{fm}^{-3}$ ($T$ goes from 1 MeV to about 60 MeV for  $\eta$=0  and to about 30 MeV for $\eta$=0.4). 
When the strong interaction is turned on,  temperature increments are further enhanced compared with the zero strong coupling constant case.  In fact, we find that for  $\alpha_c$ = 0.47 the final temperature is about 10\% larger than for $\alpha_c = 0$, for both $n_B=0.32 \, \mathrm{fm}^{-3}$ and $n_B =0.96 \, \mathrm{fm}^{-3}$.

The neutrino and antineutrino energy loss per baryon can be seen in Fig. \ref{Energ-NS-eta0}. The largest emissivities are attained during the first nanosecond after deconfinent.  Thereafter, they decline by several orders of magnitude in a timescale of $\sim 10^{-8} - 10^{-7} \, \mathrm{s}$. The largest values of the emissivities per baryon  range between  $10^{9} - 10^{12} \, \mathrm{MeV/s}$ for neutrinos and  $10^{6} - 10^{9} \, \mathrm{MeV/s}$  for antineutrinos. Notice that the emissivity follows the same trend as the temperature: it increases with $\alpha_c$,  decreases with $\eta$,  and increases with the baryon number density.

The relevance of the different weak interaction processes in a cold NS is analyzed in Fig. \ref{ratesNS}. In all cases, the nonleptonic process $u + d \rightarrow u + s$ dominates the rate until matter reaches chemical equilibrium.

The electron capture reactions have a smaller contribution to the rate, but they are the most important processes that emit neutrinos. However, once matter reaches equilibrium, their contribution to the total rate decays steeply.
Notice that the contribution of the decay of $s$ and $d$ quarks to the total rate is always negligible. After $10^{-9}-10^{-8}$ s, chemical equilibrium is maintained essentially  by the two nonleptonic processes $u + d \leftrightarrow u + s$. 

Finally, we calculate the total energy released by each baryon of quark matter in the form of neutrinos $({\cal E}_{\nu_{e}})$ and antineutrinos $({\cal E}_{\bar{\nu}_e})$.  Both, ${\cal E}_{\nu_{e}}$ and  ${\cal E}_{\bar{\nu}_e}$, can be obtained by integrating the corresponding emissivities with time, since the initial time of hadron deconfinement (at region 2 in Fig. \ref{fig:flame}) until the moment when chemical equilibrium is attained (when the baryon enters region 4 in Fig. \ref{fig:flame}). Our results are shown in Table \ref{tab:addlabelTeNS1} and show that each baryon releases $\sim 6-60 \, \mathrm{MeV}$ in neutrinos and  $\sim 0.03-0.5 \, \mathrm{MeV}$ in antineutrinos depending on the initial strange quark abundance and the initial baryon number density. For a typical NS with $10^{58}$ baryons, we can estimate an order of magnitude of the energy released if the whole star were converted into quark matter.  This value is just a rough estimate because e.g. we don't consider that the density of matter changes along the star. Anyway, it gives a hint of the ``chemical''  energy released by the hadron-quark conversion and suggests that it could be sufficient to power a gamma ray burst. Notice that this estimate doesn't take into account the additional amount of (gravitational) energy that would be released by the rearrangement of the stellar configuration after the hadron-quark conversion.

\section{Results for hot neutron stars with trapped neutrinos} 
\label{repns}

\begin{figure*}[tbh]
\begin{center}
\includegraphics[scale=0.22]{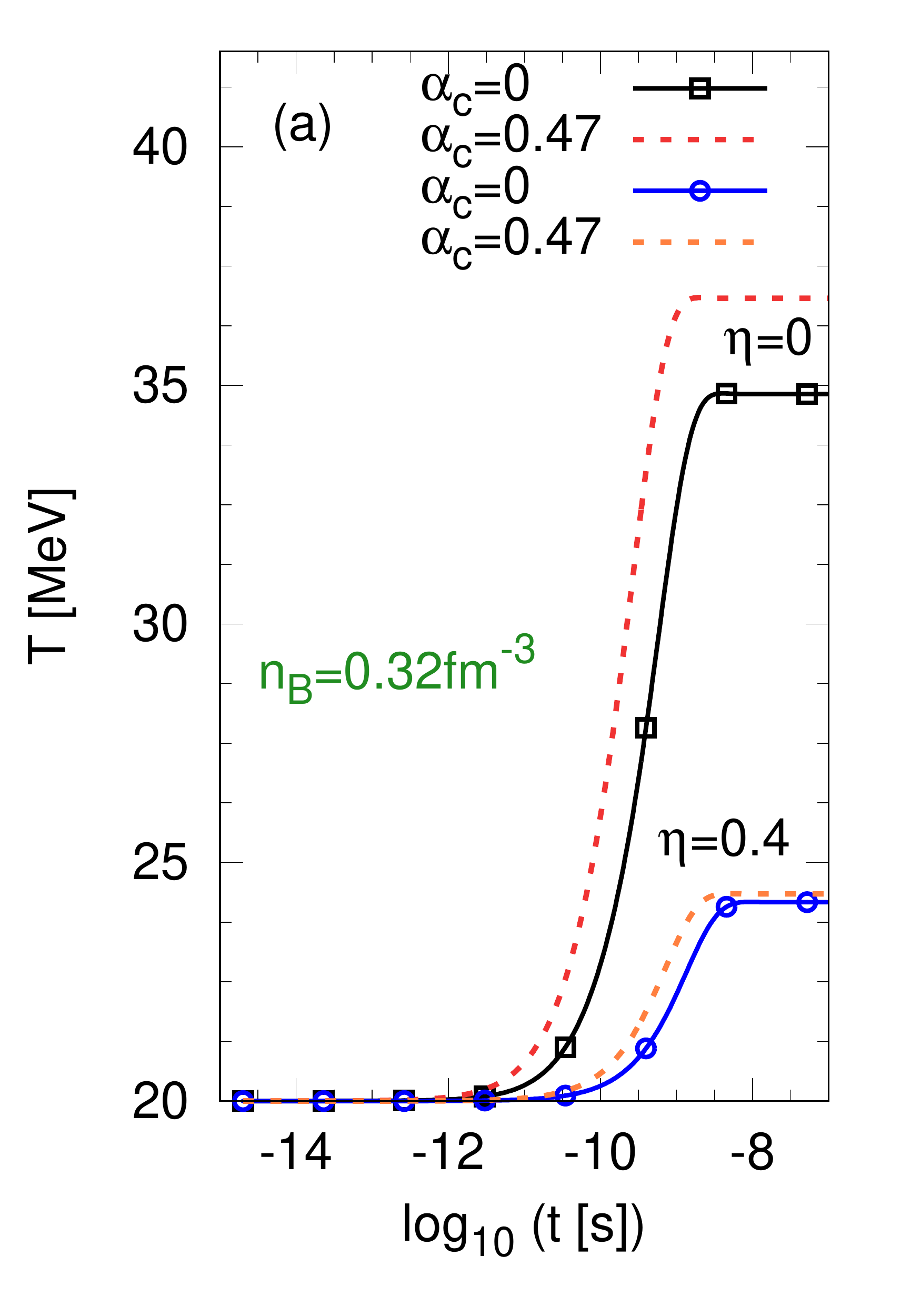}
\includegraphics[scale=0.22]{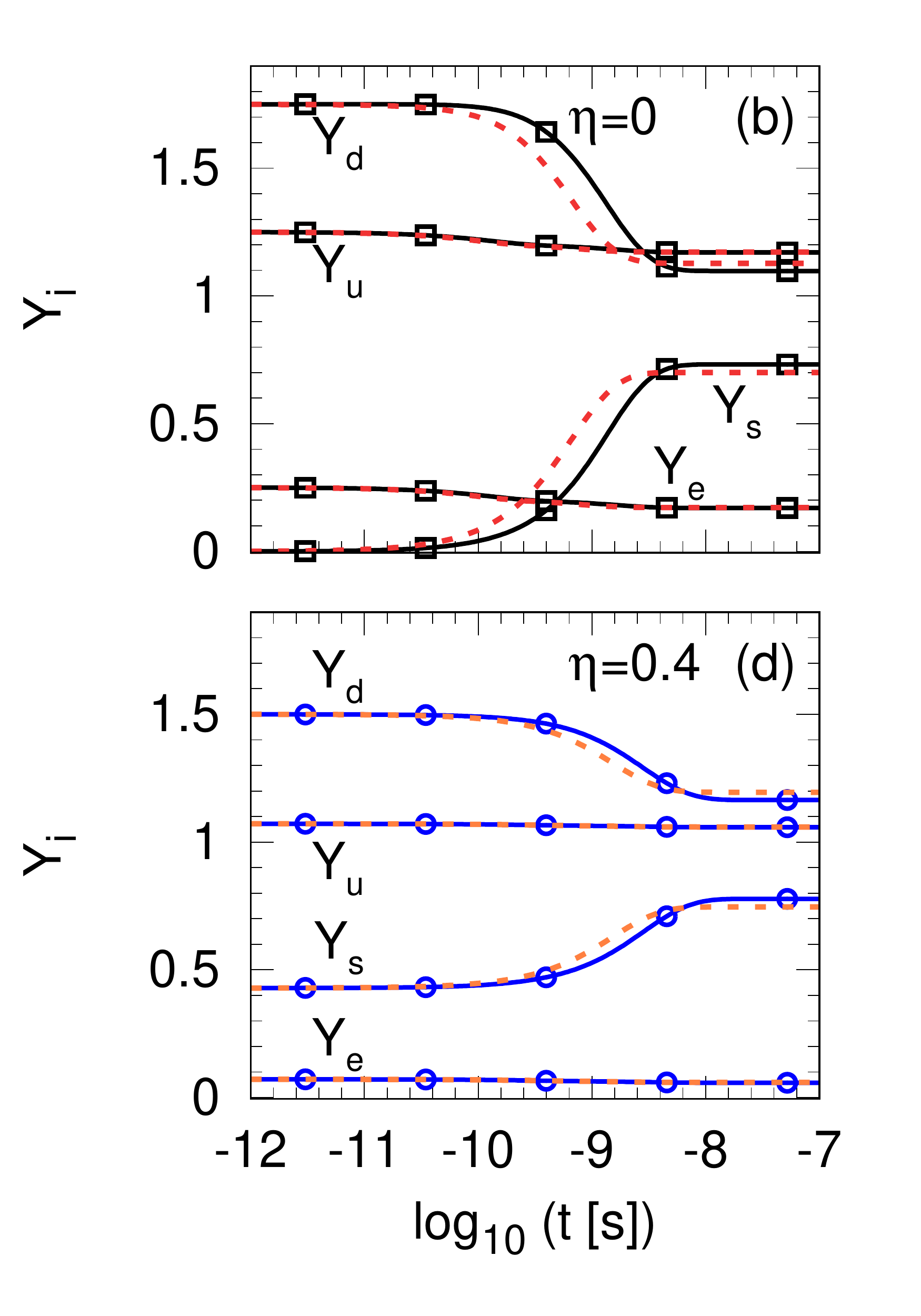}
\includegraphics[scale=0.22]{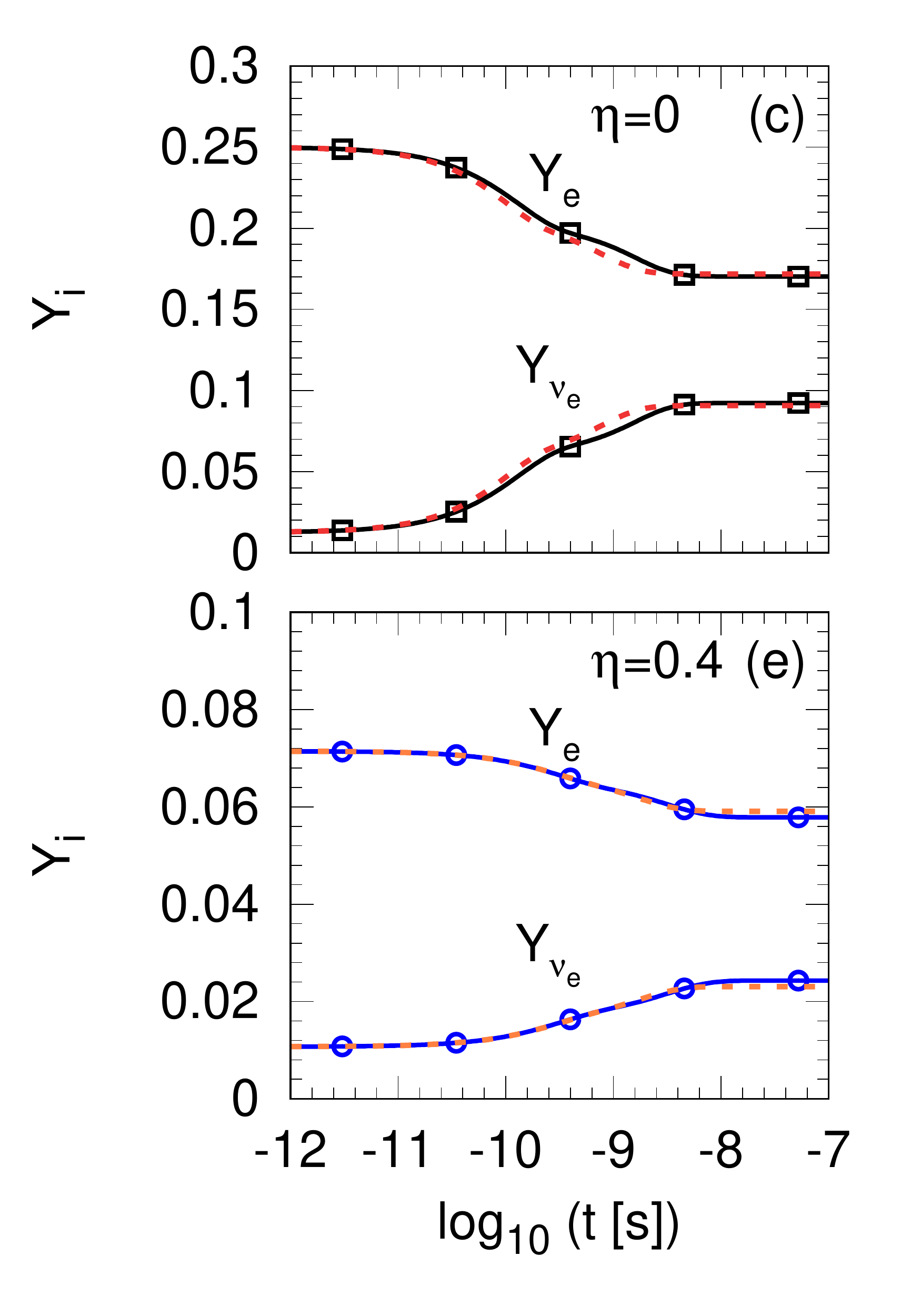}
\end{center}
\vspace{-0.6cm}
\caption{Time evolution of the temperature $T$ and the particle abundances $Y_i$ in hot NS matter with initial temperature $T_i = 20 \, \mathrm{MeV}$. The initial composition is described by the parameters $\eta=0$ and $\eta=0.4$ (see Table \ref{tab:dec_pns}) and we use $n_B=0.32 \, \mathrm{fm}^{-3}$.}
\label{TY-PNS20-032}
\end{figure*}

\begin{figure*}[tbh]
\begin{center}
\includegraphics[scale=0.22]{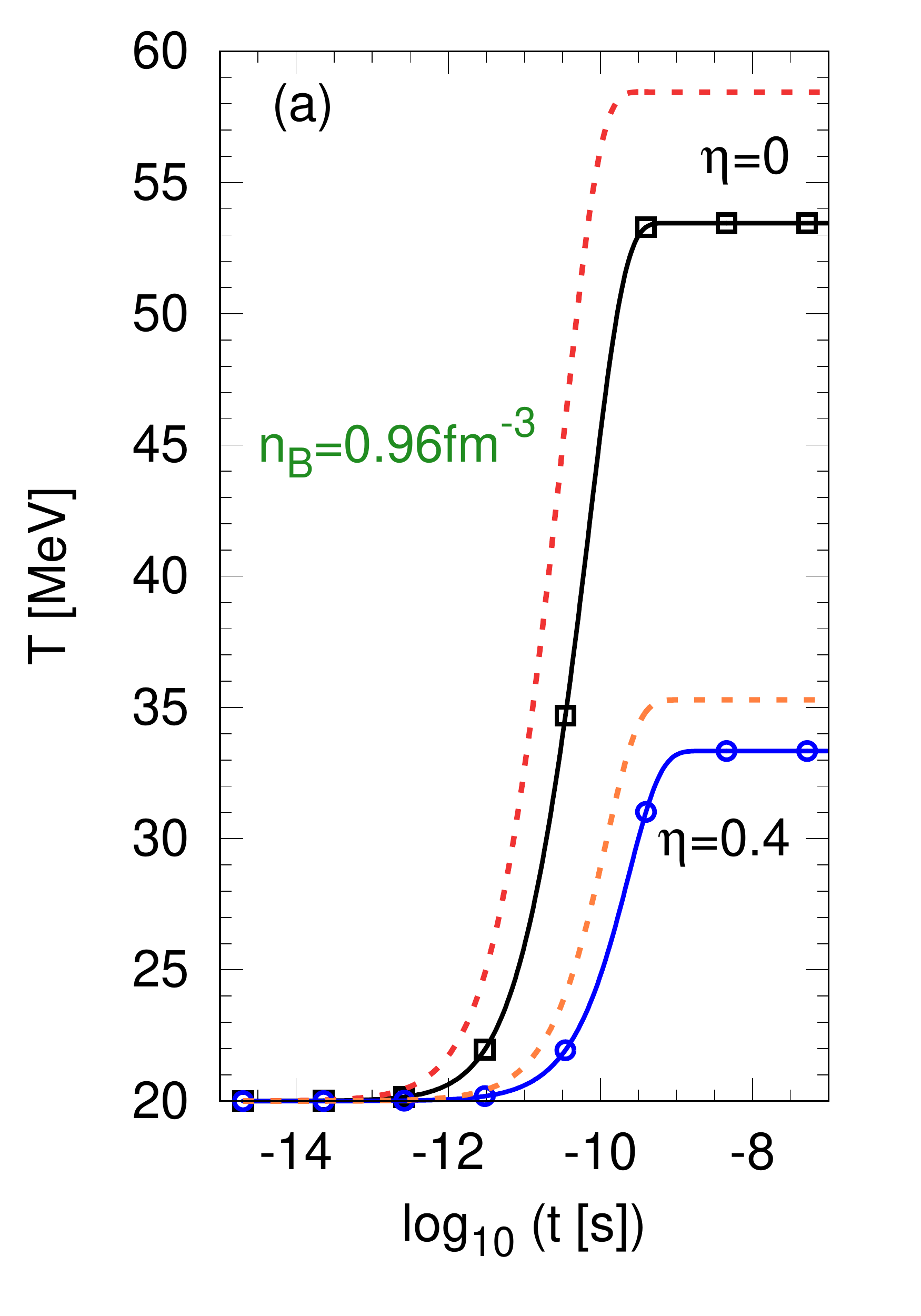}
\includegraphics[scale=0.22]{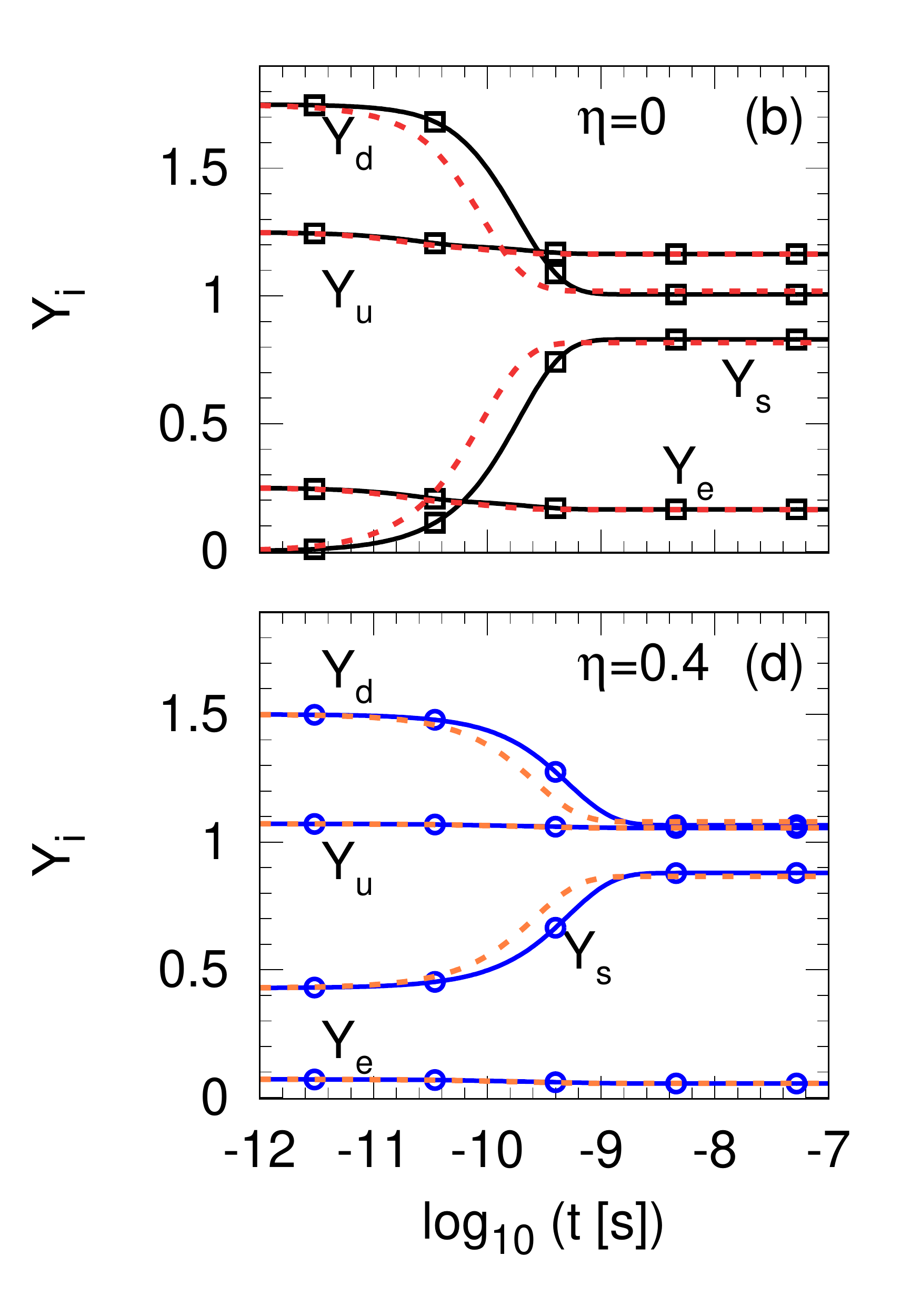}
\includegraphics[scale=0.22]{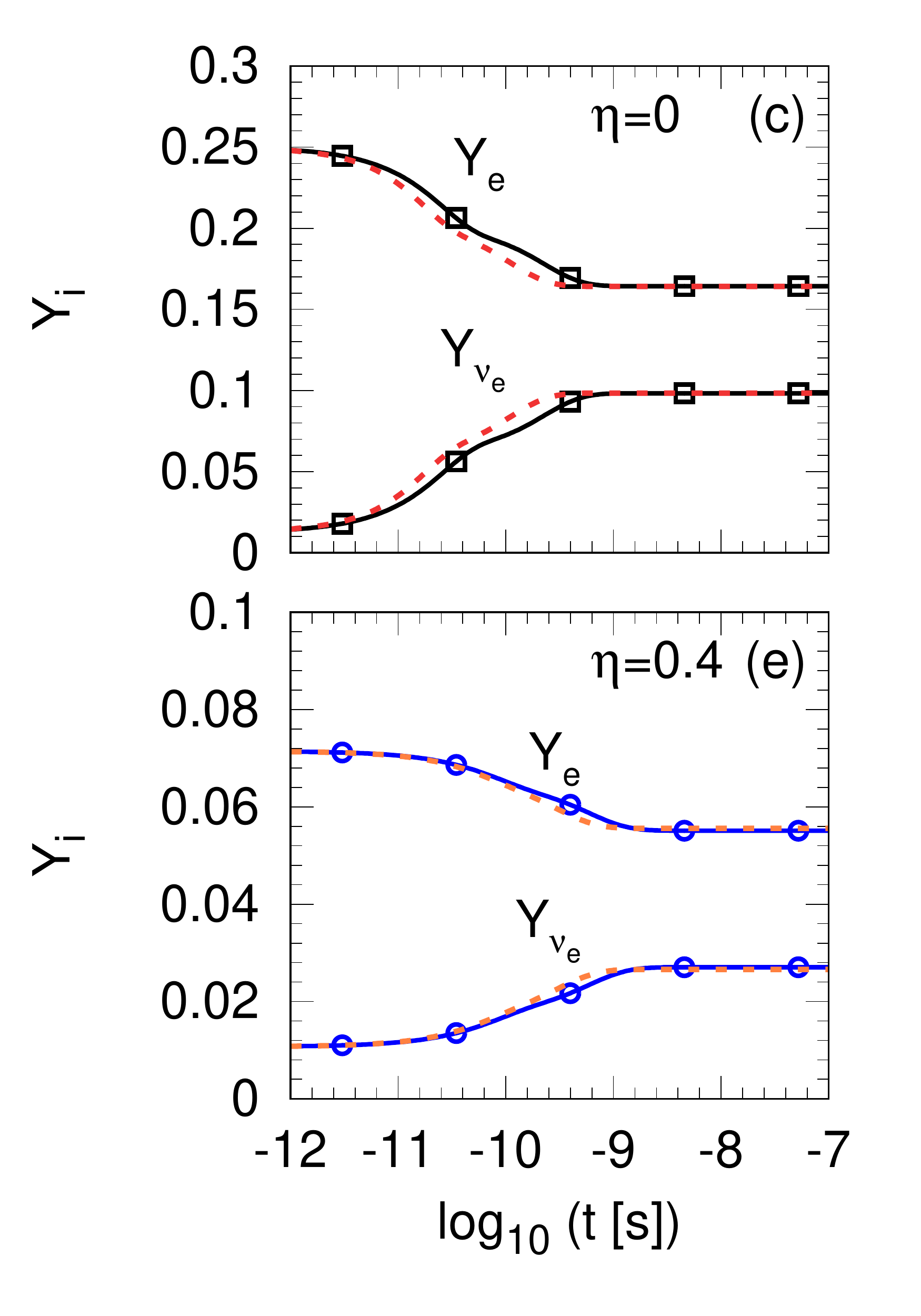}
\end{center}
\vspace{-0.6cm}
\caption{Same as in Fig. \ref{TY-PNS20-032} but for $n_B =0.96 \, \mathrm{fm}^{-3}$.}
\label{TY-PNS20-096}
\end{figure*}

\begin{figure*}[tbh]
\begin{center}
\includegraphics[scale=0.22]{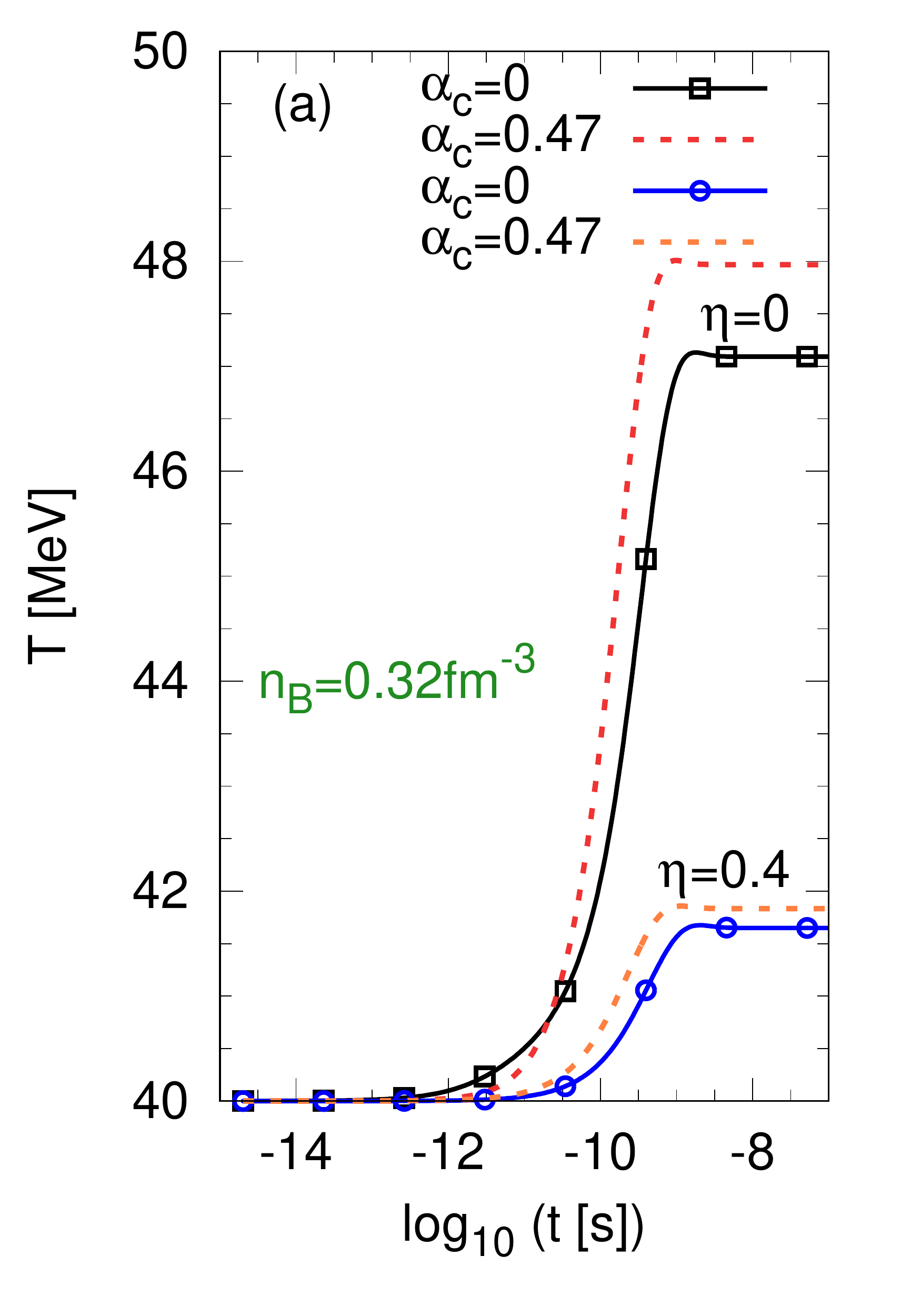}
\includegraphics[scale=0.22]{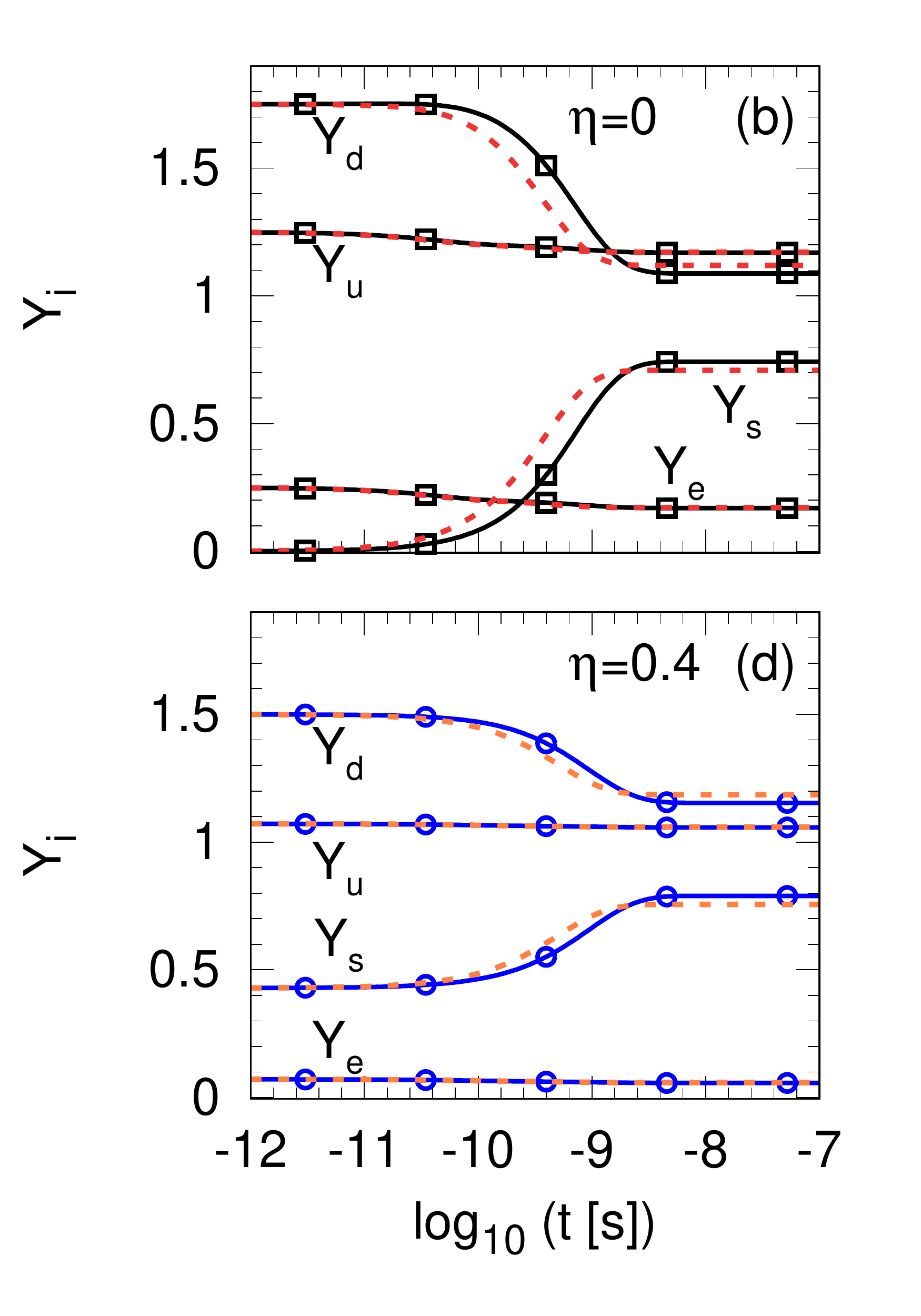}
\includegraphics[scale=0.22]{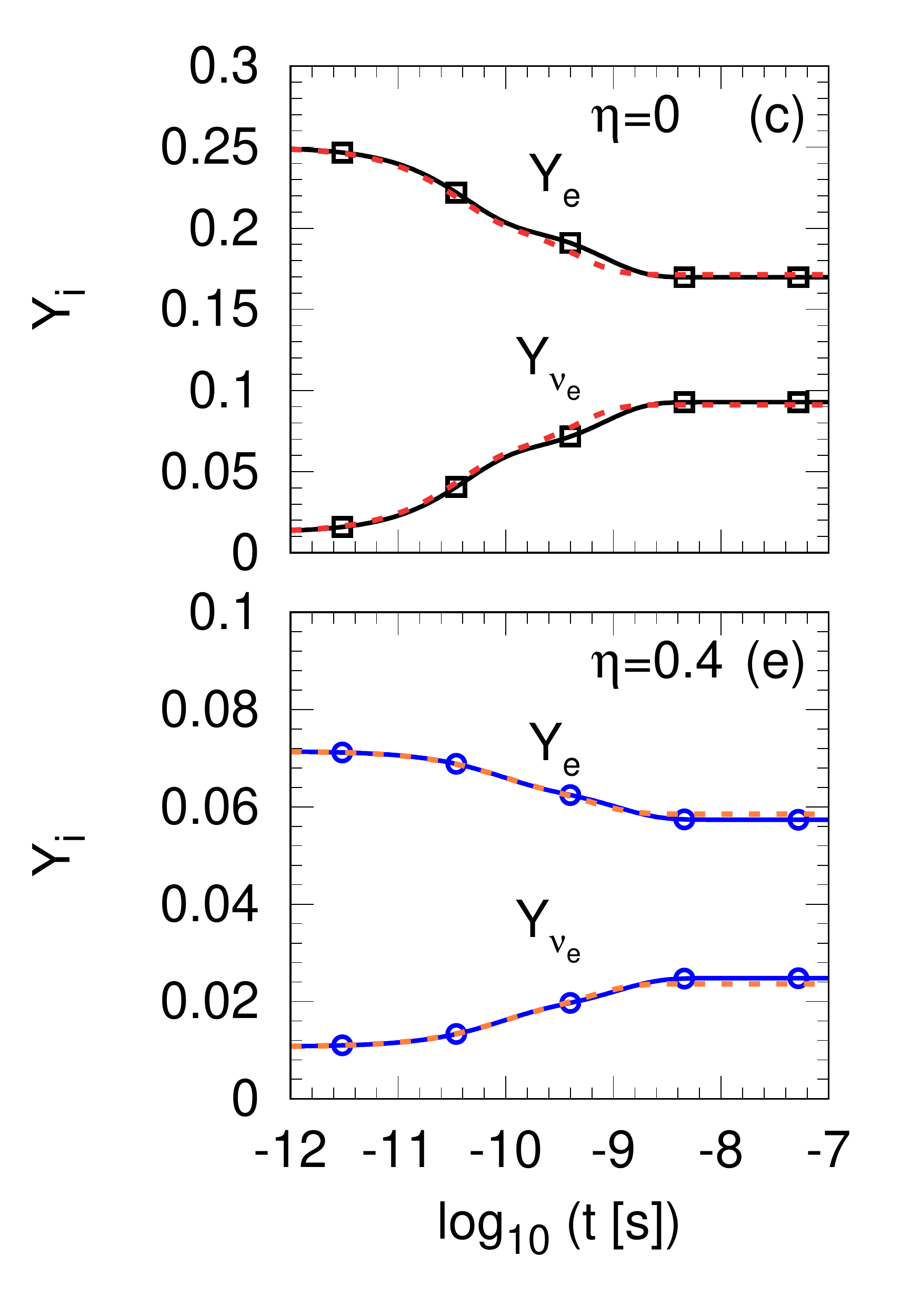}
\end{center}
\vspace{-0.6cm}
\caption{Time evolution of the temperature $T$ and the particle abundances $Y_i$ in hot leptonized NS matter with initial temperature $T_i = 40 \, \mathrm{MeV}$ and $n_B=0.32 \, \mathrm{fm}^{-3}$. }
\label{TY-PNS40-032}
\end{figure*}

\begin{figure*}[tbh]
\begin{center}
\includegraphics[scale=0.22]{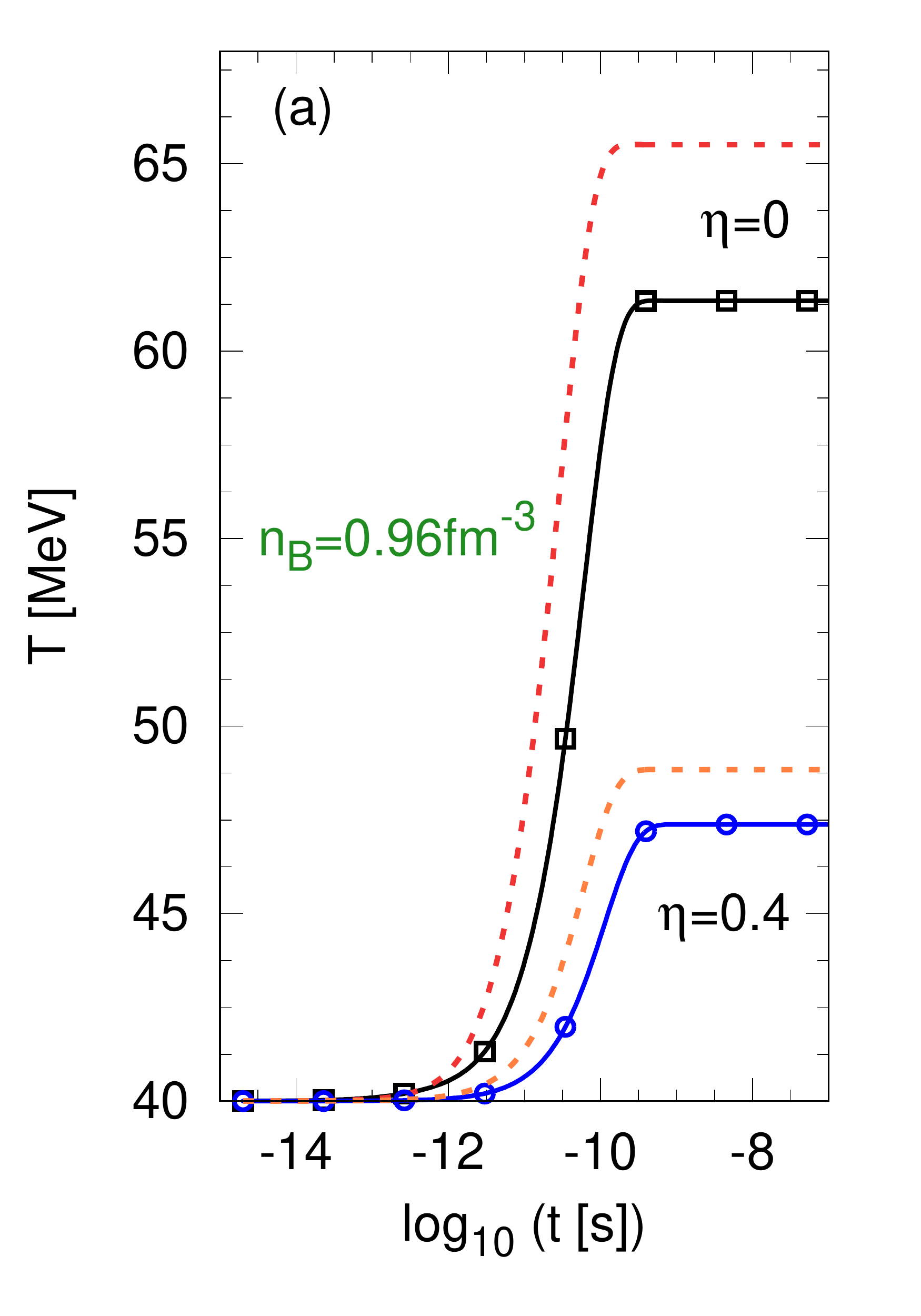}
\includegraphics[scale=0.22]{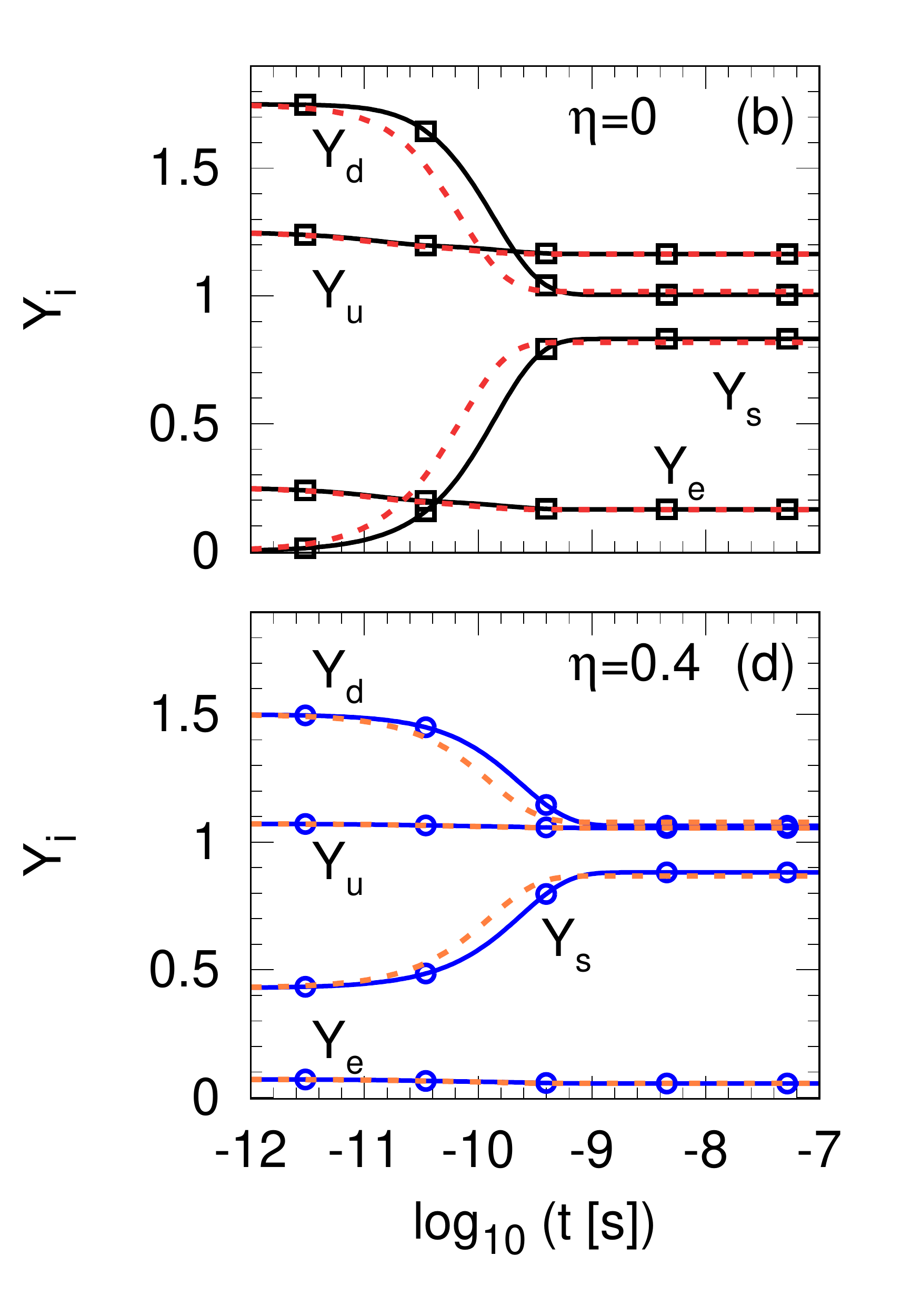}
\includegraphics[scale=0.22]{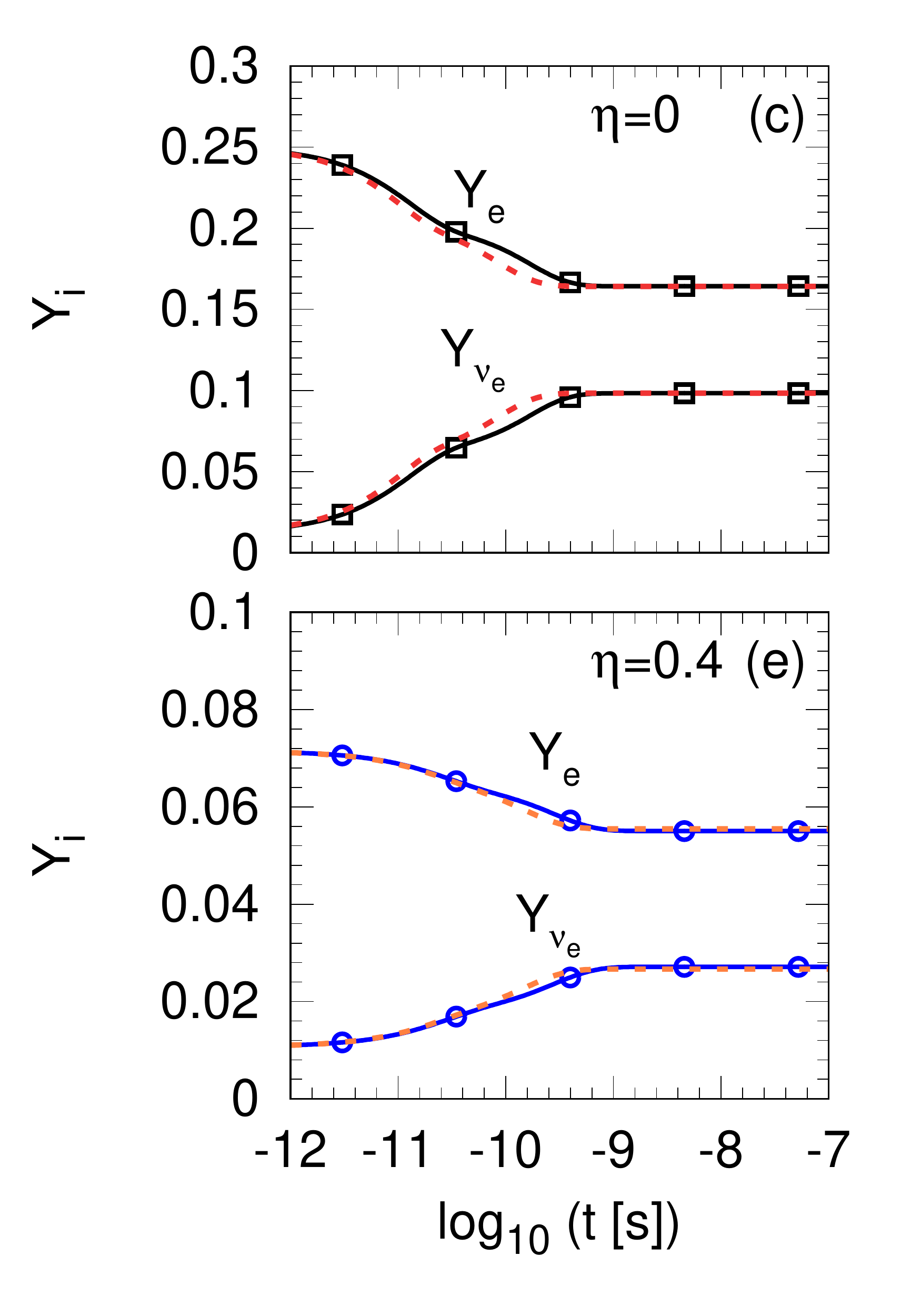}
\end{center}
\vspace{-0.6cm}
\caption{Same as in Fig. \ref{TY-PNS40-032} but for $n_B =0.96 \, \mathrm{fm}^{-3}$.}
\label{TY-PNS40-096}  
\end{figure*}

%
\begin{figure}[tb]
\begin{center}
\includegraphics[scale=0.31,angle=270]{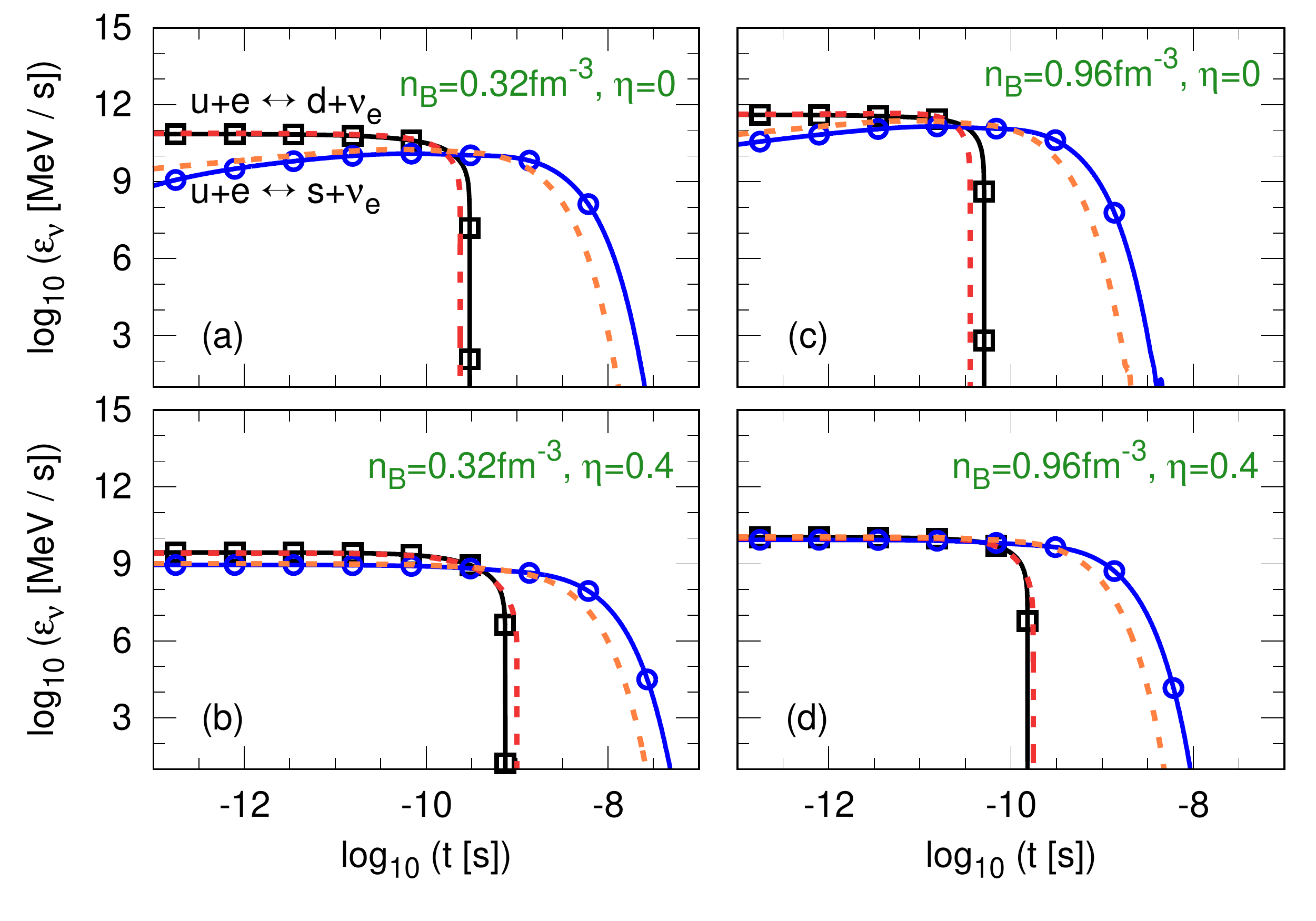} 
\end{center}
\vspace{-0.6cm}
\caption{Time evolution of the neutrino energy loss rate per baryon $\varepsilon_\nu$ for hot NS matter with initial temperature $T_i= 20\, \mathrm{MeV}$. We assume  $\alpha_c$ = 0 (solid lines) and $\alpha_c$ = 0.47 (dashed lines).}
\label{Energ-PNS20}
\end{figure}

%
\begin{figure}[tb]
\begin{center}
\includegraphics[scale=0.31,angle=270]{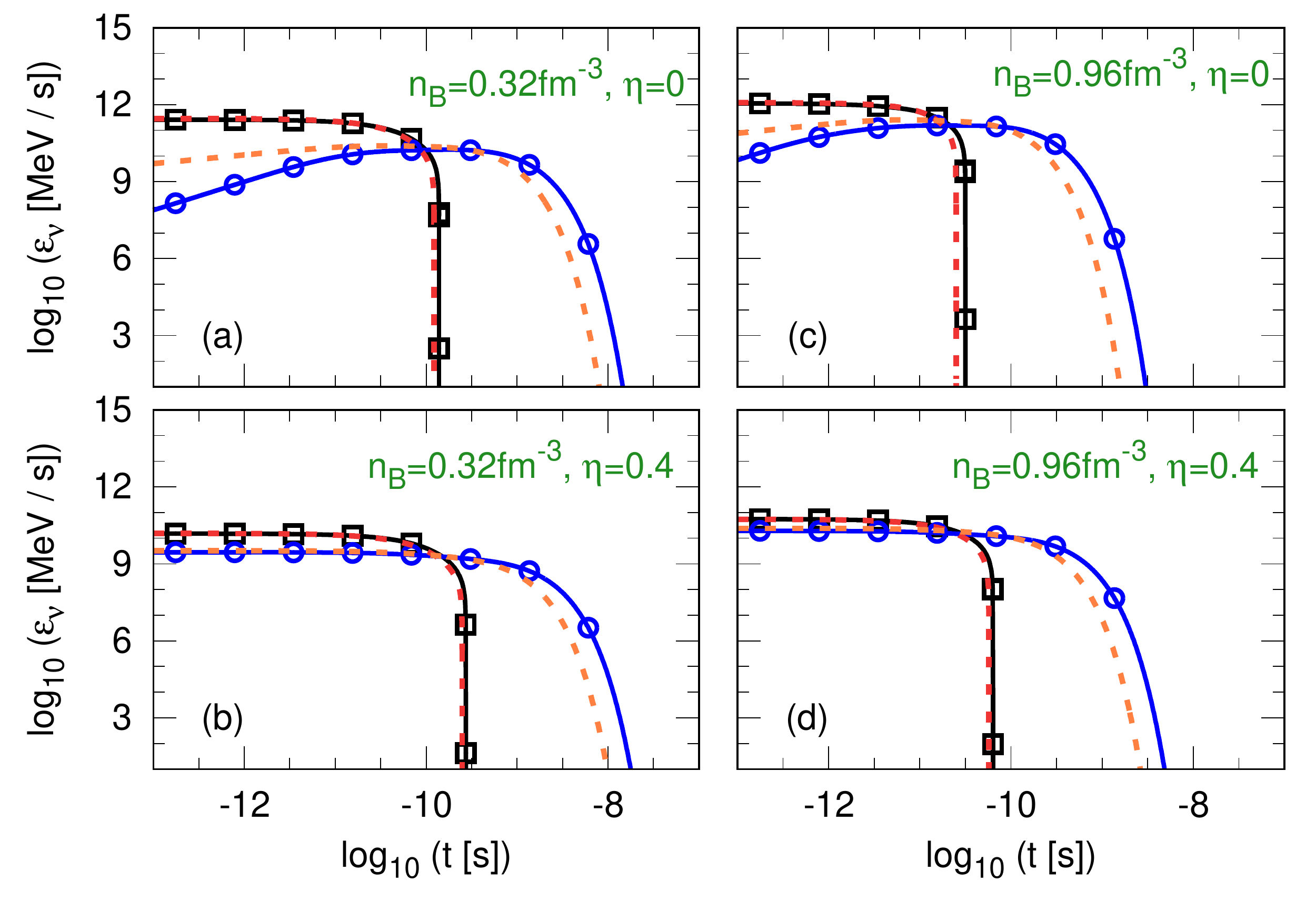} 
\end{center}
\vspace{-0.6cm}
\caption{Same as in Fig.  \ref{Energ-PNS20} but for $T_i =40 \, \mathrm{MeV}$.}
\label{Energ-PNS40}
\end{figure}

%
\begin{figure*}[tb]
\begin{center}
\includegraphics[scale=0.22]{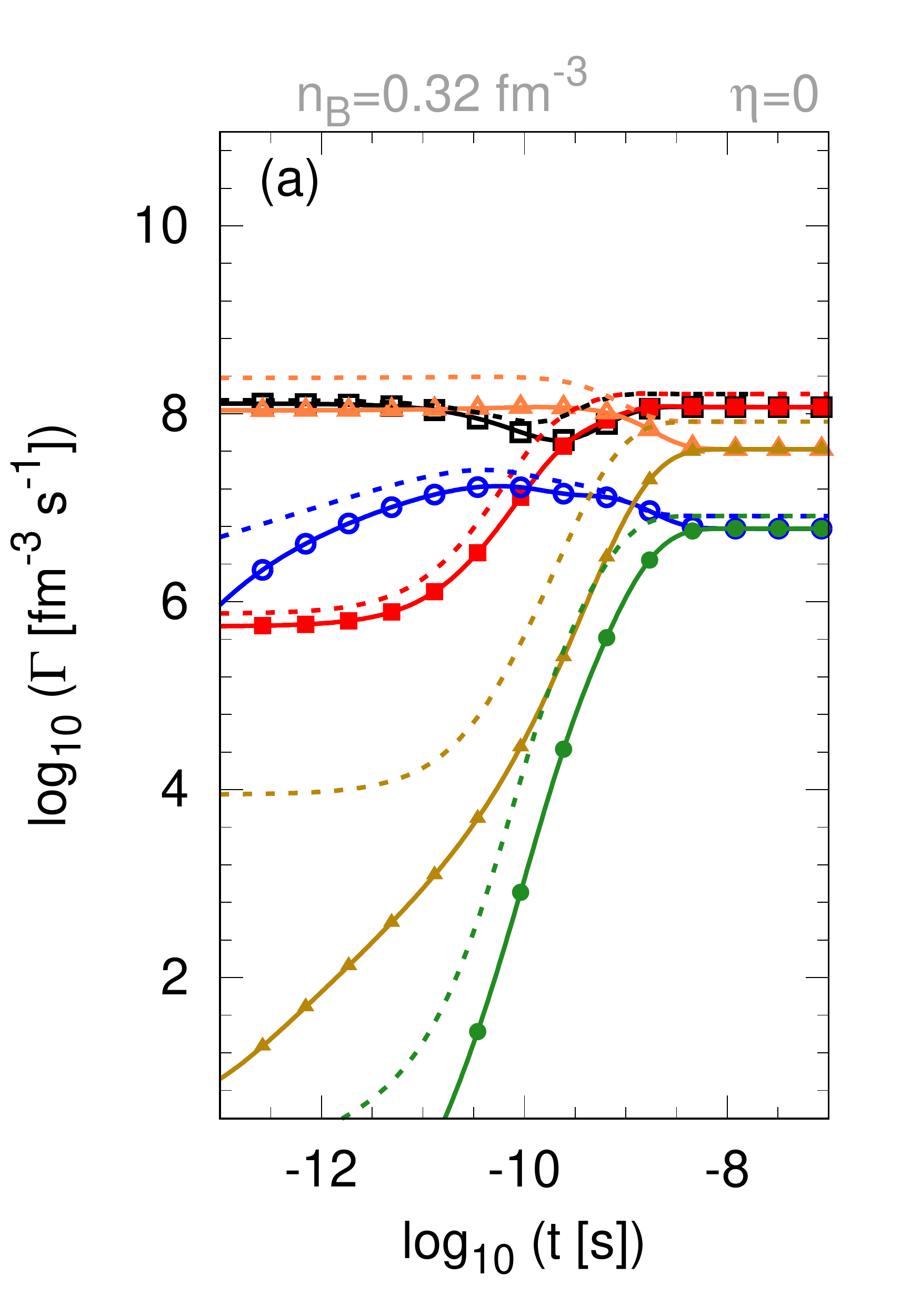}
\includegraphics[scale=0.22]{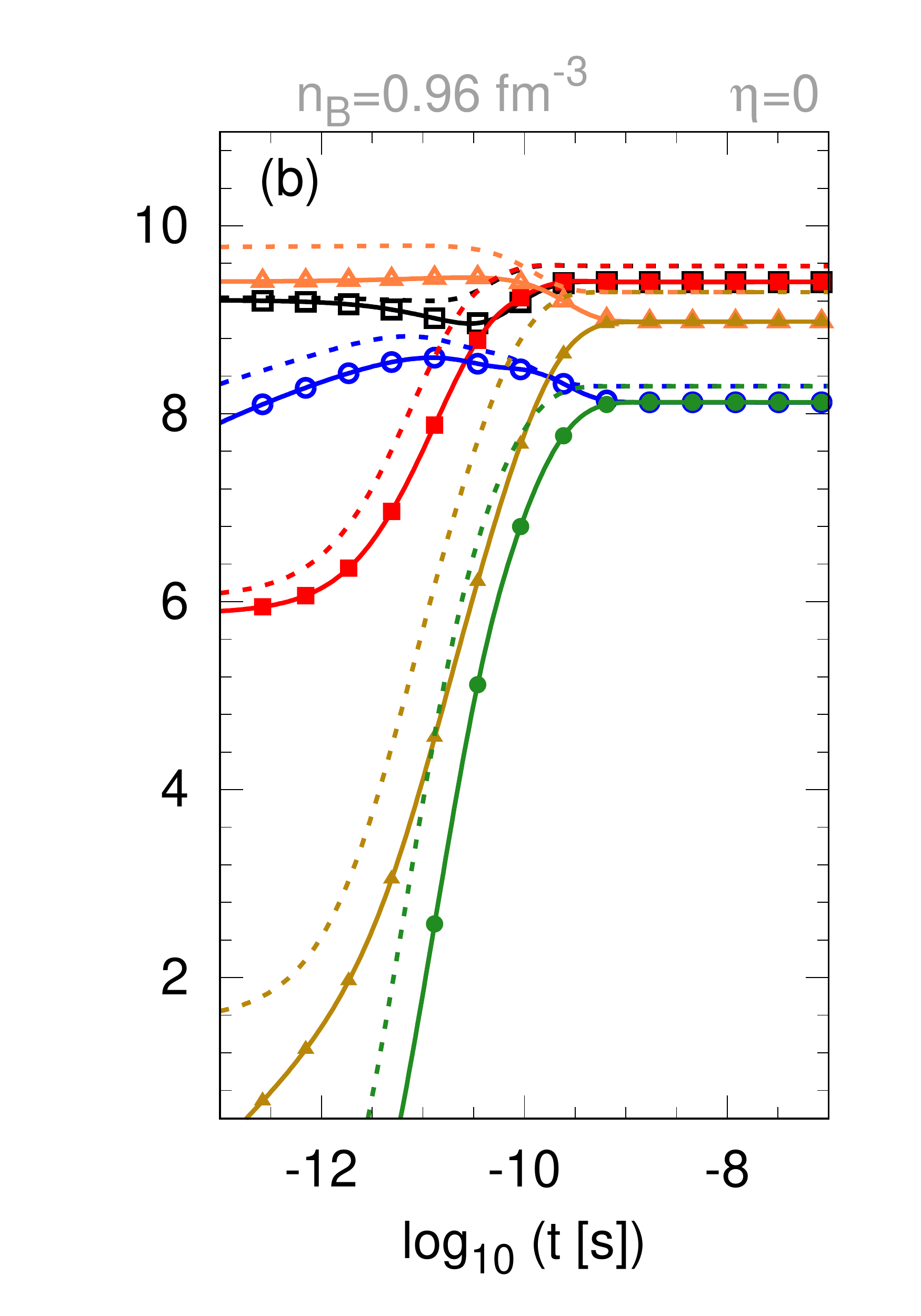}
\includegraphics[scale=0.22]{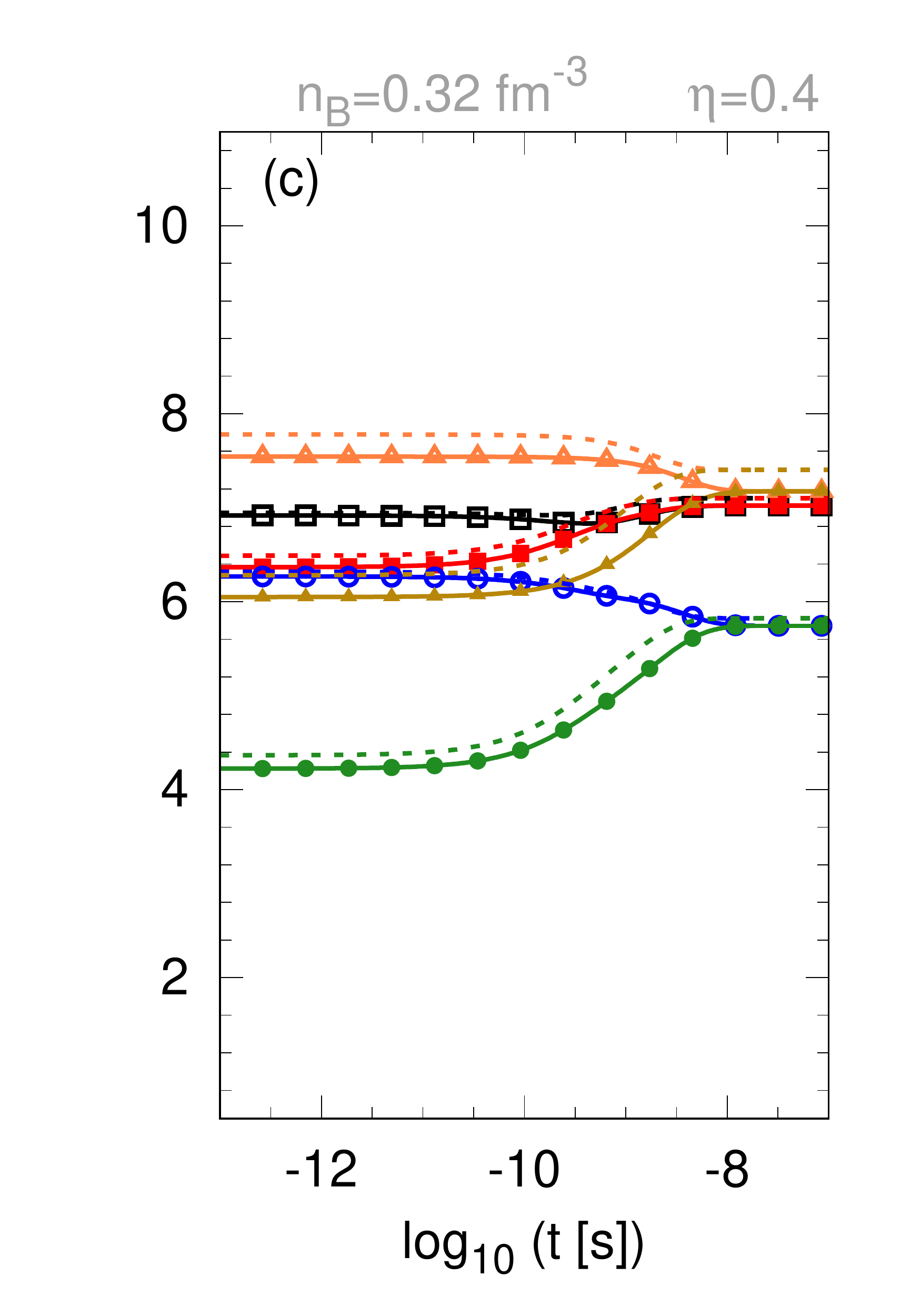}
\includegraphics[scale=0.22]{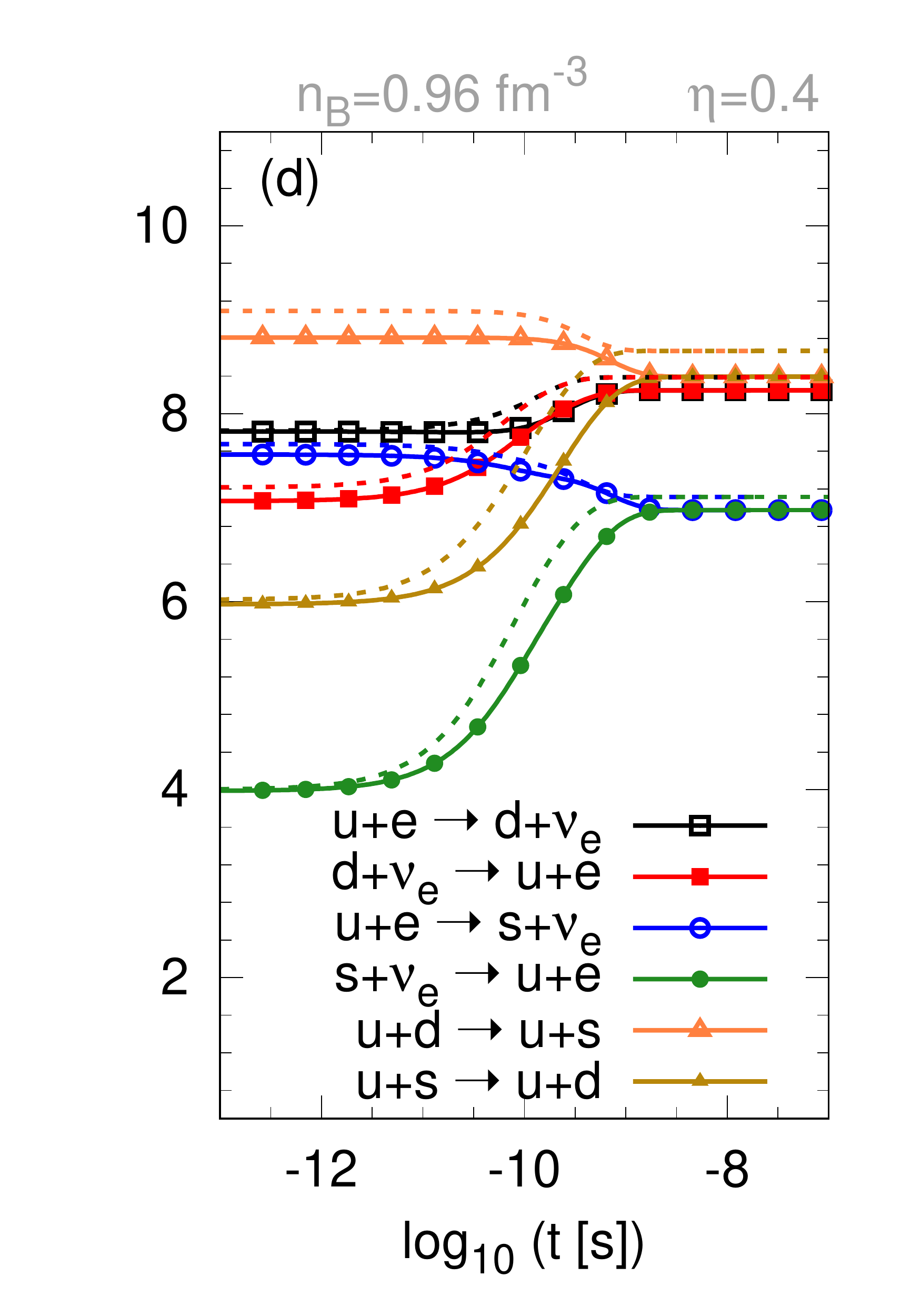}
\end{center}
\vspace{-0.6cm}
\caption{We show $\log_{10} ( \Gamma_i \mathrm{[fm^{-3} s^{-1}]} )$ as a function of time for all the relevant processes in a hot NS. The values of $n_B$ and $\eta$ are specified at the top of each panel. We used $T_i= 20\, \mathrm{MeV}$ and $ \alpha_c= 0$ (solid lines) and $\alpha_c$ = 0.47 (dashed lines).}
\label{ratesPNS20}
\end{figure*}
%
%
\begin{figure*}[tb]
\begin{center}
\includegraphics[scale=0.22]{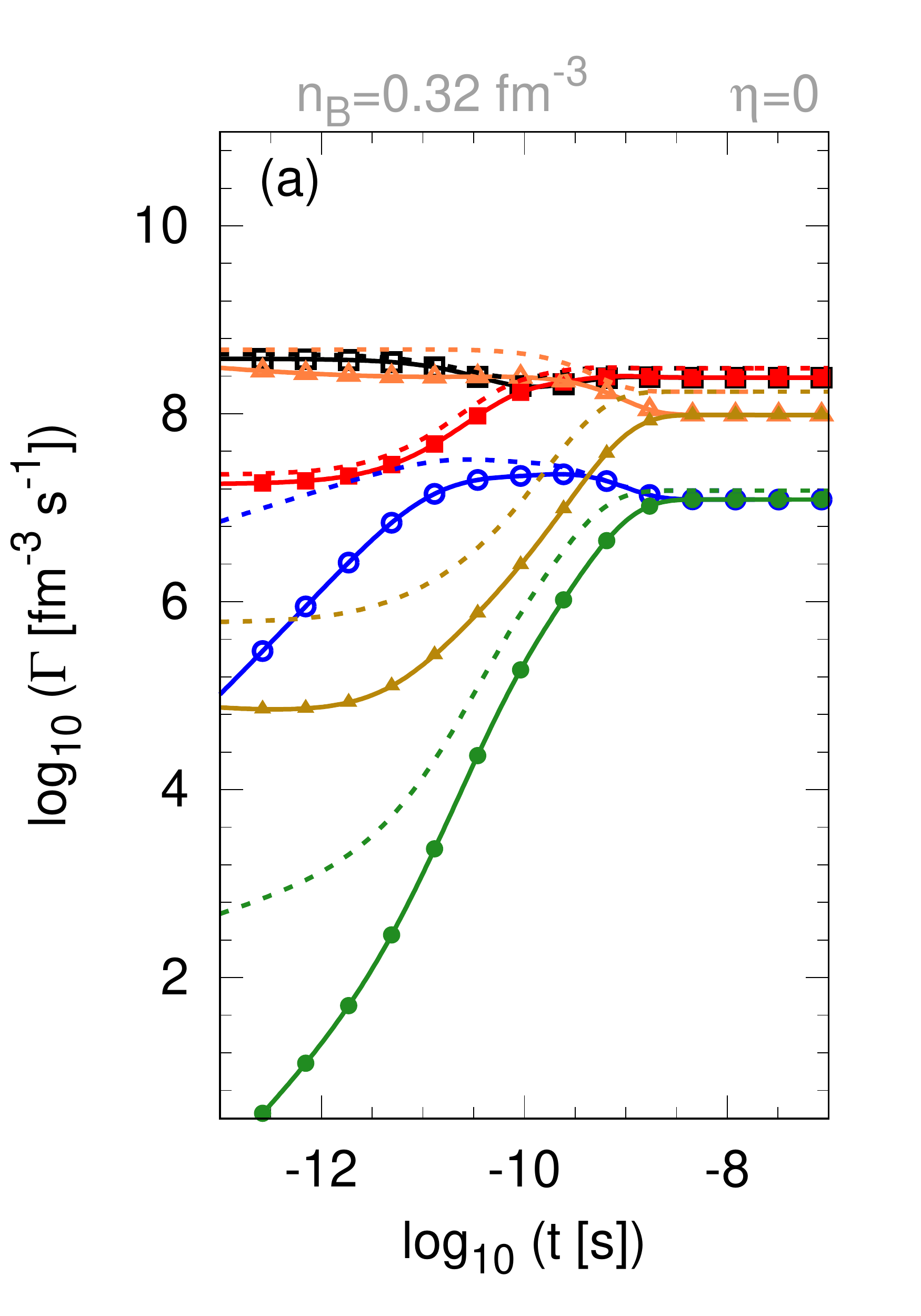}
\includegraphics[scale=0.22]{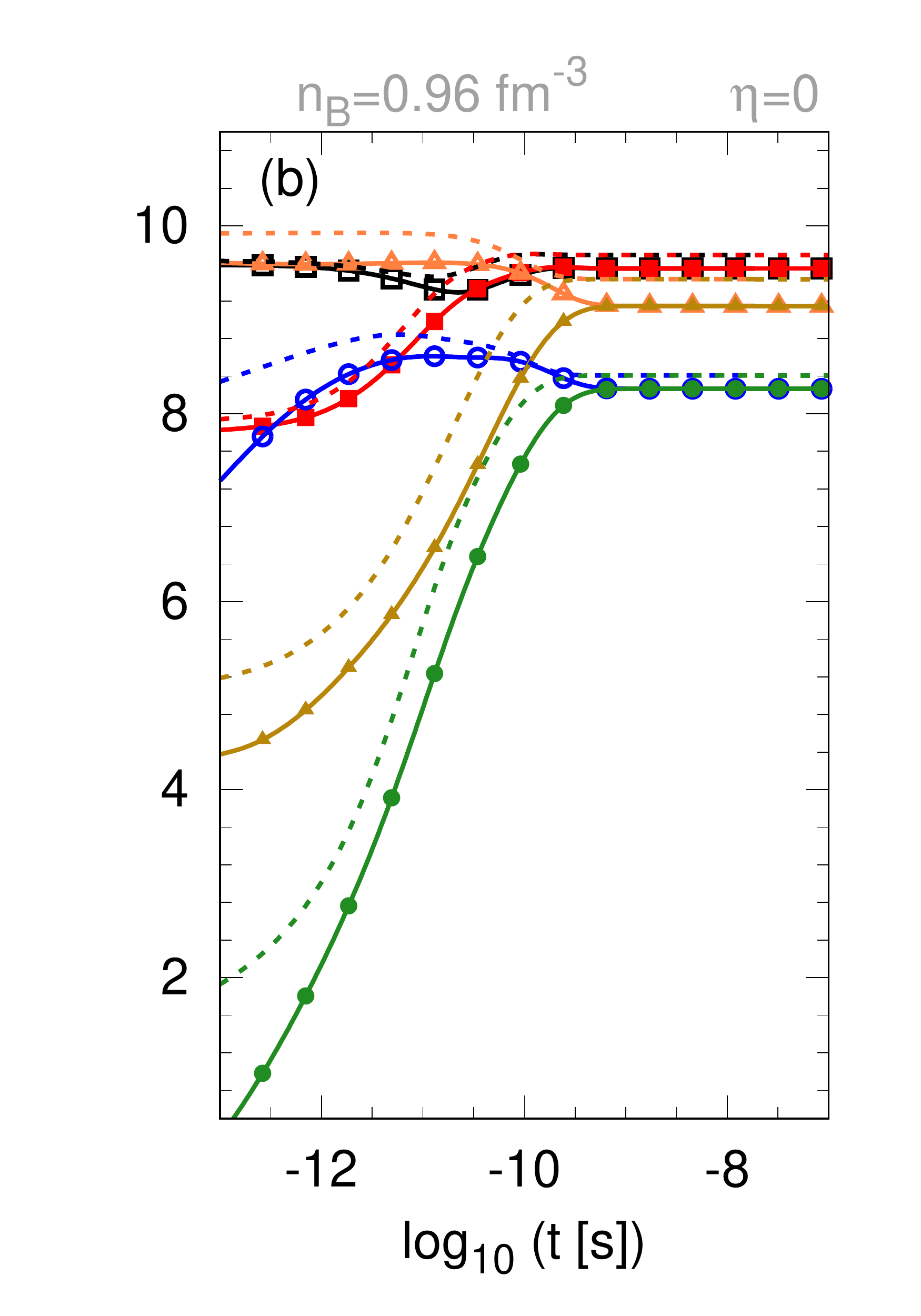}
\includegraphics[scale=0.22]{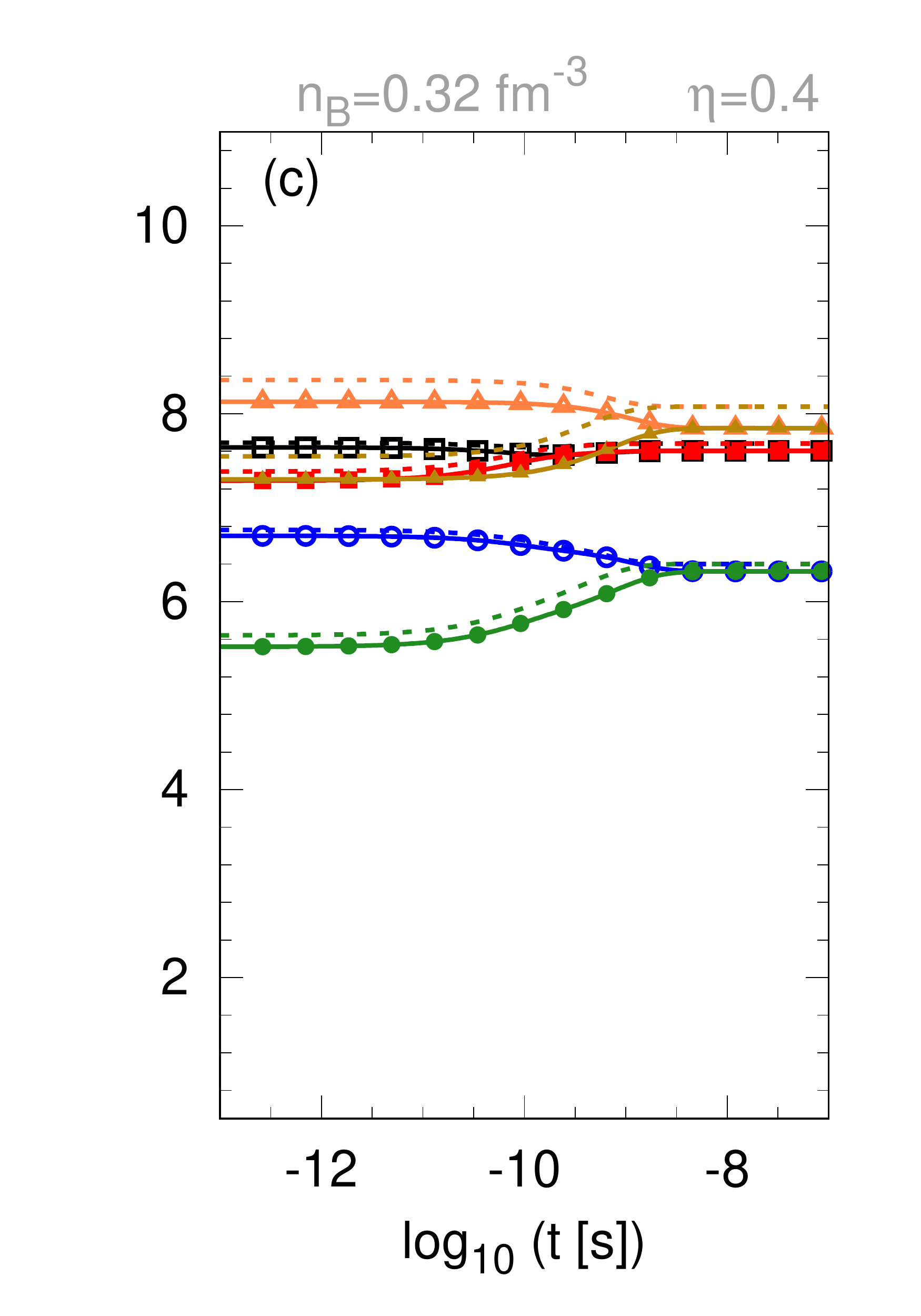}
\includegraphics[scale=0.22]{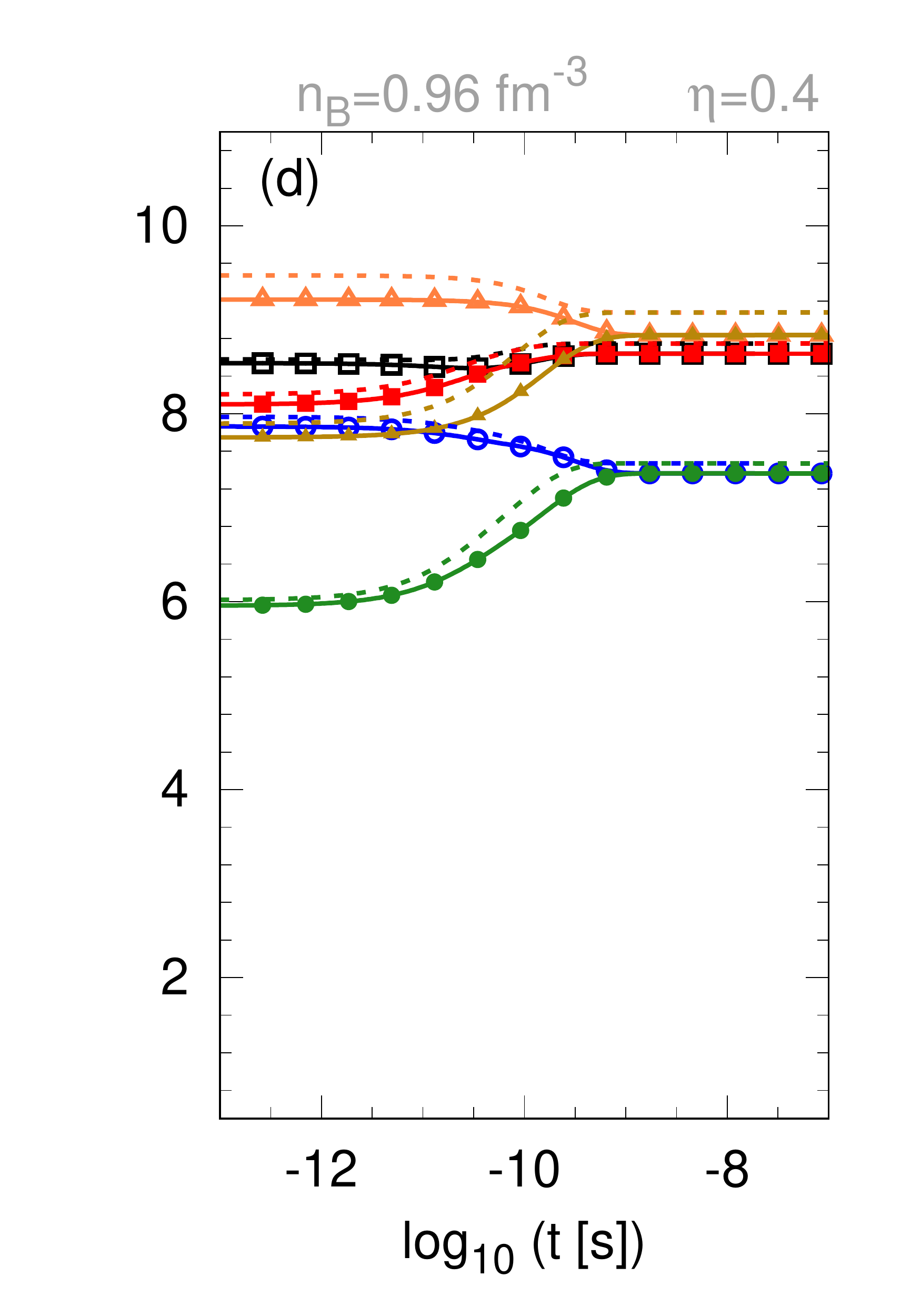}
\end{center}
\vspace{-0.6cm}
\caption{Same as in Fig. \ref{ratesPNS20} but for an initial temperature $T_i = 40 \, \mathrm{MeV}$.}
\label{ratesPNS40}
\end{figure*}

We assume now that the hadron-quark combustion process occurs in a hot NS, such as in a proto NS born in the aftermath of a core collapse supernova or in the compact object formed after a merging event in a binary system.  Due to the energy released by weak reactions, quark matter just behind the combustion front becomes even hotter in $\sim 1$ nanosecond. Additionally, hadron matter ahead the front is already hot. As a consequence, neutrinos are expected to be trapped in quark matter and, therefore,  the reverse reactions ($\nu$-captures) III and IV are allowed. On the other hand,  we have shown that the contribution of the decay of $s$ and $d$ quarks to the total rate is always negligible. Thus, the equations for the evolution of the quark abundances are now:
\begin{eqnarray}
&& \frac{dY_{u}}{dt}=\frac{1}{n_{B}}\left[\Gamma_{\text{III}}^{rev}-\Gamma_{\text{III}}^{dir}+\Gamma_{\text{IV}}^{rev}-\Gamma_{\text{IV}}^{dir}\right],\nonumber\\
&& \frac{dY_{d}}{dt}=\frac{1}{n_{B}}\left[\Gamma_{\text{III}}^{dir}-\Gamma_{\text{III}}^{rev}-\Gamma_{\text{V}}^{dir}+\Gamma_{\text{V}}^{rev}\right].\nonumber\\
\end{eqnarray}
As before, electric charge neutrality and baryon number conservation give:
\begin{eqnarray*}
&& Y_{s}=3-Y_{u}-Y_{d},\\
&& Y_{e}=Y_{u}-1.
\end{eqnarray*}
For the case of a phase transition taking place in a hot leptonized NS, there is an additional constraint coming from lepton number conservation because of the fact that during the short duration of the phase transition, neutrinos are trapped. As a consequence, the lepton abundance verifies:
\begin{eqnarray}
Y_{L}=Y_{e}+Y_{\nu_e} = \text{constant}.
\end{eqnarray}
The evolution of temperature is obtained from Eq. (\ref{evolution_of_T}).

Our results are shown in Figs. \ref{TY-PNS20-032}--\ref{ratesPNS40}.  To fix the initial values of the particle number densities we have considered $\xi=1.4$, $\eta= 0$ and $0.4$, and $\kappa$= 0.01 where the latter determines the initial value of the neutrino number density (see Table \ref{tab:dec_pns}). Notice that now we are using two different initial temperatures, $T_i = 20$ and $40 \, \mathrm{MeV}$. The results for the temperature and the abundances of $u$, $d$, $s$ quarks have some similarities with the cold NS case studied before. In particular, it is valid the same analysis about the effect of the strong coupling constant and the presence of finite strangeness in quark matter. However, the difference between the final and the initial temperatures ($\Delta T \equiv T_f - T_i$) is smaller than in the NS case (see Figs. \ref{TY-PNS20-032},  \ref{TY-PNS20-096}, \ref{TY-PNS40-032}, \ref{TY-PNS40-096} and Table \ref{tab:addlabelTePNS}). 

The most significant differences with respect to the case of cold NS matter are related to the evolution of the electron and neutrino abundances, as shown in the right panel of Figs. \ref{TY-PNS20-032}, \ref{TY-PNS20-096}, \ref{TY-PNS40-032}, and \ref{TY-PNS40-096}. Due to neutrino trapping, the lepton abundance $Y_L$ is assumed to be constant during the transition. The electron abundance decreases as in the case of cold NSs, but now it doesn't go to zero. The neutrino abundance increases up to about 0.1 neutrinos per baryon for $\eta$= 0. We also find that if quark matter has an initial strangeness $\eta$= 0.4, the final abundance of neutrinos tends to be smaller ($\sim 0.02-0.03$ neutrinos per baryon).

The net neutrino energy loss per baryon is shown in  Figs. \ref{Energ-PNS20} and \ref{Energ-PNS40} for the processes $u + e^{-} \leftrightarrow d + {\nu_e}$ and $u + e^{-} \leftrightarrow s + {\nu_e}$.   The emissivity is high during $\sim 10^{-9} \, \mathrm{s}$ and is followed by a steep decline to a value several orders of magnitude smaller in a timescale of $\sim 10^{-8}\, \mathrm{s}$. The maximum value of the emissivity per baryon is between  $10^{9} \, \mathrm{MeV}/\mathrm{s}$ and  $10^{12} \, \mathrm{MeV}/\mathrm{s}$  as for cold NSs. Initially, most neutrinos are emitted as a consequence of the $u + e^{-} \leftrightarrow d + {\nu_e}$ process, but around $t \approx 10^{-10}$ s the emissivity of this process falls significantly because the direct and reverse processes attain equilibrium around this time (see the point in Figs. \ref{ratesPNS20} and \ref{ratesPNS40} at which $\Gamma_{\text{III}}^{rev}$ and $\Gamma_{\text{III}}^{dir}$ become equal). On the other hand, the emissivity of the process   $u + e^{-} \leftrightarrow s + {\nu_e}$ remains active for a longer time because processes involving $s$ quarks attain equilibrium later (see curves for the processes $\mathrm{IV}$ and $\mathrm{V}$ in Figs. \ref{ratesPNS20} and \ref{ratesPNS40}).  In fact, for $t \gtrsim 10^{-10}$ s, the neutrino emission is dominated by the process   $u + e^{-} \leftrightarrow s + {\nu_e}$. 

In Table  \ref{tab:addlabelTePNS} we show the total energy per baryon released by quark matter in the form of neutrinos, which has been obtained through time integration of the total neutrino emissivity. The total energy per baryon is in the range ${\cal E}_{\nu_{e}} =  2-44 \, \mathrm{MeV}$  for different initial conditions. As for cold NSs, we estimate the order of magnitude of the energy released if the whole star is converted into quark matter. For a typical star with $10^{58}$ baryons we find ${\cal E}_{total} \approx 0.4-7 \times 10^{53} \, \mathrm{erg}$. {We emphasize that neutrinos produced within the flame stay trapped in the hot quark matter core of the star until it cools and deleptonizes.  As a consequence, they  are released in a diffusion timescale which is of the order of tens of seconds \cite{Pons:2001ar}. }

Finally, in Figs. \ref{ratesPNS20} and \ref{ratesPNS40} we show the reaction rates for all the  processes in a hot NS.  The nonleptonic process $u+d \rightarrow u+s$ is dominant most of the time {in most cases} but the process $u + e^{-} \leftrightarrow d + {\nu_e}$ has a significant contribution in the scenario of low initial strangeness, and becomes dominant near chemical equilibrium.

\begin{table*}[tbh]
  \centering
\caption{Total energy per baryon ${\cal E}_{\nu_{e}}$ released by quark matter in a hot NS in the form of neutrinos. We also show ${\cal E}_{\nu_{e}} \times 10^{58}$   in order to have a rough estimate of the energy release in a typical NS with $10^{58}$ baryons. The initial and final values of the temperature are also presented.  We assumed $\alpha_c$= 0.  }   
    \begin{tabular*}{\linewidth}{c @{\extracolsep{\fill}} ccc|cccc}
    \hline
     $n_B$ & $\xi$ & $\eta$  & $\kappa$  & $T_i$  & $T_f$  & ${\cal E}_{\nu_{e}}$ & ${\cal E}_{total}$ \\
\,[fm$^{-3}$]& & & &[MeV]&[MeV]&[MeV]&[$10^{53}$ ergs] \\
    \hline \hline  
0.32 & 1.4 & 0  & 0.01& 20& 34.82& 28.62 & 4.59  \\ %
0.32 & 1.4 & 0  & 0.01& 40& 47.09& 28.95 & 4.64  \\
0.32 & 1.4 & 0.4& 0.01& 20& 24.17& 2.77  & 0.44  \\ %
0.32 & 1.4 & 0.4& 0.01& 40& 41.65& 3.19  & 0.51  \\
0.96 & 1.4 & 0  & 0.01& 20& 53.45 & 44.02 & 7.05  \\ %
0.96 & 1.4 & 0  & 0.01& 40& 61.34 & 44.24 & 7.09  \\
0.96 & 1.4 & 0.4& 0.01& 20& 33.34 & 4.80  & 0.77  \\ %
0.96 & 1.4 & 0.4& 0.01& 40& 47.37 & 5.12  & 0.82  \\
    \hline
    \end{tabular*}   
  \label{tab:addlabelTePNS}
\end{table*}

%
\section{Summary and Conclusions}
\label{sec:conclusions}

In this work, we have performed a detailed analysis of a flame that converts hadronic matter into quark matter in a compact star (see Fig. \ref{fig:flame}).  We focused on a small portion of just deconfined quark matter which is initially at the deconfinement region of the flame and followed its evolution as it approaches equilibrium by means of weak interactions. For quark matter, we employed the MIT bag model at finite temperature including the effect of the finite mass of strange quarks  and QCD corrections to the first order in the coupling constant $\alpha_c$ (see Sec. \ref{sec:MIT}). 

The time evolution of quark matter was described by means of  the Boltzmann equation (Sec. \ref{sec:boltzmann}) using the reaction rates  of  all  the  relevant  weak  interaction processes (Sec. \ref{Sec-rate}).  These reaction rates have  already been  calculated in the literature (see e.g. \cite{Dai:1993nq,Dai:1995uj,dai1995conversion,Madsen:1993xx,Anand:1997vk,Lai:2008zza,Lai:2008zz}) but we have introduced some improvements here. For example, in Refs. \cite{Dai:1993nq, Dai:1995uj, dai1995conversion, Anand:1997vk, Lai:2008zza, Lai:2008zz} neutrinos in hot NSs are taken as completely degenerate and those in cold NSs as completely non-degenerate. Although this is a reasonable approximation we have generalized the results, i.e. we treated the neutrinos without making any assumption about their degeneracy. Also, we have treated the equation of state in more detail. In some previous works (e.g. Refs. \cite{Dai:1993nq,Dai:1995uj,dai1995conversion,Anand:1997vk,Lai:2008zza,Lai:2008zz}) the equation of state was considered at $T=0$ even for matter at finite temperature. Again, this is a plausible approximation when quarks are degenerate, but we have generalized this issue by considering the full expressions at finite temperature. 

In Secs. \ref{rens} and \ref{repns} we have solved the Boltzmann equation employing the rates obtained in Sec. \ref{Sec-rate}. In order to close the system of equations, we used the condition of electric charge neutrality, baryon number conservation, lepton number conservation (only in hot NSs) and the first law of thermodynamics. 
Previous works have adopted a similar approach \cite{Dai:1993nq, Dai:1995uj, dai1995conversion, Anand:1997vk, Lai:2008zza, Lai:2008zz}, but we have improved the description in this issue as well. In particular, we have included a much more detailed description of the initial conditions of quark matter {with several combinations of density, temperature, strangeness and neutrino trapping, and we have included the neutrino energy loss in our equations}. To keep the analysis as general as possible,  we didn't consider any specific hadronic EOS but we adopted  a set of parameters $\xi$, $\eta$ and $\kappa$ that encode the effect of the initial $u$ to $d$ ratio, the initial  strangeness, and  neutrino trapping  of quark matter in the deconfinement region (Sec. \ref{sec:initial_conditions}  and Table \ref{tab:dec_pns}). 

Our results were presented in Sec. \ref{rens} and \ref{repns}, and show that after quark deconfinement there is a significant increase in the temperature $T$ and in the strange quark abundance $Y_s$ in a timescale of $\sim 10^{-9} \, \mathrm{s}$. The abundances of  quarks $u$, $d$ and electrons decrease.  The increase in $T$ strongly depends on the initial strangeness of hadronic matter.  In fact, in cold NSs the final temperature for vanishing initial strangeness may be twice the value attained  in the case of large initial strangeness.  In hot NSs, the difference is also significant, although smaller. This occurs because  just deconfined matter  with larger strangeness is closer to chemical equilibrium than matter without strangeness and releases less energy. We also find that transitions at  higher densities present a more drastic temperature rise, and that  temperature increments are further enhanced when the strong coupling constant is non vanishing. 

We have also analysed the relevance of the different weak interaction processes. We find that the nonleptonic process $u + d \rightarrow u + s$ is always dominant in cold NSs, but in hot NSs the process $u + e^{-} \leftrightarrow d + {\nu_e}$ becomes relevant, and in some cases dominant,  near chemical equilibrium.  The rates for the other processes are orders of magnitude smaller.

Concerning the neutrino energy loss per baryon, we find that it is high during the first nanosecond  and it is followed by a steep decline to a value several orders of magnitude smaller in a timescale of $\sim  10^{-8} -10^{-7} \, \mathrm{s}$.
The maximum value of the neutrino emissivity per baryon is between  10$^{9}  \, \mathrm{MeV}/\mathrm{s}$ and  10$^{12} \, \mathrm{MeV}/\mathrm{s}$ for both  cold and hot NSs. 
For  a flame in a cold NSs the neutrino emission is dominated by the process $u + e^{-} \leftrightarrow d + {\nu_e}$.  The antineutrino emission is dominated by the decay $d \rightarrow u + e^{-} + {\bar{\nu}_e}$ and $s \rightarrow u + e^{-} + {\bar{\nu}_e}$ but it is  orders of magnitude smaller than for neutrinos. 
In a hot NSs, most neutrinos are emitted initially as a consequence of the $u + e^{-} \leftrightarrow d + {\nu_e}$ process, but around $t \approx 10^{-10}$ s  the emissivity of this process falls significantly. On the other hand, the emissivity of the process   $u + e^{-} \leftrightarrow s + {\nu_e}$ grows significantly as the abundance of $s$ quarks becomes large. For $t \gtrsim 10^{-10}$ s, the neutrino emission is dominated by the process   $u + e^{-} \leftrightarrow s + {\nu_e}$.
{Our results for the neutrino emissivity change the picture with respect to previous works.  For example, in Ref. \cite{Anand:1997vk}, the largest value of the neutrino emissivity is $\sim 10^9$ MeV/s for matter with zero initial strangeness and $n_B \approx 0.4 \, \mathrm{fm}^{-3}$. Our results for similar conditions are significantly larger ($\sim 10^{11}$ MeV/s), which is probably due to the fact that  analytic approximations for the rates and the emissivities are used in Ref. \cite{Anand:1997vk}. On the other hand, the neutrino emisivity shown in Fig. 4 of Ref. \cite{Lai:2008zz} for  $n_B \approx 0.4 \, \mathrm{fm}^{-3}$, zero initial strangeness and initial temperature of 20 MeV, is in agreement with our results for the same conditions; in particular, the largest value of the neutrino emissivity is $\sim 10^{11}$ MeV/s.  However, since  we considered here many initial conditions that were not analyzed in Ref. \cite{Lai:2008zz}, we obtain a wide range of results, including much lower neutrino emissivities in the case of large initial strangeness and larger ones for zero strangeness and large densities. }

Finally, we have integrated in time the total neutrino emissivity and obtained the total energy per baryon ${\cal E}_{\nu_{e}}$ released by quark matter in the form of neutrinos. 
${\cal E}_{\nu_{e}}$  represents the total energy per baryon released by a fluid element  initially located inside the deconfinement zone of the flame, during the time that it moves inside the decay region, until it attains chemical equilibrium at the end of the flame. The value of  ${\cal E}_{\nu_{e}}$ is strongly dependent on the initial conditions, because they determine how far is just deconfined matter from equilibrium. If hadronic matter has a large strangeness, ${\cal E}_{\nu_{e}}$ is  $\sim 6-10 \, \mathrm{MeV}$ for cold NSs and $\sim 3-5 \, \mathrm{MeV}$ for hot NSs. In the scenario of hadronic matter with zero strangeness ${\cal E}_{\nu_{e}}$ is much larger: $\sim 40 -60 \, \mathrm{MeV}$ for cold NSs and $\sim 30-40 \, \mathrm{MeV}$ for hot NSs.
These are very large numbers that may lead to observable astrophysical consequences. As a rough estimate we  considered the ignition of  $10^{58}$ baryons and found that the conversion of a whole  NS would emit  around $\sim 10^{53}$ erg in neutrinos. Notice that this is only the ``chemical'' energy associated with the weak decay of quarks. Additional energy is expected from the rearrangement of the remnant object provided that the star survives the explosion. In fact, since the total energy of the combustion is of the same order of the gravitational binding energy of the compact object,  the conversion process may have enough energy to disrupt the star. 
In the case of a hot proto NS, these neutrinos can be absorbed by  matter just behind the shock wave that travels along the external layers of the progenitor star, and help to a successful core collapse supernova explosion. In the case of a binary NS merger the combustion energy may have a role in  the hypermassive object that forms after the fusion. Notice also that the liberated energy is of the order of the energy of a gamma ray burst (GRB), indicating that models of GRBs involving the hadronic matter to quark matter conversion in a NS deserve further study.

\acknowledgements
G. Lugones acknowledges the Brazilian agency Conselho Nacional de Desenvolvimento Cient\'{\i}fico e Tecnol\'ogico (CNPq) for financial support. 

\onecolumngrid

\appendix

\section{Calculation of reaction rates and emissivities} 
\label{reaction_rates}

\subsection{Reaction rate for $d$ quark decay: $d \rightarrow u+e^{-}+\bar{\nu}_{e}$}\label{subsd--uenu}
%
Using $\left\langle |\mathcal{M}|^2 \right\rangle$ for process $I$ given in Table \ref{table:reactions}, the reaction rate reads 
\begin{eqnarray}
\Gamma_{\text{I}} & = & 6\int \prod_{i=1}^{4} \left[\frac{d^{3}p_{i}}{\left(2\pi\right)^{3}}\right] \left\langle |\mathcal{M}|^2 \right\rangle \delta^4(P_3 - P_1 - P_2 - P_4) \times {\cal S}_I = \nonumber \\
 & = & 6\times64G_{F}^{2}\mbox{cos}^{2}\ \theta_{C}\int\prod_{i=1}^{4}\left[\frac{d^{3}p_{i}}{(2\pi)^3 2 E_i}\right]  (2\pi)^4  \delta^4(P_3 - P_1 - P_2 - P_4)    (P_1\cdot P_2) (P_3 \cdot P_4)  \times {\cal S}_I , 
\end{eqnarray}
where  $i=$1, 2, 3, 4 represent $u$, $e$, $d$ and $\bar{\nu}_{e}$  respectively, and  ${\cal S}_I = f(E_3) [1-f(E_1)] [1-f(E_2)] [1-f(E_4)]$. We write the phase space element as $d^3 p_i=p_i^2 dp_{i}d\Omega_i=p_i E_i dE_i d\Omega_i$, where $p_i=\sqrt{E_{i}^2 -m_i^2}$ and  $d\Omega_i$ is an element of solid angle. Thus, we have
\begin{eqnarray}
\Gamma_{\text{I}} =  \frac{3 G_F^2 \cos^2\theta_C}{2\pi^5}   \int \prod_{i=1}^{4} dE_i  f(E_3) [1-f(E_1)] [1-f(E_2)] [1-f(E_4)]  \delta(E_3-E_1-E_2-E_4) I(E_1, E_2, E_3, E_4) ,
\end{eqnarray}
where the angular integral $I(E_1, E_2, E_3, E_4)$ is
\begin{eqnarray}
I(E_1, E_2, E_3, E_4) & = & \frac{1}{16 \pi^3}\int\prod_{i=1}^{4}p_id\Omega_{i}\left(P_1\cdot P_2\right)\left(P_3\cdot P_4\right)   \delta(\mathbf{p}_3-\mathbf{p}_1-\mathbf{p}_2-\mathbf{p}_4)
\label{Iint},
\end{eqnarray}
being $\mathbf{p}_{i}$ the vector momentum of the $i$-species.  The  angular integral $I(E_1, E_2, E_3, E_4)$ can be calculated following \cite{Wadhwa:1995dv}.

The integral for $\Gamma_{\text{I}}$ can be considerably simplified if we consider degenerate matter. 
Inside a compact star, the density is very large, and therefore it is a good approximation to consider that quarks and electrons (particles 1, 2 and 3) are degenerate. Thus, the process $d  \rightarrow u+e^{-}+\bar{\nu}_{e}$ involves particles $u$, $e$ and $d$ that are very close to their respective Fermi surfaces. 
In view of this, we can simplify the expression for $\left\langle |\mathcal{M}|^2 \right\rangle$  by replacing the particle momenta and energies by the corresponding Fermi momenta and chemical potentials. Notice that we are not using this approximation in the Fermi blocking factors, nor in the delta function. With this approximation we obtain:
\begin{eqnarray}
\Gamma_{\text{I}} & = & \frac{3 G_F^2}{2\pi^{5}} \cos^2 \theta_C\int_{m_4}^{\infty} dE_4 \frac{A(E_4)I(\mu_1, \mu_2, \mu_3, E_4)}{e^{(\mu_4-E_4)/T}  + 1 } , 
\end{eqnarray}
where $A(E_4)$ is given by
\begin{eqnarray}
A(E_4)& = & 
 \int_{m_1}^{\infty}\frac{dE_1}{e^{(\mu_1-E_1)/T} + 1}    
 \int_{m_2}^{\infty}\frac{dE_2}{e^{(\mu_2-E_2)/T} + 1}  
 \int_{m_3}^{\infty} dE_3\frac{\delta(E_3-E_1-E_2-E_4)} {e^{(E_3-\mu_3)/T} + 1} \approx \nonumber\\
&\approx &  \frac{(\mu_1 + \mu_2 - \mu_3 + E_4)^2 + \pi^2 T^2 }{ 2 [e^{(\mu_1 + \mu_2 - \mu_3 + E_4)/T} + 1]} .
\label{Aint}
\end{eqnarray}

Recalling the labeling of the particles, and assuming massless antineutrinos, we get
\begin{eqnarray}
\Gamma_{\text{I}} & = & \frac{3 G_F^2}{2\pi^{5}} \cos^2 \theta_C\int_{0}^{\infty} dE_{\bar{\nu}_{e}} \frac{(\mu_u + \mu_e - \mu_d + E_{\bar{\nu}_{e}})^2 + \pi^2 T^2 }{ 2 [e^{(\mu_u + \mu_e - \mu_d + E_{\bar{\nu}_{e}})/T} + 1]}  \times \frac{I(\mu_u, \mu_e, \mu_d, E_{\bar{\nu}_{e}} )}{e^{(\mu_{\bar{\nu}_{e}} - E_{\bar{\nu}_{e}})/T}  + 1 } .
\label{Gamma1Ap}
\end{eqnarray}
%

\subsection{Reaction rate for $s$ quark decay: $s  \rightarrow u+e^{-}+\bar{\nu}_{e}$}
\label{erate2} 
%
The reaction rate for this process can be obtained straightforwardly if we assume that $s$ quarks are degenerate. This is a good approximation if hadronic matter contains hyperons, because in this case, the just deconfined quark matter will contain a significant initial fraction of $s$ quarks. However, if the hadronic phase is composed by nucleons and leptons (no hyperons), the just deconfined quark matter will contain \textit{initially} only $u$ and $d$ quarks and leptons (no strange quarks).  In this case, the process $s  \rightarrow u+e^{-}+\bar{\nu}_{e}$ doesn't occur at the beginning of the conversion. However, the nonleptonic process $u+d \leftrightarrow u+s$ is possible and it produces more and more $s$ quarks. This happens very fast, and soon the $s$ quark degenerate sea gets populated. Thus, even in this case, it is a reasonable approximation to treat the $s$ quarks as degenerate, and the reaction rate can be obtained by replacing $\mu_d$ with $\mu_s$ and $\cos \theta_{C}$ with $\sin \theta_{C}$ in Eq. \eqref{Gamma1Ap}:
\begin{eqnarray}
\Gamma_{\text{II}} & = & \frac{3 G_F^2}{2\pi^{5}} \sin^2 \theta_C\int_{0}^{\infty} dE_{\bar{\nu}_{e}} \frac{(\mu_u + \mu_e - \mu_s + E_{\bar{\nu}_{e}})^2 + \pi^2 T^2 }{ 2 [e^{(\mu_u + \mu_e - \mu_s + E_{\bar{\nu}_{e}})/T} + 1]}  \times \frac{I(\mu_u, \mu_e, \mu_s, E_{\bar{\nu}_{e}} )}{e^{(\mu_{\bar{\nu}_{e}} - E_{\bar{\nu}_{e}})/T}  + 1 } .
\label{Gamma2}
\end{eqnarray}

\subsection{Reaction rate for the process $u+e^{-} \leftrightarrow d+\nu_{e}$}\label{uednu}   
%
Using $\left\langle |\mathcal{M}|^2 \right\rangle$ given in Table \ref{table:reactions}, the reaction rate for the direct process  $u+e^{-}  \rightarrow d+\nu_{e}$ reads: 
\begin{eqnarray}
\Gamma_{\text{III}}^{dir} & = & 6\int \prod_{i=1}^{4} \left[\frac{d^{3}p_{i}}{\left(2\pi\right)^{3}}\right] \left\langle |\mathcal{M}|^2 \right\rangle \delta^4(P_1 + P_2 - P_3 - P_4) \times {\cal S}_{III} = \nonumber \\
 & = & 6\times64G_{F}^{2}\mbox{cos}^{2}\ \theta_{C}\int\prod_{i=1}^{4}\left[\frac{d^{3}p_{i}}{(2\pi)^3 2 E_i}\right]  (2\pi)^4  \delta^4(P_1 + P_2 - P_3 - P_4)    (P_1\cdot P_2) (P_3 \cdot P_4)  \times {\cal S}_{III} ,
\label{Gamma3_b}  
\end{eqnarray}
where $i=$1, 2, 3, 4 represent $u$, $e^-$, $d$ and $\nu_{e}$  respectively, and ${\cal S}_{III} = f(E_1)f(E_2) [1-f(E_3)] [1-f(E_4)]$. 
Similarly to Subsection \ref{subsd--uenu} , we obtain  
\begin{eqnarray}
\Gamma_{\text{III}}^{dir} & = & \frac{3 G_F^2 \cos^2\theta_C}{2\pi^5}   \int \prod_{i=1}^{4} dE_i  f(E_1)f(E_2) [1-f(E_3)] [1-f(E_4)]  \delta(E_1 + E_2 - E_3 - E_4) J(E_1, E_2, E_3, E_4) ,\label{GammaIII}
\end{eqnarray}
where the angular integral $J$ is
\begin{eqnarray}
J(E_1, E_2, E_3, E_4) = \frac{1}{16\pi^{3}} \int \prod_{i=1}^{4} p_i d\Omega_{i} (P_1\cdot P_{2})(P_{3}\cdot P_{4}) \delta(\mathbf{p}_1+\mathbf{p}_{2}-\mathbf{p}_{3}-\mathbf{p}_{4}) ,
\label{JintA}
\end{eqnarray}
and can be solved following \cite{Wadhwa:1995dv}.

For matter with degenerate quarks and electrons we can simplify the expression for $\Gamma_{\text{III}}^{dir} $ by replacing the particle momenta and energies with the corresponding Fermi momenta and chemical potentials:
\begin{eqnarray}
\Gamma_{\text{III}}^{dir}  & = & \frac{3 G_F^2}{2\pi^5}\cos^2 \theta_C \int_{m_4}^{\infty} dE_4 \frac{A(E_4)J(\mu_1, \mu_2, \mu_3, E_4)}{e^{(\mu_4-E_4)/T} + 1},
\label{Gamma3-1}
\end{eqnarray}
where $A(E_4)$ is:
\begin{eqnarray}
A(E_4)&=&
\int_{m_1}^{\infty} \frac{dE_1}{e^{(E_1-\mu_1)/T} + 1} 
\int_{m_2}^{\infty} \frac{dE_2}{e^{(E_2-\mu_2)/T} + 1} 
\int_{m_3}^{\infty} dE_3 \frac{\delta(E_1+E_2-E_3-E_4)}{e^{(\mu_3-E_3)/T} + 1}  \nonumber\\
&\approx &  \frac{(\mu_1 + \mu_2 - \mu_3 - E_4)^2 + \pi^2 T^2 }{ 2 [e^{(-\mu_1 - \mu_2 + \mu_3 + E_4)/T} + 1]} .
\label{Ae4-3}
\end{eqnarray}
Recalling the labeling of the particles, and assuming massless neutrinos, we get
\begin{eqnarray}
\Gamma_{\text{III}}^{dir} & = & \frac{3 G_F^2}{2\pi^{5}} \cos^2 \theta_C\int_{0}^{\infty} dE_{\nu_e} 
\frac{(\mu_u + \mu_e - \mu_d - E_{\nu_e})^2 + \pi^2 T^2 }{ 2 [e^{(-\mu_u - \mu_e + \mu_d + E_{\nu_e})/T} + 1]}  \times 
\frac{J(\mu_u, \mu_e, \mu_d, E_{\nu_e} )}{e^{(\mu_{\nu_e} - E_{\nu_e})/T}  + 1 } .
\label{Gamma3-2}
\end{eqnarray}
The reaction rate $\Gamma_{\text{III}}$ for the reverse process $d+\nu_{e}  \rightarrow u+e^{-}$ is given by
\begin{eqnarray}
\Gamma_{\text{III}}^{rev} = e^{-(\mu_u + \mu_e -\mu_d - \mu_{\nu_e})/T} \Gamma_{\text{III}}^{dir} .
\end{eqnarray}
%
\subsection{Reaction rate for the process $u+e^{-}  \leftrightarrow s+\nu_{e}$}
%
For the direct process $u+e^{-}  \rightarrow s+\nu_{e}$ the reaction rate can be obtained by replacing  $\mu_d$ with $\mu_s$ and $\cos \theta_{C}$ with $\sin \theta_{C}$ in Eq. \eqref{Gamma3-2}:
\begin{eqnarray}
\Gamma_{\text{IV}}^{dir} & = & \frac{3 G_F^2}{2\pi^{5}} \sin^2 \theta_C\int_{0}^{\infty} dE_{\nu_e} 
\frac{(\mu_u + \mu_e - \mu_s - E_{\nu_e})^2 + \pi^2 T^2 }{ 2 [e^{(-\mu_u - \mu_e + \mu_s + E_{\nu_e})/T} + 1]}  \times 
\frac{J(\mu_u, \mu_e, \mu_s, E_{\nu_e} )}{e^{(\mu_{\nu_e} - E_{\nu_e})/T}  + 1 } .
\label{Gamma4}
\end{eqnarray}
As before, we are assuming here that all quarks and electrons are degenerate. Initially, this is not a good approximation for $s$-quarks. However, as mentioned before, the abundance of $s$ quarks increases by the nonleptonic process $u+d \leftrightarrow u+s$. This happens very fast, and soon the $s$ quark degenerate sea gets populated.

As in the previous case, the reverse process $ s+\nu_{e} \rightarrow u+e^{-}$ is given by 
\begin{eqnarray}
\Gamma_{\text{IV}}^{rev} = e^{-(\mu_u + \mu_e -\mu_s - \mu_{\nu_e})/T} \Gamma_{\text{IV}}^{dir} .
\end{eqnarray}

\subsection{Reaction rate for the nonleptonic process $u_1 + d \leftrightarrow u_2 + s$}

Using $\left\langle |\mathcal{M}|^2 \right\rangle$ for the direct process $V$ given in Table \ref{table:reactions} and considering that $i$=1, 2, 3, 4 represent $u_1$, $d$, $u_2$ and $s$ respectively, we can write the direct reaction rate.  According to  \cite{Madsen:1993xx}, there must be a factor $1/2$ to take into account that only left-handed helicity states of the $u_1$ quarks couple to the $W^-$, which mediates the transformation, thus:
\begin{eqnarray}
\Gamma_{\text{V}}^{dir}& = & \frac{6 \times 6}{2} \int\prod_{i=1}^{4}\left[\frac{d^{3}p_{i}}{\left(2\pi\right)^{3}}\right] \left\langle |\mathcal{M}|^2 \right\rangle \delta^4(P_1 + P_2 - P_3 - P_4) \times {\cal S}_{V} = \nonumber \\
  &=& 18 \times64G_{F}^{2}\mbox{sin}^{2}\ \theta_{C}\ \mbox{cos}^{2}\ \theta_{C}\int\prod_{i=1}^{4}\left[\frac{d^{3}p_{i}}{\left(2\pi\right)^{3}2E_{i}}\right]\left(2\pi\right)^{4}\delta^{4}\left(P_1+P_{2}-P_{3}-P_{4}\right) \left(P_1\cdot P_{2}\right)\left(P_{3}\cdot P_{4}\right)
\times {\cal S}_{V},  
\end{eqnarray}
where ${\cal S}_{V} = f\left(p_1\right)f\left(p_{2}\right)\left[1-f\left(p_{3}\right)\right]\left[1-f\left(p_{4}\right)\right]$. 

The latter expression is the same as  Eq. \eqref{Gamma3_b} but now it is multiplied by $6 \sin^2 \theta_C$ and  all particles are massive. Therefore,  replacing the correct indices  we obtain straightforwardly
\begin{eqnarray}
\Gamma_{\text{V}}^{dir} & = & \frac{9 G_F^2}{2\pi^{5}}  \sin^2 \theta_C \cos^2 \theta_C\int_{m_s}^{\infty} dE_{\nu_e} 
\frac{( \mu_d  - E_s)^2 + \pi^2 T^2 }{ 2 [e^{( \mu_d  - E_s)/T} + 1]}  \times 
\frac{J(\mu_u, \mu_d, \mu_u, E_s )}{e^{(\mu_s - E_s)/T}  + 1 }.
\label{Gamma5-Ap}
\end{eqnarray}

The reaction rate for the reverse process $u+s \rightarrow u+d$ is given by 
\begin{eqnarray}
\Gamma_{\text{V}}^{rev} = e^{-(\mu_d - \mu_s)/T} \Gamma_{\text{V}}^{dir} .
\end{eqnarray}
%

\subsection{The Antineutrino Emissivity Rates for the Decay of $d$ and $s$ Quarks }
%
In this section we calculate the antineutrino emissivity for the  decay reactions $I$ and $II$. Let us focus first on $d  \rightarrow u+e^{-}+\bar{\nu}_{e}$. The rate for emitting antineutrinos of energy $E_4$ is
\begin{eqnarray}
\varepsilon_{\text{I}}=6\int\prod_{i=1}^{4}\left[\frac{d^{3}p_{i}}{\left(2\pi\right)^{3}}\right]E_{4} \left\langle |\mathcal{M}|^2 \right\rangle \delta^4(P_3 - P_1 - P_2 - P_4)
\times {\cal S}_{I} , 
\end{eqnarray}
where $i=$1, 2, 3, 4 represent $u$, $e^-$, $d$ and $\bar{\nu}_{e}$ respectively. Replacing the transition rate $\left\langle |\mathcal{M}|^2 \right\rangle$ given in Subsection \ref{subsd--uenu}, we have
\begin{eqnarray}
\varepsilon_{\text{I}} & = & 6\times64G_{F}^{2}\mbox{cos}^{2}\ \theta_{C}\int\prod_{i=1}^{4}\left[\frac{d^{3}p_{i}}{\left(2\pi\right)^{3}2E_{i}}\right]E_{4}\left(2\pi\right)^{4}\delta^{4}\left(P_{3}-P_1-P_{2}-P_{4}\right) \left(P_1\cdot P_{2}\right)\left(P_{3}\cdot P_{4}\right)  \times {\cal S}_{I}.
\label{emi1}
\end{eqnarray}
The integral in the above equation is similar to the one already calculated in Subsection \ref{subsd--uenu}. We immediately obtain the antineutrino emissivity as 
\begin{eqnarray}
\varepsilon_{\text{I}} = \frac{3G_F^2}{2\pi^5}  \cos^{2}\theta_{C} \int_{-\infty}^{\mu_{\bar{\nu}_{e}}}\frac{(\mu_{u}+\mu_{e}-\mu_{d}+E_{\bar{\nu}_{e}})^{2}+\pi^{2}T^{2}}{2[e^{(\mu_{u}+\mu_{e}-\mu_{d}+E_{\bar{\nu}_{e}})/T}+1]} \frac{I(\mu_{u},\mu_{e},\mu_{d},E_{\bar{\nu}_{e}})}{e^{(\mu_{\bar{\nu}_{e}}-E_{\bar{\nu}_{e}})/T}+1} E_{\bar{\nu}_{e}}dE_{\bar{\nu}_{e}} .
\end{eqnarray}
The calculation of the emissivity rate due to process $s  \rightarrow u+e^{-}+\bar{\nu}_{e}$ is performed in a similar fashion. We just have to replace $\mu_d$ with $\mu_{s}$ and $\mbox{cos}\ \theta_{C}$ with $\mbox{sin}\ \theta_{C}$:
\begin{eqnarray}
\varepsilon_{\text{II}} = \frac{3G_F^2}{2\pi^5}  \sin^{2}\theta_{C} \int_{-\infty}^{\mu_{\bar{\nu}_{e}}}\frac{(\mu_{u}+\mu_{e}-\mu_{s}+E_{\bar{\nu}_{e}})^{2}+\pi^{2}T^{2}}{2[e^{(\mu_{u}+\mu_{e}-\mu_{s}+E_{\bar{\nu}_{e}})/T}+1]} \frac{I(\mu_{u},\mu_{e},\mu_{s},E_{\bar{\nu}_{e}})}{e^{(\mu_{\bar{\nu}_{e}}-E_{\bar{\nu}_{e}})/T}+1} E_{\bar{\nu}_{e}}dE_{\bar{\nu}_{e}} .
\end{eqnarray}
%
\subsection{The Neutrino Emissivity Rates for Electron Capture Processes}\label{emis2}
%
Now, we calculate the neutrino emissivity for reactions $III$ and $IV$.  Let us focus first in the reaction $u+e^{-}  \rightarrow d+\nu_{e}$. The rate for emitting neutrinos is 
\begin{eqnarray}
\varepsilon_{\text{III}} = 6\int\prod_{i=1}^{4}\left[\frac{d^{3}p_{i}}{\left(2\pi\right)^{3}}\right] E_{4} \left\langle |\mathcal{M}|^2 \right\rangle \delta^4(P_1 + P_2 - P_3 - P_4)
\times {\cal S}_{III}
\end{eqnarray}
where $i=$1, 2, 3, 4 represent $u$, $e^-$, $d$ and $\nu_{e}$  respectively. Replacing the transition rate $\left\langle |\mathcal{M}|^2 \right\rangle$, we have
\begin{eqnarray}
\varepsilon_{\text{III}}&=&6\times64G_{F}^{2}\mbox{cos}^{2}\ \theta_{C}\int\prod_{i=1}^{4}\left[\frac{d^{3}p_{i}}{\left(2\pi\right)^{3}2E_{i}}\right]E_{4}\left(2\pi\right)^{4}\delta^{4}\left(P_1+P_{2}-P_{3}-P_{4}\right)
\left(P_1\cdot P_{2}\right)\left(P_{3}\cdot P_{4}\right)  \times {\cal S}_{III} .
\end{eqnarray}
The integral of the above equation is essentially the same as the one calculated in Subsection \ref{uednu}. We immediately obtain: 
\begin{eqnarray}
\varepsilon_{\text{III}} = \frac{3G_F^2}{2\pi^5} \sin^{2}\theta_{C} \int_{-\infty}^{\mu_{\nu_{e}}}\frac{(\mu_{u}+\mu_{e}-\mu_{d}-E_{\nu_{e}})^{2}+\pi^{2}T^{2}}{2[e^{(-\mu_{u}-\mu_{e}+\mu_{d}+E_{\nu_{e}})/T}+1]}   \frac{J(\mu_{u},\mu_{e},\mu_{d},E_{\nu_{e}})}{e^{(\mu_{\nu_{e}}-E_{\nu_{e}})/T}+1}E_{\nu_{e}}dE_{\nu_{e}}.
\end{eqnarray}
Similarly, the emissivity $\varepsilon_{\text{IV}}$ for $u+e^{-}  \rightarrow s+\nu_{e}$ is obtained replacing $\mu_d$ by $\mu_s$ and $\mbox{cos}\ \theta_{C}$ by $\mbox{sin}\ \theta_{C}$ in the
previous formula,
\begin{eqnarray}
\varepsilon_{\text{IV}} = \frac{3G_F^2}{2\pi^5} \sin^{2}\theta_{C} \int_{-\infty}^{\mu_{\nu_{e}}}\frac{(\mu_{u}+\mu_{e}-\mu_{d}-E_{\nu_{e}})^{2}+\pi^{2}T^{2}}{2[e^{(-\mu_{u}-\mu_{e}+\mu_{d}+E_{\nu_{e}})/T}+1]}   \frac{J(\mu_{u},\mu_{e},\mu_{d},E_{\nu_{e}})}{e^{(\mu_{\nu_{e}}-E_{\nu_{e}})/T}+1}E_{\nu_{e}}dE_{\nu_{e}}.
\end{eqnarray}
Finally the total neutrino and antineutrino emissivities in a cold NS are
\begin{eqnarray}
\varepsilon &=&\varepsilon_{\text{III}}+\varepsilon_{\text{IV}}\nonumber\\\bar{\varepsilon}&=&\varepsilon_{\text{I}}+\varepsilon_{\text{II}}.
\end{eqnarray}
For hot quark matter with trapped neutrinos, the total neutrino emissivity is
\begin{eqnarray}
\varepsilon = \varepsilon_{\text{III}} (1-e^{-\xi_{d}})+\varepsilon_{\text{IV}} (1-e^{-\xi_{s}}),
\end{eqnarray}
where $\xi_{d}= (\mu_u +\mu_e - \mu_d-\mu_{\nu_e})/T$ and  $\xi_{s}= (\mu_u +\mu_e - \mu_s-\mu_{\nu_e})/T$. 

\subsection{Calculation of the integral of Eq. (\ref{Aint})}\label{appendix_A}
%
Introducing the change of variables, $x_i  =  -(E_i-\mu_i) / T$ for $i = 1, 2$, $x_3  =  (E_3-\mu_3) / T$, $x  =  -(E_4+\mu_4) / T$, $\xi_1 = (\mu_1+\mu_2-\mu_3-\mu_4) /T$, and using 
$\delta(E_3-E_1 - E_2 - E_4) = \delta(x_1 + x_2 + x_3 + x - \xi_1) / T$, we can write
\begin{eqnarray}
A(E_4)& = & 
 \int_{m_1}^{\infty}\frac{dE_1}{e^{(\mu_1-E_1)/T} + 1}    
 \int_{m_2}^{\infty}\frac{dE_2}{e^{(\mu_2-E_2)/T} + 1}  
 \int_{m_3}^{\infty} dE_3\frac{\delta(E_3-E_1-E_2-E_4)} {e^{(E_3-\mu_3)/T} + 1} = \nonumber\\  
&=& T^{2}\int_{-\infty}^{(\mu_1-m_1)/T}\frac{dx_1}{1+e^{x_1}}\int_{-\infty}^{\left(\mu_2-m_2\right)/T}\frac{dx_2}{1+e^{x_2}}
\int_{-(\mu_3-m_3)/T}^{\infty}\frac{dx_3}{1+e^{x_3}} \delta(x_1 + x_2 + x_3 + x - \xi_1) \nonumber\\
&\approx & T^2 \int_{-\infty}^{\infty}\frac{dx_1}{1+e^{x_1}}\int_{-\infty}^{\infty}\frac{dx_2}{1+e^{x_2}}\left(\frac{1}{1+e^{-x-x_1-x_2+\xi_1}}\right)\nonumber\\
&\approx &T^{2}\int_{-\infty}^{\infty}\frac{dx_1}{1+e^{x_1}}\left(\frac{-x-x_1+\xi_1}{e^{-x-x_1+\xi_1}-1}\right) =\frac{T^{2}}{2}  \frac{( \xi_1-x )^2 + \pi^2 }{ (e^{-x + \xi_1} + 1)}.
\label{Atotal}
\end{eqnarray}
In the above calculation  we have replaced the upper integration limits  $(\mu_1 - m_1)/T$ and $(\mu_2 - m_2)/T$ by  $ +\infty$. The reason is that  $(\mu_i - m_i)/T$ is always large enough in the range of densities and temperatures of interest, and  the Fermi blocking factors $(1+\exp(x_1))^{-1}$ and $(1+\exp(x_2))^{-1}$ tend to zero very fast for $x_1, x_2 > 0$ (see e.g. \cite{{Wadhwa:1995dv},{Anand:1997vk}}).

\section{Comparison between approximated and exact rates}\label{Comparison_rates}

\begin{figure*}[tb]
\begin{center}
\includegraphics[scale=0.22]{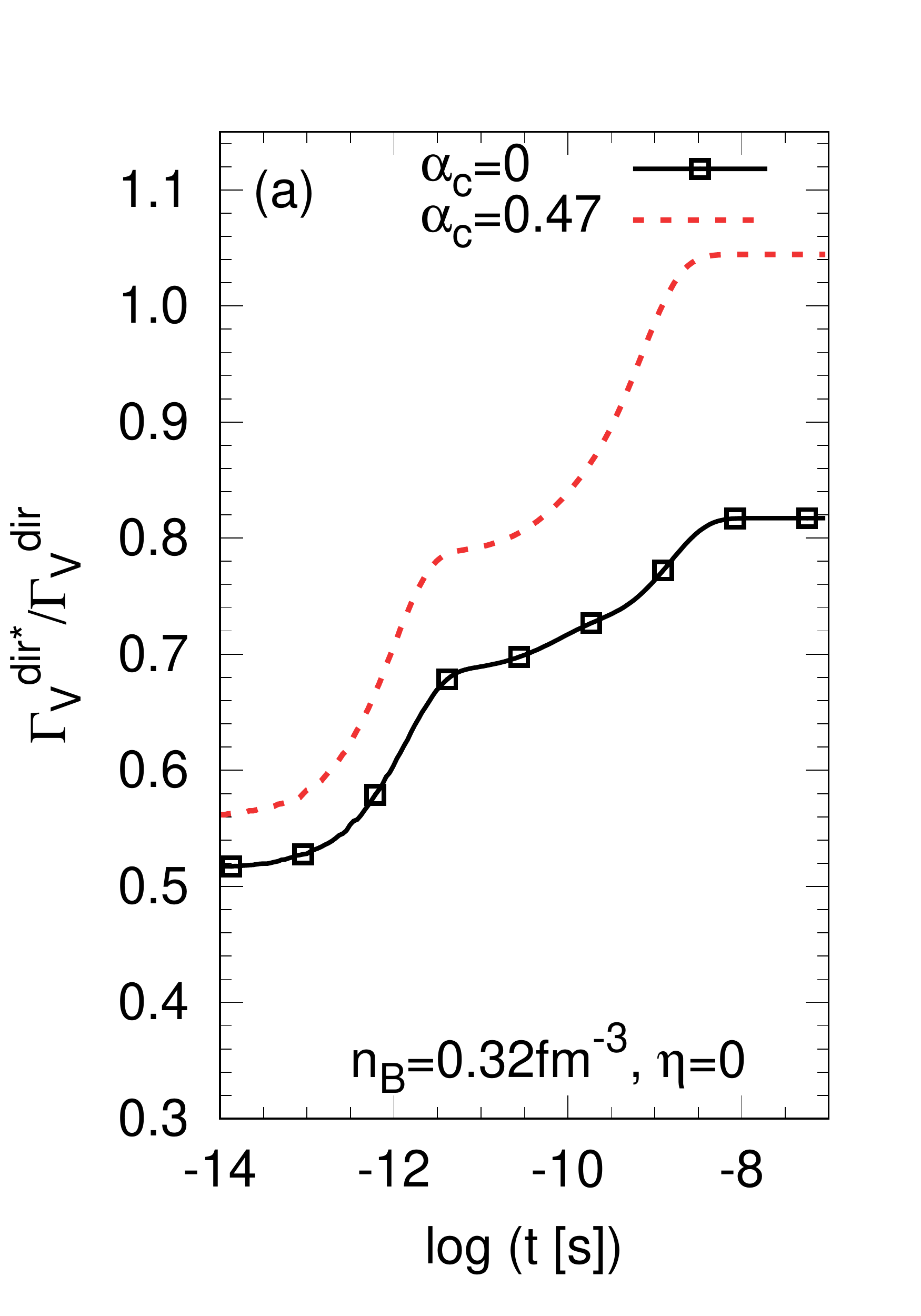}
\includegraphics[scale=0.22]{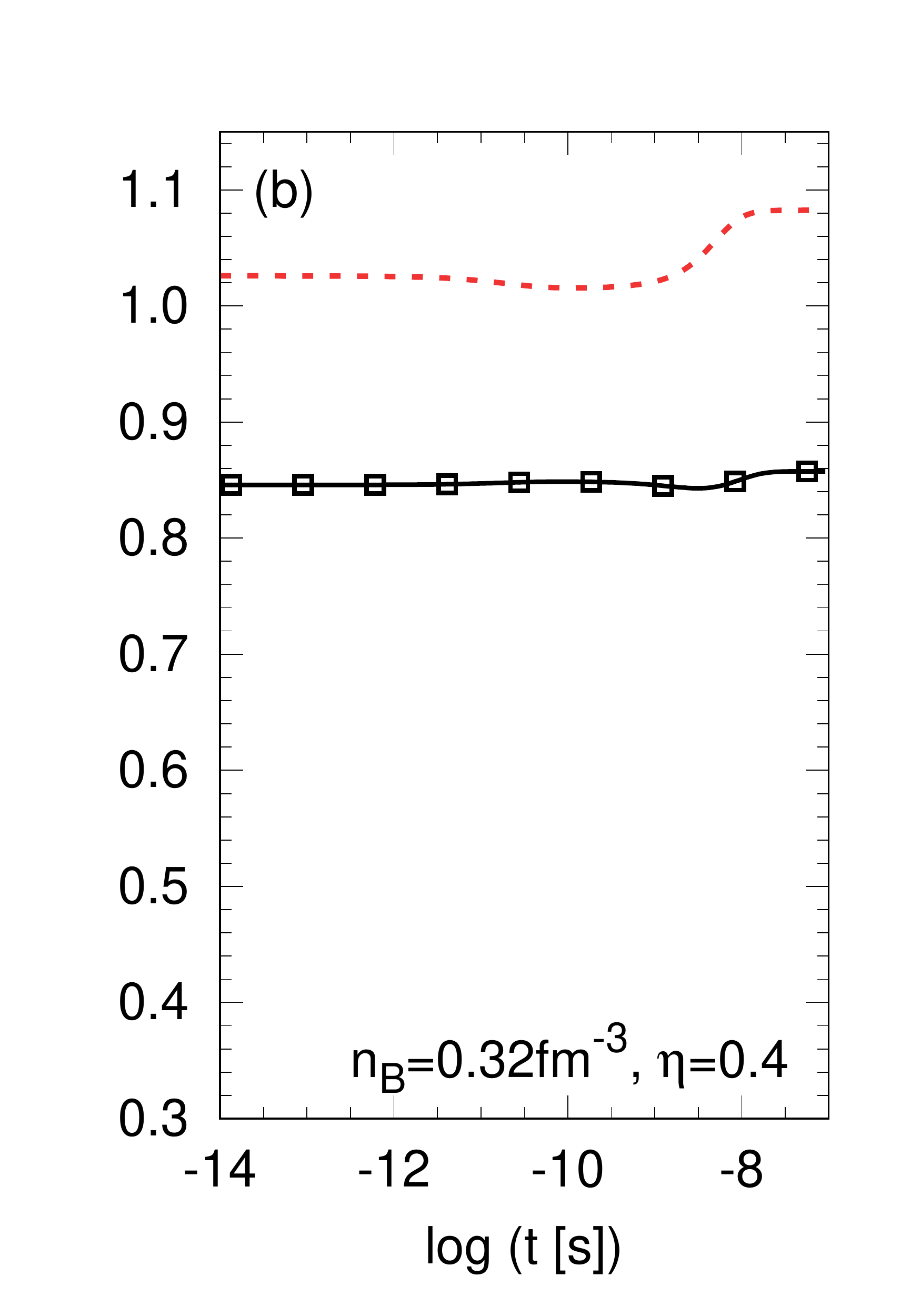}
\includegraphics[scale=0.22]{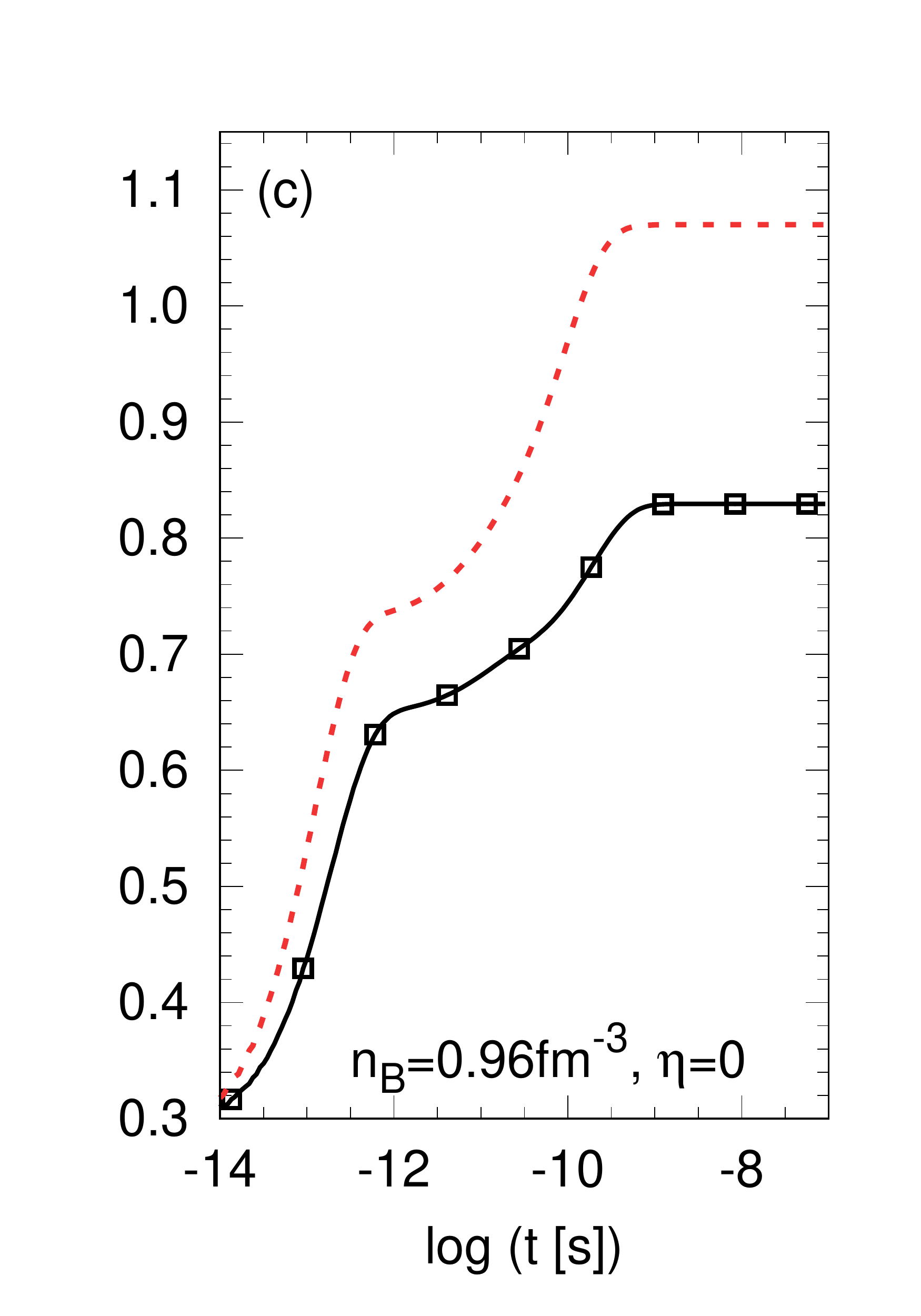}
\includegraphics[scale=0.22]{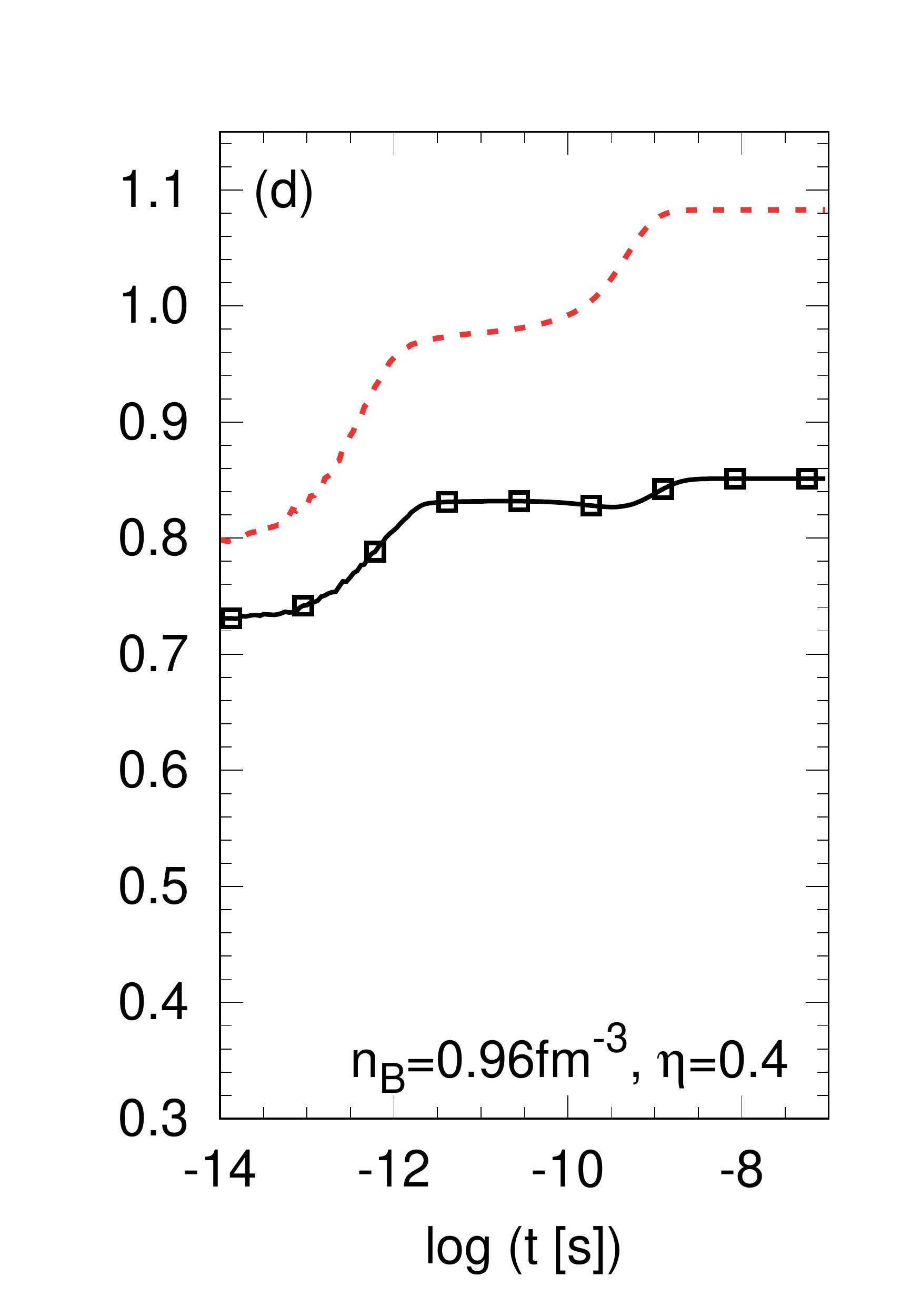}
\end{center}
\vspace{-0.6cm}
\caption{{We show $\Gamma_{V}^{dir*}/\Gamma_{V}^{dir}$ as a function of time for the nonleptonic process for cold NSs, where $\Gamma_{V}^{dir*}$ is the rate calculated by Madsen in \cite{Madsen:1993xx} and $\Gamma_{V}^{dir}$ is the one shown in Eq. (\ref{Gamma5}). The values of $n_B$ and $\eta$ are specified in the figure. We assumed $\alpha_c = 0$ (solid lines) and $\alpha_c$ = 0.47 (dashed lines).}}
\label{rate/rateNS}
\end{figure*}

\begin{figure*}[tb]
\begin{center}
\includegraphics[scale=0.22]{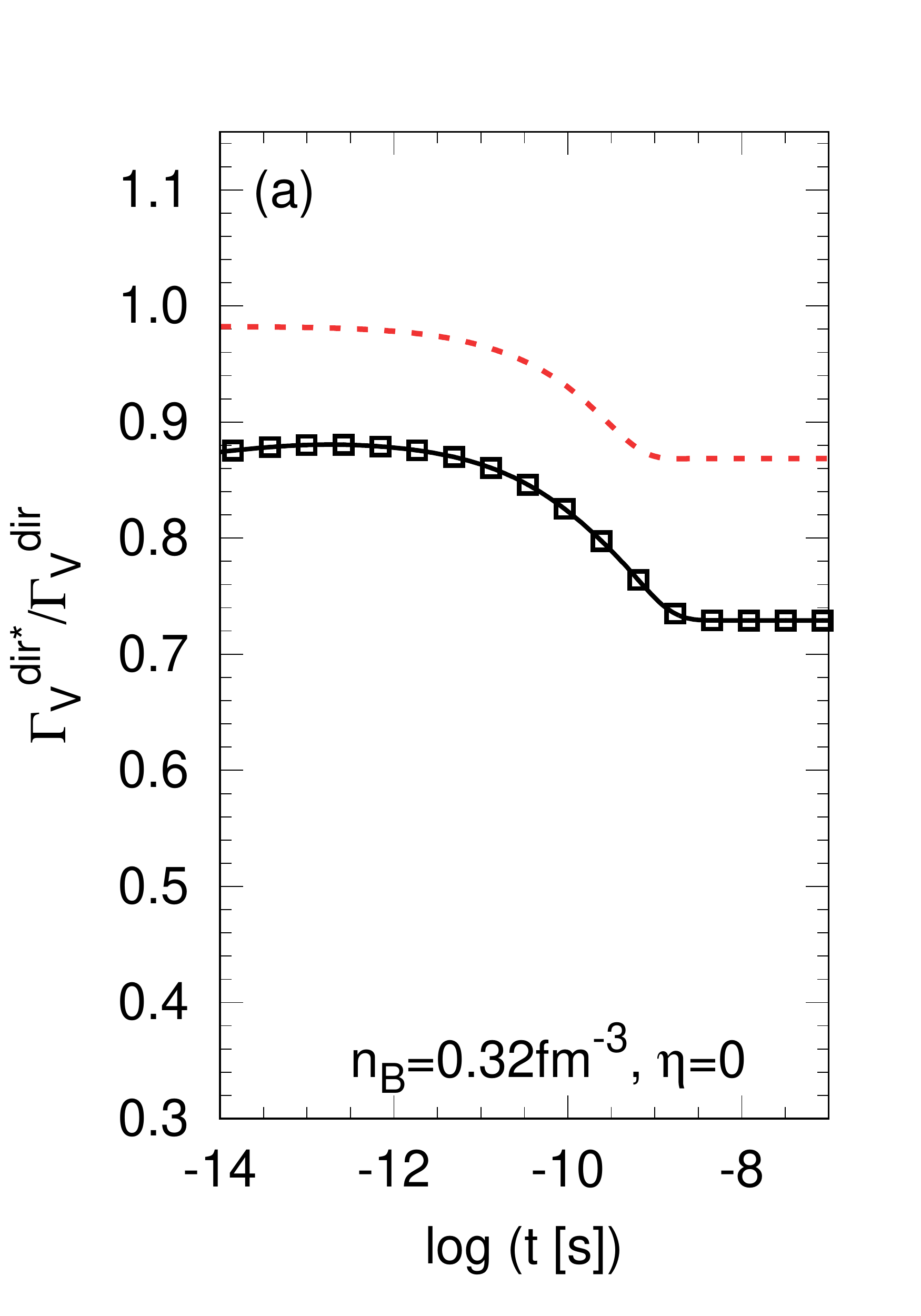}
\includegraphics[scale=0.22]{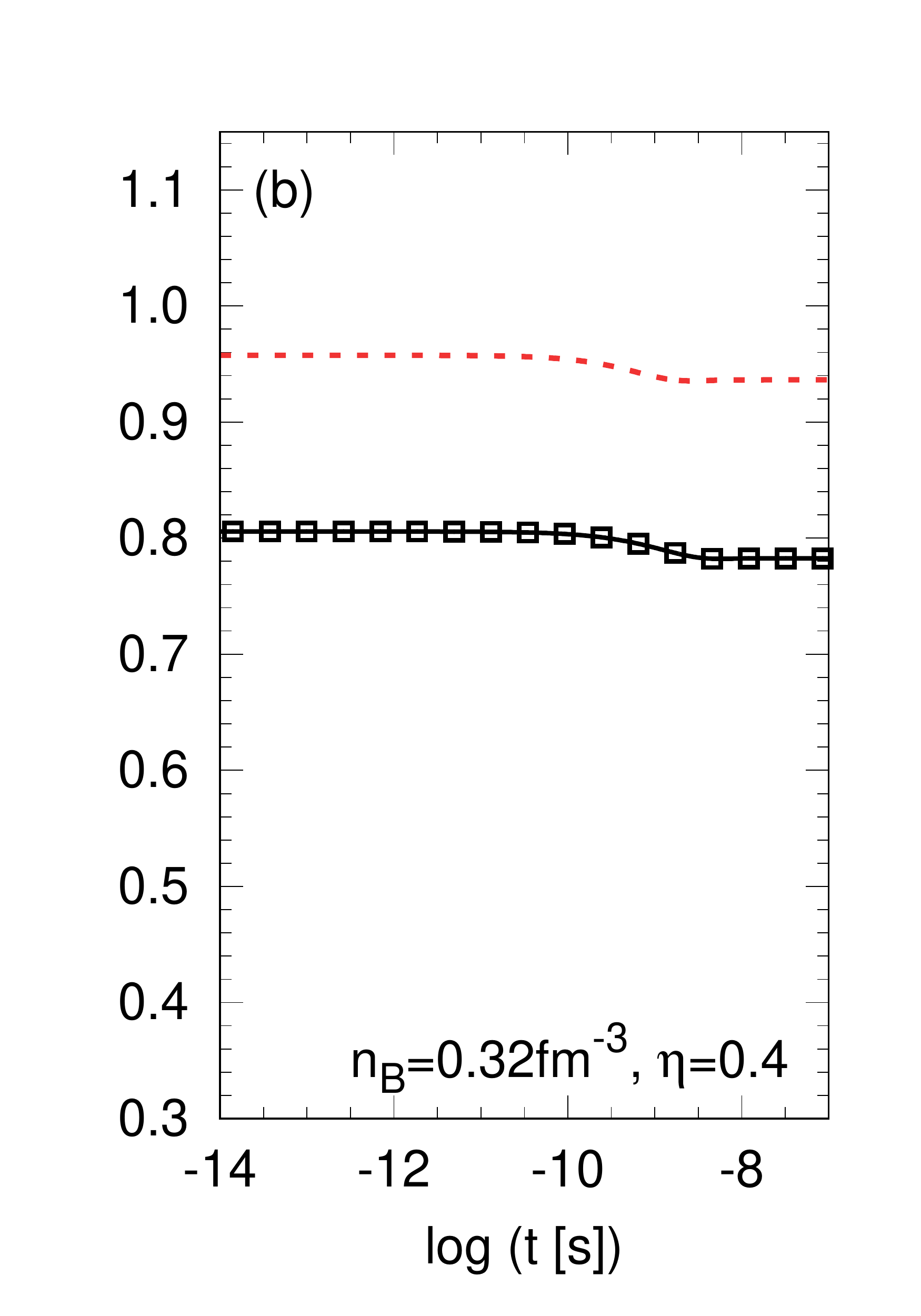}
\includegraphics[scale=0.22]{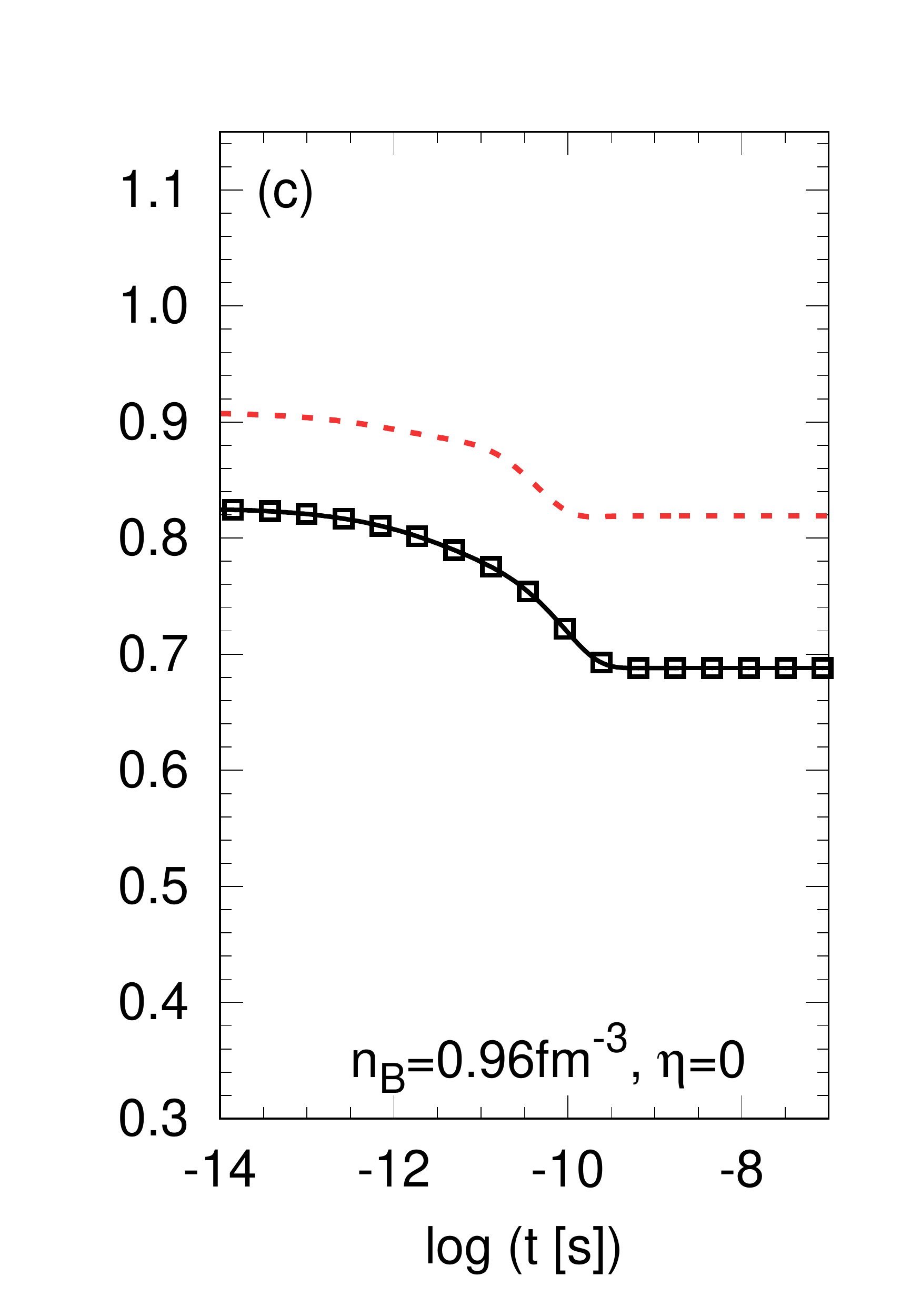}
\includegraphics[scale=0.22]{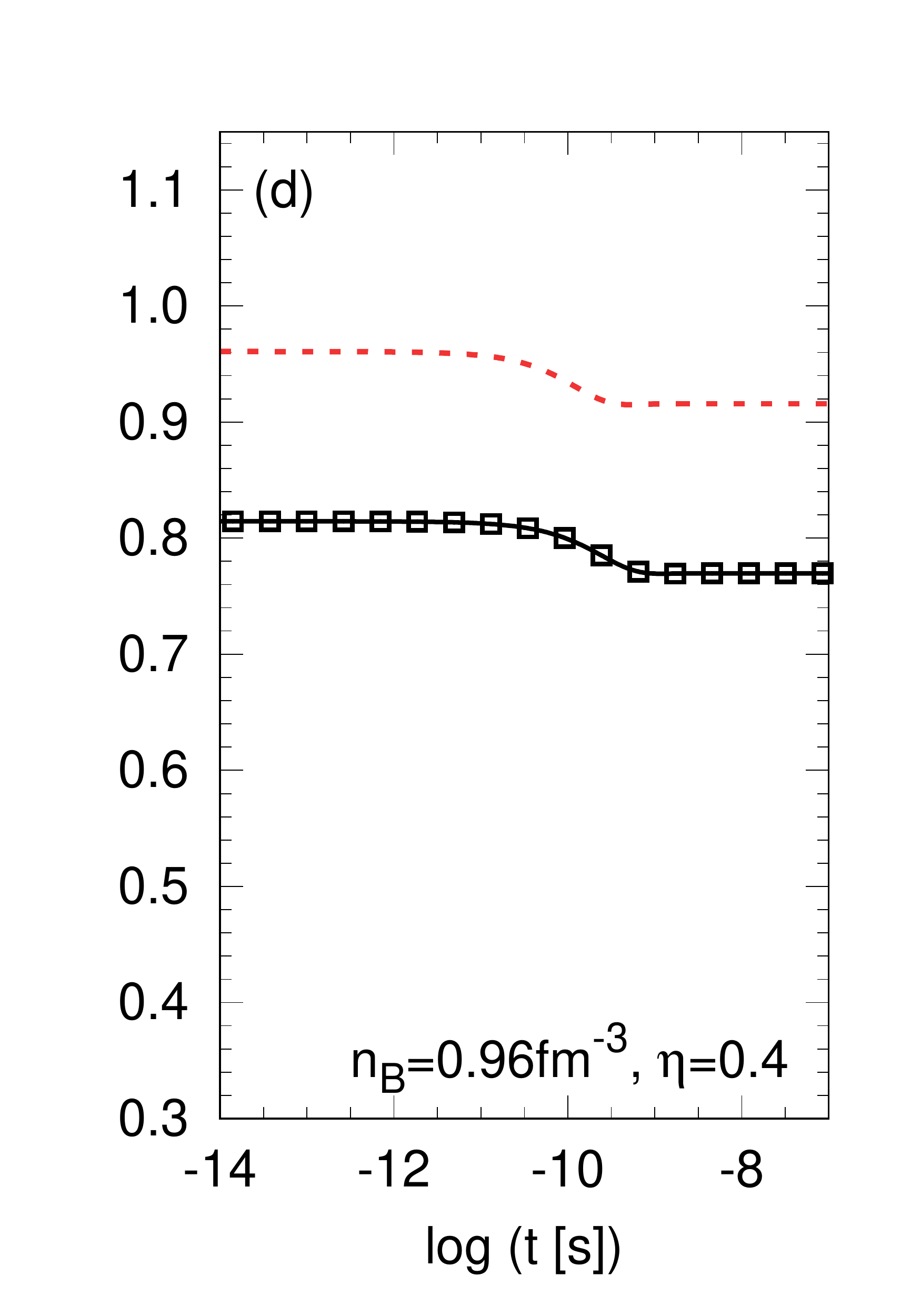}
\end{center}
\vspace{-0.6cm}
\caption{{$\Gamma_{V}^{dir*}/\Gamma_{V}^{dir}$ as a function of time for all the relevant process in a hot NS for $T_i= 20\, \mathrm{MeV}$. We used $\alpha_c = 0$ (solid lines) and $\alpha_c$ = 0.47 (dashed lines).}}
\label{rate/ratePNS20}
\end{figure*}

\begin{figure*}[tb]
\begin{center}
\includegraphics[scale=0.22]{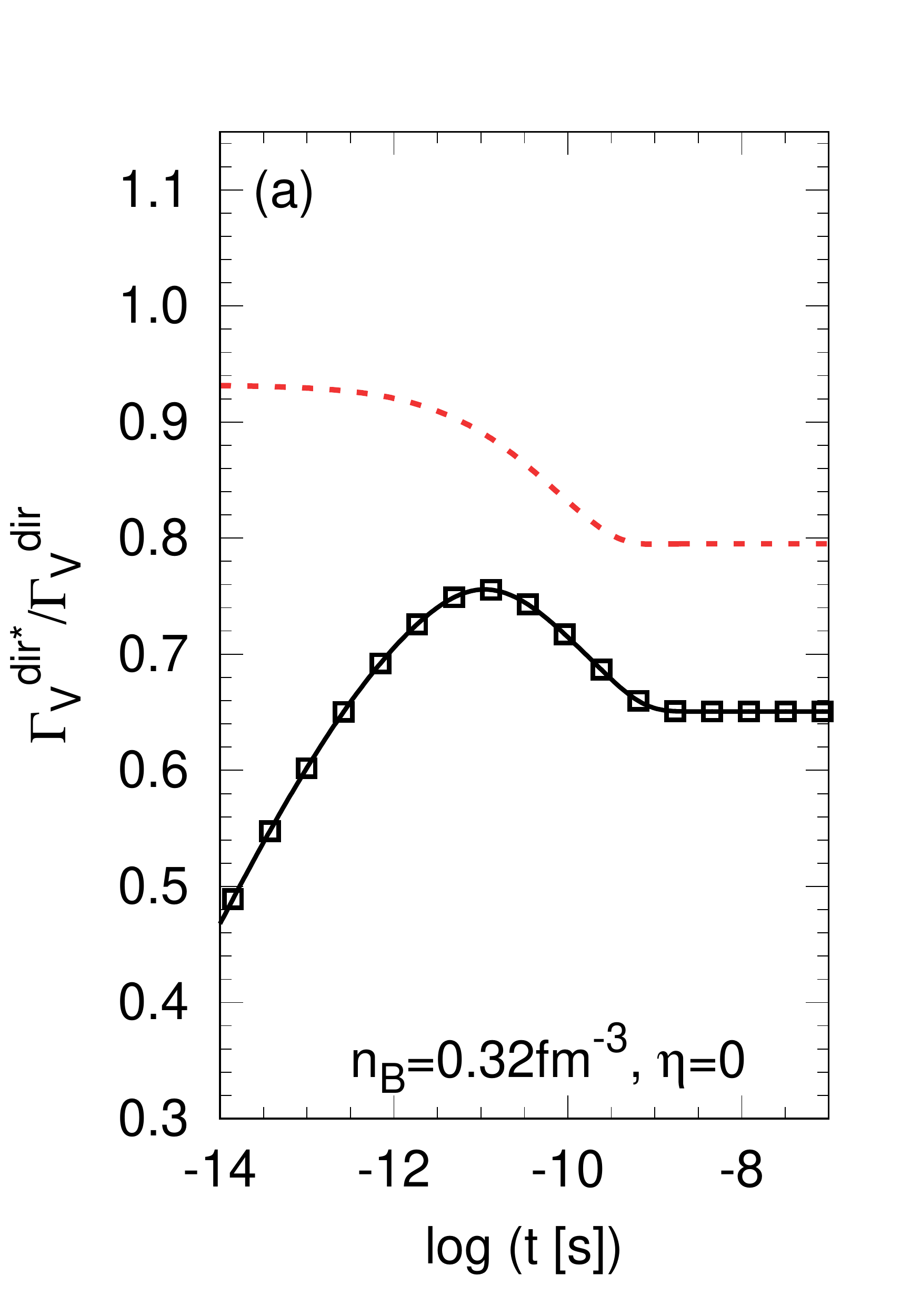}
\includegraphics[scale=0.22]{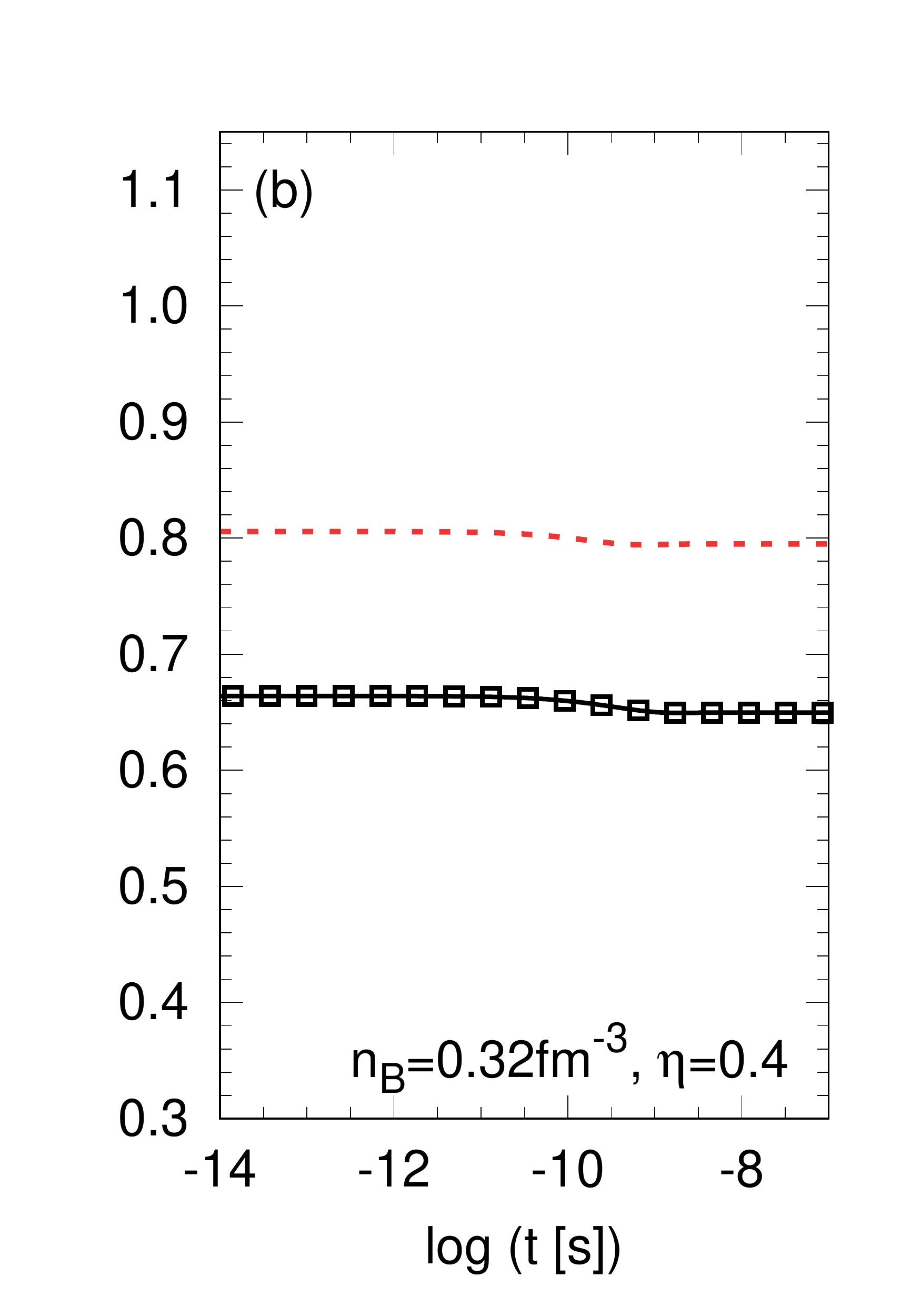}
\includegraphics[scale=0.22]{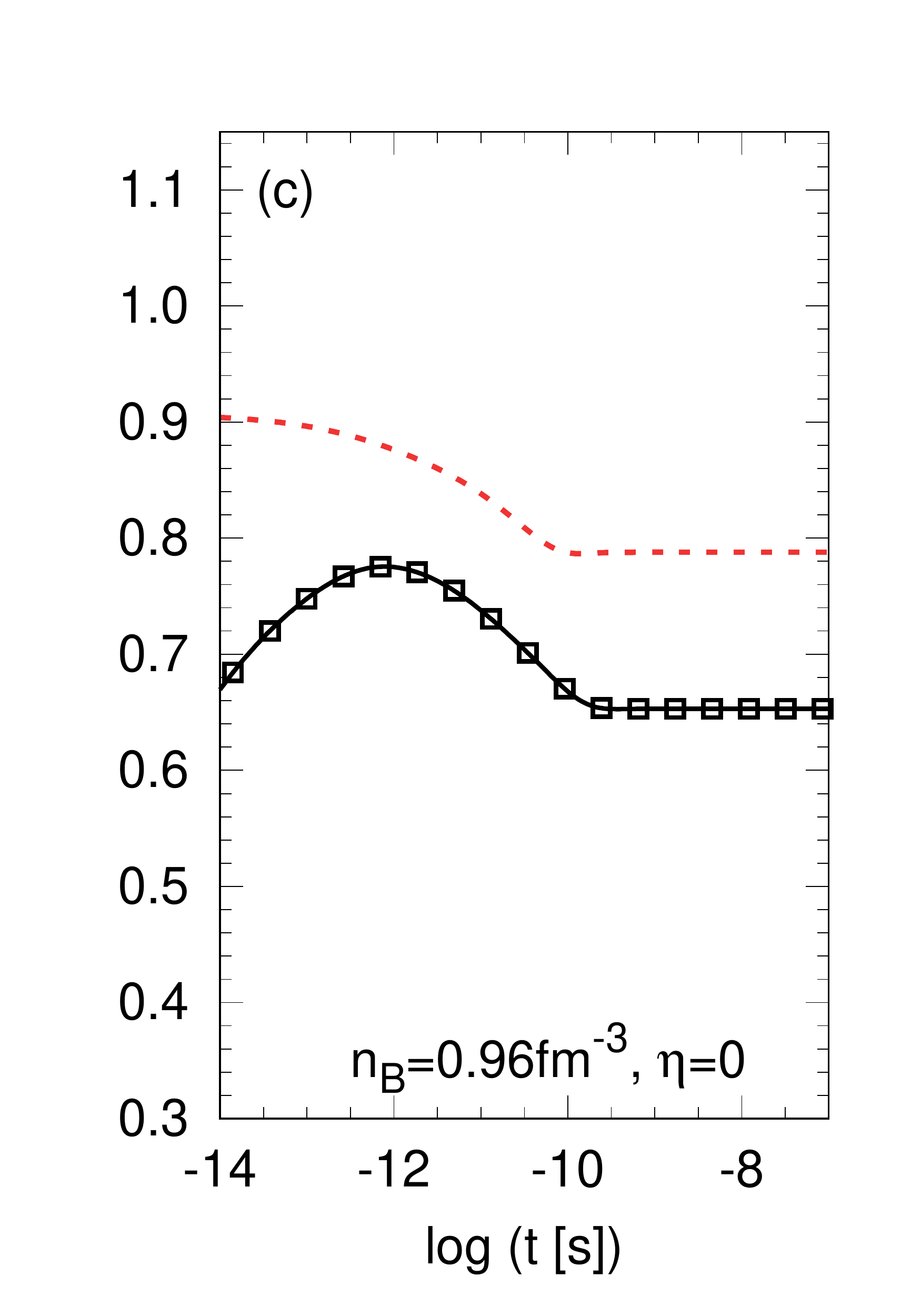}
\includegraphics[scale=0.22]{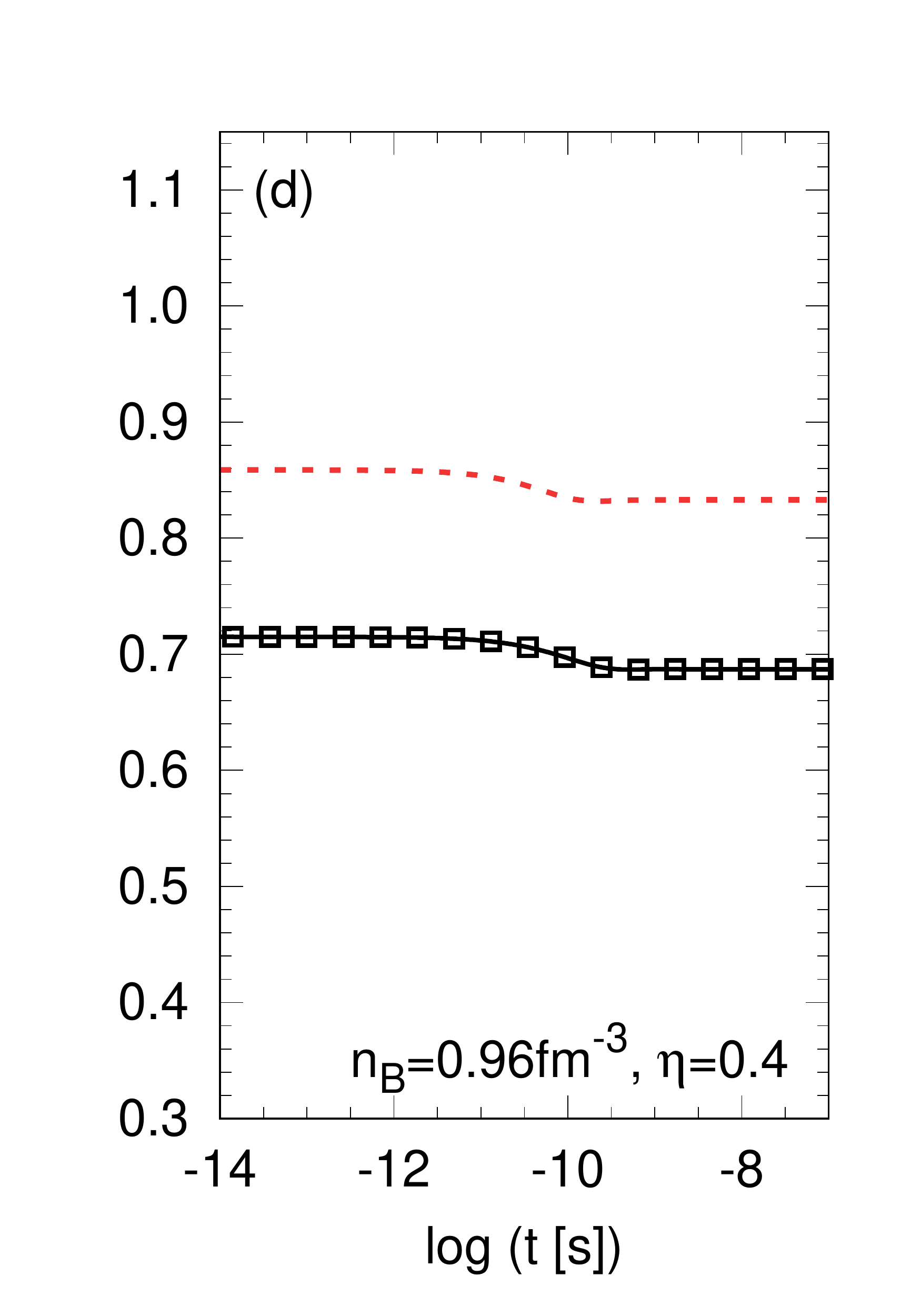}
\end{center}
\vspace{-0.6cm}
\caption{{Same as in Fig. \ref{rate/ratePNS20} but for an initial temperature $T_i= 40\, \mathrm{MeV}$}.}
\label{rate/ratePNS40}
\end{figure*}

{In this work we have used the approximate rates presented in the previous appendix, which assume quarks and electrons as degenerate for evaluating the matrix elements. In order to assess the error committed due to this assumption, we present here a comparison between the approximate and the exact rates \cite{Madsen:1993xx} for the dominant reaction, which in our case is the nonleptonic one, $u+d \rightarrow u+ s$. The largest errors occur at low temperatures and low strangeness, and because of that, they affect mainly the cold NSs scenario in the first $10^{-10}$ s after deconfinement. Once quark matter is hot or once the strangeness is large enough,  the approximation is very close to the exact calculation. The error is below $30 \%$ in almost all results of Figs. \eqref{rate/ratePNS20} and \eqref{rate/ratePNS40}, in Fig. \eqref{rate/rateNS} for $\eta =0.4$, and in Fig. \eqref{rate/rateNS} for $\eta =0$ at late times. }

%

%
%

%
%

\twocolumngrid
\bibliography{INSPIRE-Cite}

\end{document}